\documentclass[%
reprint,
nofootinbib,
 amsmath,amssymb,
 aip,
 jcp,
 superscriptaddress,
 floatfix,
 a4paper,
 longbibliography
]{revtex4-1}

\usepackage{graphicx}
\usepackage{dcolumn}
\usepackage{bm}
\usepackage{comment}
\usepackage{color}

\begin{document}

\title{Holstein polaron transport from numerically "exact" real-time quantum dynamics simulations}

\author{Veljko Jankovi\'c}%
 \email{veljko.jankovic@ipb.ac.rs}
\affiliation{%
 Institute of Physics Belgrade, University of Belgrade, Pregrevica 118, 11080 Belgrade, Serbia
}%
\begin{abstract}
Numerically "exact" methods addressing the dynamics of coupled electron--phonon systems have been intensively developed.
Nevertheless, the corresponding results for the electron mobility $\mu_\mathrm{dc}$ are scarce, even for the one-dimensional (1d) Holstein model.
Building on our recent progress on single-particle properties, here, we develop the momentum-space hierarchical equations of motion (HEOM) method to evaluate real-time two-particle correlation functions of the 1d Holstein model at finite temperature.
We compute numerically "exact" dynamics of the current--current correlation function up to real times sufficiently long to capture the electron's diffusive motion and provide reliable results for $\mu_\mathrm{dc}$ in a wide range of model parameters.
In contrast to the smooth ballistic-to-diffusive crossover in the weak-coupling regime, we observe a temporally limited slow-down of the electron on intermediate time scales already in the intermediate-coupling regime, which translates to a finite-frequency peak in the optical response.
Our momentum-space formulation lowers the numerical effort with respect to existing HEOM-method implementations, while we remove the numerical instabilities inherent to the undamped-mode HEOM by devising an appropriate hierarchy closing scheme.
Still, our HEOM remains unstable at too low temperatures, for too strong electron--phonon coupling, and for too fast phonons.
\end{abstract}

\maketitle

\section{Introduction}
The electron--phonon interaction governs the transport of charge (and energy) in systems ranging from semiconductors,~\cite{NatRevMater.6.560,AdvMater.33.2007057,Jacoboni,Rossibook,Mahanbook,Alexandrov-Devreese-book} organic molecular crystals and polymers,~\cite{ChemRev.107.926,AdvFunctMater.25.1915,Baessler-Koehler-book} to molecular aggregates relevant for photosynthesis.~\cite{PhysRep.567.1,RevModPhys.90.035003,Photosynthetic-excitons-book,May-Kuhn-book,Valkunas-Abramavicius-Mancal-book}
The simplest model of such diverse systems is the Holstein model,~\cite{AnnPhys.8.325} in which an electron is locally and linearly coupled to phonons.
Well-established transport theories are formulated as perturbative expansions from either the limit of vanishing coupling (Boltzmann-like~\cite{Mahanbook,Rossibook,RepProgPhys.83.036501} or Redfield-like~\cite{REDFIELD19651,May-Kuhn-book,Valkunas-Abramavicius-Mancal-book} theories) or vanishing electronic bandwidth (small-polaron/Lang--Firsov,~\cite{SovPhysJETP.16.1301} Marcus,~\cite{RevModPhys.65.599,May-Kuhn-book,Valkunas-Abramavicius-Mancal-book} or F\"{o}rster~\cite{Foerster1964,May-Kuhn-book,Valkunas-Abramavicius-Mancal-book} theories).
However, in many instances, the energy scales representative of the electron motion, phonons, electron--phonon interaction, and thermal fluctuations are all comparable to one another.~\cite{NatRevMater.6.560,AnnuRevPhysChem.66.305,AnnuRevCondensMatterPhys.3.333,AnnuRevPhysChem.66.69}
This circumstance calls for the development of methods beyond standard transport theories.

Such methods are typically formulated under physically motivated approximations.
Examples include the cumulant expansion,~\cite{PhysRevB.105.224304,PhysRevB.105.224305,arxiv.2212.13846,JChemPhys.142.094106,JChemPhys.157.095103} dynamical mean-field theory,~\cite{PhysRevB.56.4494,PhysRevB.63.153101,PhysRevLett.91.256403,PhysRevB.74.075101,PhysRevLett.129.096401} polaron transformation-based approaches,~\cite{JChemPhys.54.4843,JChemPhys.67.5818,JChemPhys.72.2763,PhysRevB.69.075212,jcp.128.114713,PhysRevB.79.235206,JPhysChemB.115.5312,AnnPhys.391.183,PhysRevB.99.104304,PhysRevX.10.021062} momentum-average approximation,~\cite{PhysRevLett.97.036402,PhysRevB.74.245104,PhysRevLett.107.076403} and kinetic Monte Carlo approaches.~\cite{ChemSci.12.2276,kassal36}
The approximate methods are generally computationally efficient and can thus be combined with electronic-structure methods to provide first-principles results on systems large enough that a direct comparison with experimental results is sensible.~\cite{ApplPhysLett.85.1535,NewJPhys.12.023011,PhysRevResearch.1.033138,JPhysChemB.124.8610,PhysRevB.104.085203,npjComputMater.8.63,npjComputMater.8.228}
While the agreement between numerical and experimental results justifies the approximations introduced, it does not fully reveal their domain of validity. 
This can be unveiled by comparison to results produced by numerically "exact" methods, which do not lean on any approximation beyond those in the Hamiltonian.
Since they are computationally intensive, numerically "exact" methods are usually applied to model Hamiltonians only.

The numerically "exact" approaches used to study interacting electron--phonon models may be roughly divided into: (i) quantum Monte Carlo (QMC) methods;~\cite{PhysRevB.27.6097,PhysRevB.30.1671,PhysRevLett.81.2514,PhysRevB.62.6317,Phys-Usp.48.887,PhysRevLett.114.146401,JChemPhys.156.204116,PhysRevB.107.184315} (ii) wavefunction-based methods considering the electron and phonons as a closed system, such as exact diagonalization (ED)-based techniques,~\cite{Prelovsek2013,PhysRevB.60.1633,PhysRevB.69.064302,PhysRevB.100.094307,PhysRevB.103.054304,PhysRevB.106.174303} the density-matrix renormalization group (DMRG),~\cite{PhysRevB.57.6376,PhysRevB.60.14092,WIREsComputMolSci.12.e1614,JChemTheoryComput.14.5027,JPhysChemLett.11.4930,JChemTheoryComput.18.6437,PhysRevB.102.165155,PhysRevB.106.155129} thermo-field dynamics,~\cite{JChemPhys.145.224101,SciRep.7.9127,WIREsComputMolSci.11.e1539}
and the hierarchy of Davydov's ans\"{a}tze;~\cite{JChemPhys.158.080901,WIREsComputMolSci.12.1589,JChemPhys.147.214102,AnnalenderPhysik.529.1600367,PhysChemChemPhys.19.25996}
(iii) methods leaning on the theory of open quantum systems, such as the hierarchical equations of motion (HEOM)~\cite{JPhysSocJpn.75.082001,PhysRevE.75.031107,JChemPhys.130.234111,JChemPhys.153.020901}, its generalizations~\cite{JChemPhys.148.014103,MolPhys.116.780} and hybridizations~\cite{JChemPhys.146.174105,JChemPhys.146.214105} with the stochastic Schr\"{o}dinger equation.~\cite{PhysLettA.235.569,JChemPhys.134.034902}

Many of the above-mentioned methods deliver practically exact results on ground-state or equilibrium finite-temperature properties of the Holstein model.~\cite{PhysRevB.57.6376,PhysRevLett.81.2514,PhysRevB.60.1633,PhysRevB.102.165155}
Some of them have recently been used to examine the model's single-particle properties (the electronic spectral function or linear absorption spectra).~\cite{PhysRevB.100.094307,PhysRevB.102.165155,PhysRevB.103.054304,PhysRevB.105.054311,PhysRevB.106.155129,PhysRevB.106.174303,JChemPhys.146.174105,JChemPhys.134.034902,PhysRevB.60.14092,JChemTheoryComput.14.5027}
However, much more demanding numerically "exact" evaluations of two-particle correlation functions at finite temperature, such as the ac and dc electrical conductivity, have started only recently.~\cite{PhysRevLett.114.146401,PhysRevB.106.155129,JChemPhys.156.204116,JChemPhys.146.214105,JPhysChemLett.11.4930,JChemTheoryComput.18.6437,JChemPhys.132.081101,JChemPhys.142.094106,JChemPhys.143.194106,JChemPhys.156.244102,PhysRevB.107.184315}
Each class of the above-mentioned numerically "exact" methods encounters certain issues in computations of two-particle quantities.
(i) QMC methods are usually formulated directly in the thermodynamic limit, but require the numerical analytical continuation to reconstruct real-frequency spectra.
A combination of statistical errors and uncertainties in the analytical continuation may lead to final results whose errors are comparable to the results themselves.
Real-time QMC simulations witness a progressive development of the infamous sign (or phase) problem, which limits their applicability to relatively short-time dynamics.  
(ii) ED-based methods are applied to small clusters and typically require artificial broadening parameters to construct real-frequency spectra.
The DMRG equations can be propagated only up to relatively short times,~\cite{PhysRevB.102.165155,PhysRevB.106.155129} which may not be long enough to reliably estimate the dc mobility.~\cite{PhysRevB.106.155129}
(iii) The HEOM method treats small clusters, but can, in principle, capture the full decay of correlation functions.~\cite{JChemPhys.132.081101,JChemPhys.142.174103,JChemPhys.143.194106,JChemPhys.156.244102}
Being based on a formally exact expression for the electronic reduced density matrix (RDM), the HEOM method is undoubtedly numerically "exact".
However, when employed on a finite system and truncated at a finite depth, the HEOM with undamped phonon modes suffers from numerical instabilities,~\cite{JChemPhys.150.184109} which practically limit the maximum propagation time~\cite{JPhysChemLett.6.3110} and whose overcoming requires further algorithmic developments.~\cite{JChemPhys.150.184109,JChemPhys.153.204109}

In this study, we provide numerically "exact" results for the electron mobility within the 1d Holstein model.
In contrast to the best currently available results, which are obtained by performing the numerical analytical continuation of imaginary-axis QMC data~\cite{PhysRevLett.114.146401} (possibly combined with real-time QMC data on rather short time scales~\cite{PhysRevB.107.184315}), our results entirely follow from real-time computations. 
We extend the momentum-space HEOM we developed in Ref.~\onlinecite{PhysRevB.105.054311} to follow the time evolution of finite-temperature two-particle correlation functions up to very long (practically infinite) real times.
In addition, our momentum-space HEOM also enables us to obtain highly accurate results for imaginary-time correlation functions, which are the central quantities in QMC simulations.
Our imaginary-axis results help us establish the minimum chain length and hierarchy depth needed to obtain the results representative of the thermodynamic limit.
The high quality of our real-time results is ensured by checking that different sum rules (e.g., the optical sum rule---OSR) are satisfied with a high accuracy.
We lower the numerical effort with respect to existing HEOM implementations by exploiting the translational symmetry, which reduces the number of independent dynamical variables in the formalism, and noting that the totally symmetric phonon mode does not contribute to the time evolution of correlation functions.
We avoid numerical instabilities in a wide range of parameter space by devising a specific closing of the hierarchy.
Still, the numerical instabilities inherent to the undamped-mode HEOM prevent us from obtaining results at low temperatures, for strong electron--phonon coupling, and when electronic dynamics is much slower than phonon dynamics (the so-called antiadiabatic regime).

The paper is organized as follows.
Section~\ref{Sec:theoretical_considerations} specifies the model and introduces our momentum-space HEOM method for the current--current correlation function.
Technical details are presented in Appendices~\ref{App:real-time-heom}--\ref{App:OSR_finite_N}.
In Sec.~\ref{Sec:Numerical_results}, we provide a number of numerical examples testing critical points of our methodology, and present our main results concerning temperature-dependent dc mobility.
Section~\ref{Sec:Conclusion} is devoted to a summary and prospects for future work.

\section{Theoretical framework}
\label{Sec:theoretical_considerations}
\subsection{Model and definitions}
\label{SSec:model_definitions}
We consider the Holstein model on the 1d lattice comprising $N$ sites with periodic boundary conditions.
In the momentum space, its Hamiltonian reads as
\begin{equation}
\label{Eq:def_H}
\begin{split}
    H&=H_\mathrm{e}+H_\mathrm{ph}+H_\mathrm{e-ph}\\&=\sum_k\varepsilon_k |k\rangle\langle k|+\sum_q\omega_q b_q^\dagger b_q+\sum_q V_qB_q.
\end{split}
\end{equation}
The electronic and phononic wave numbers $k$ and $q$ may assume any of the $N$ allowed values in the first Brillouin zone $-\pi<k,q\leq\pi$.
The Hamiltonian $H_\mathrm{e}$ describes an electron in a free-electron band whose dispersion $\varepsilon_k=-2J\cos(k)$ originates from the nearest-neighbor electronic hopping of amplitude $J$.
The Hamiltonian $H_\mathrm{ph}$ describes an optical-phonon branch with dispersion $\omega_q$ such that $\omega_{q=0}\neq 0$.
The interaction term $H_{\mathrm{e-ph}}$ is characterized by its strength $g$ and contains the purely electronic operator $V_q=\sum_k |k+q\rangle\langle k|$, which increases the electronic momentum by $q$, and the purely phononic operator $B_q=\frac{g}{\sqrt{N}}\left(b_q+b_{-q}^\dagger\right)$, which decreases the phononic momentum by $q$.
In the following, we set the lattice constant $a_l$, the elementary charge $e_0$, $\hbar$, and $k_B$ to unity. 

We focus on the dynamics of the current--current correlation function
\begin{equation}
\label{Eq:def_C_jj_t}
    C_{jj}(t)=\frac{1}{Z}\mathrm{Tr}\{ j(t)j(0)e^{-\beta H}\},
\end{equation}
where the current operator reads as
\begin{equation}
\label{Eq:def_j}
    j=-2J\sum_k\sin(k)|k\rangle\langle k|,
\end{equation}
$j(t)=e^{iHt}je^{-iHt}$, while $Z=\mathrm{Tr}\:e^{-\beta H}$ is the partition sum at temperature $T=\beta^{-1}$.
Its Fourier transform $C_{jj}(\omega)=\int_{-\infty}^{+\infty}dt\:e^{i\omega t}C_{jj}(t)$ [with $C_{jj}(-t)=C_{jj}(t)^*$] determines the frequency-dependent (or dynamical) mobility
\begin{equation}
\label{Eq:def_Re_mu_ac}
\begin{split}
    \mathrm{Re}\:\mu_\mathrm{ac}(\omega)&=\frac{1-e^{-\beta\omega}}{2\omega}C_{jj}(\omega)\\&=\frac{C_{jj}(\omega)-C_{jj}(-\omega)}{2\omega},
\end{split}
\end{equation}
where the second equality follows from the fluctuation--dissipation theorem for equilibrium correlation functions, $C_{jj}(-\omega)=e^{-\beta\omega}C_{jj}(\omega)$.
The dc mobility is $\mu_\mathrm{dc}=\lim_{\omega\to 0}\mathrm{Re}\:\mu_\mathrm{ac}(\omega)$ and may be computed using only real or only imaginary part of $C_{jj}(t)$~\cite{PhysRevB.90.155104}
\begin{equation}
\begin{split}
    \mu_\mathrm{dc}&=\frac{1}{T}\int_{0}^{+\infty}dt\:\mathrm{Re}\:C_{jj}(t)\\&=-2\int_{0}^{+\infty}dt\:t\:\mathrm{Im}\:C_{jj}(t).\label{Eq:mu_from_Im_C_jj_t}
\end{split}
\end{equation}

$C_{jj}(t)$ (for $t>0$) carries information on the carrier's dynamics resulting from a sudden ($\delta$-like) perturbation of the electron--phonon equilibrium.
Its real part is proportional to the velocity--velocity anticommutator correlation function, which is the quantum counterpart of the velocity--velocity correlation function used, e.g., to study the Brownian motion.
The mean-square displacement (MSD) of the carrier's position, $\Delta x^2(t)=\langle[x(t)-x(0)]^2\rangle$, where $\langle\dots\rangle$ denotes averaging with respect to $e^{-\beta H}/Z$, grows at a rate determined by the time-dependent diffusion constant~\cite{RevModPhys.93.025003}
\begin{equation}
\label{Eq:def_mathcal_D_t}
    \mathcal{D}(t)=\frac{1}{2}\frac{d}{dt}\Delta x^2(t)=\int_0^t ds\:\mathrm{Re}\:C_{jj}(s).
\end{equation}
Within the model considered here, $\mathrm{Re}\:C_{jj}(t)$ decays to zero in the long-time limit, so that $\mathcal{D}(t)$ varies from 0 at short times to $\mathcal{D}_\infty=\int_0^{+\infty}ds\:\mathrm{Re}\:C_{jj}(s)$ at long times, where $\mathcal{D}_\infty$ is the diffusion constant related to the dc mobility by the Einstein relation (in the units we use, $\mu_\mathrm{dc}=\mathcal{D}_\infty/T$).
The electron's dynamics then exhibits a crossover from short-time ballistic dynamics, when $\Delta x^2(t)=C_{jj}(0)t^2$ and $\mathcal{D}(t)=C_{jj}(0)t$, to long-time diffusive dynamics, when $\Delta x^2(t)=2\mathcal{D}_\infty t$ and $\mathcal{D}(t)=\mathcal{D}_\infty$.
In addition to $\mathcal{D}(t)$, another quantity useful to describe this crossover is the diffusion exponent $\alpha(t)\geq 0$ defined by assuming that the power-law scaling $\Delta x^2(t)\propto t^{\alpha(t)}$ holds locally around instant $t$,~\cite{PhysRevLett.114.086601} so that
\begin{equation}
\label{Eq:def_alpha_t}
    \alpha(t)=\frac{2t\mathcal{D}(t)}{\Delta x^2(t)}.
\end{equation}
At short times, $\alpha(t)$ is close to 2, while it reaches the value of unity in the long-time diffusive limit.

\subsection{HEOM for the real-time current--current correlation function}
\label{SSec:HEOM_real_time_theory}
We formulate the HEOM method for the purely electronic operator
\begin{equation}
\label{Eq:def_iota}
    \iota(t)=\frac{1}{Z}\mathrm{Tr}_\mathrm{ph}\left\{e^{-iHt}je^{-\beta H}e^{iHt}\right\},
\end{equation}
while $C_{jj}(t)=\mathrm{Tr}_\mathrm{e}\left\{j\iota(t)\right\}$.
Since the totally symmetric phonon mode ($q=0$ mode) couples to the unit operator in the electronic subspace, it does not affect the dynamics of $\iota(t)$, while its contributions to $Z$ and $e^{-\beta H}$ cancel out after performing the partial trace over phonons.
This decoupling of $q=0$ phonon mode from the rest of phonon modes and the electronic states~\cite{JChemPhys.95.1588} somewhat lowers the number of auxiliary density operators (ADOs) $\iota_\mathbf{n}^{(n)}(t)$, which are characterized by the vector $\mathbf{n}=\{n_{qm}|q\neq 0,\:m=0,1\}$ containing $2(N-1)$ nonnegative integers $n_{qm}$ counting individual phonon absorption and emission events whose total number is $n=\sum'_{qm}n_{qm}$.
The prime on the sum indicates the omission of the $q=0$ term.
The momentum conservation implies that the ADO $\iota_\mathbf{n}^{(n)}(t)$ changes the electronic momentum by $k_\mathbf{n}=\sum'_{qm}qn_{qm}$, so that only $N$ of its $N^2$ matrix elements are nonzero.
The only nonzero matrix elements of $\iota_\mathbf{n}^{(n)}(t)$ are the ones connecting the states whose momenta differ by $k_\mathbf{n}$.
This requirement leads to a drastic reduction in the number of equations with respect to existing real-space HEOM formulations.~\cite{JChemPhys.132.081101,JChemPhys.156.244102,JPhysChemLett.6.3110}
A more detailed discussion in this direction is deferred for the last paragraph of Sec.~\ref{SSec:finite_N_D_effects}.   

The dynamics of $\iota(t)$ follows from the real-time HEOM
\begin{equation}
\label{Eq:real-time-HEOM-before-closing-paper}
    \begin{split}
        &\partial_{t}\langle k|{\iota}_\mathbf{n}^{(n)}({t})|k+k_\mathbf{n}\rangle=\\
        &-i(\varepsilon_k-\varepsilon_{k+k_\mathbf{n}}+{\mu}_\mathbf{n})\langle k|{\iota}_\mathbf{n}^{(n)}({t})|k+k_\mathbf{n}\rangle\\
        &+i\sideset{}{'}\sum_{qm}\sqrt{(1+n_{qm})c_{qm}}\:\langle k-q|{\iota}_{\mathbf{n}_{qm}^+}^{(n+1)}({t})|k+k_\mathbf{n}\rangle\\
        &-i\sideset{}{'}\sum_{qm}\sqrt{(1+n_{qm})c_{qm}}\:\langle k|{\iota}_{\mathbf{n}_{qm}^+}^{(n+1)}({t})|k+k_\mathbf{n}+q\rangle\\
        &+i\sideset{}{'}\sum_{qm}\sqrt{n_{qm}c_{qm}}\:\langle k+q|{\iota}_{\mathbf{n}_{qm}^-}^{(n-1)}({t})|k+k_\mathbf{n}\rangle\\
        &-i\sideset{}{'}\sum_{qm}\sqrt{n_{qm}}\frac{c_{q\overline{m}}}{\sqrt{c_{qm}}}\:\langle k|{\iota}_{\mathbf{n}_{qm}^-}^{(n-1)}({t})|k+k_\mathbf{n}-q\rangle,
    \end{split}
\end{equation}
where ${\mu}_\mathbf{n}=\sum'_{q}\omega_q(n_{q0}-n_{q1})$.
The ADO $\iota_\mathbf{n}^{(n)}(t)$ couples to ADOs at depths $n\pm 1$, which are characterized by vectors $\mathbf{n}_{qm}^\pm$ whose components are $\left[\mathbf{n}_{qm}^\pm\right]_{q'm'}=n_{q'm'}\pm\delta_{q'q}\delta_{m'm}$.
The coefficients $c_{qm}$ and $c_{q\overline{m}}$ are defined in Eqs.~\eqref{Eq:def_c_q0} and~\eqref{Eq:def_c_q1} of Appendix~\ref{App:real-time-heom}, where we provide a detailed derivation of Eq.~\eqref{Eq:real-time-HEOM-before-closing-paper}.

The initial condition for Eq.~\eqref{Eq:real-time-HEOM-before-closing-paper} is set by the equilibrium state of the interacting electron--phonon system, see also Eq.~\eqref{Eq:def_iota}.
In our previous publication,~\cite{PhysRevB.105.054311} we derived that the hierarchical representation of that equilibrium state can be obtained from the following imaginary-time HEOM:
\begin{equation}
    \label{Eq:im-time-HEOM-eq-paper}
    \begin{split}
        &\partial_{{\tau}}\langle k|\sigma_\mathbf{n}^{(n)}({\tau})|k+k_\mathbf{n}\rangle=\\
        &-(\varepsilon_k+{\mu}_\mathbf{n})\langle k|\sigma_\mathbf{n}^{(n)}({\tau})|k+k_\mathbf{n}\rangle\\
        &+\sideset{}{'}\sum_{qm}\sqrt{(1+n_{qm})c_{qm}}\langle k-q|\sigma_{\mathbf{n}_{qm}^+}^{(n+1)}({\tau})|k+k_\mathbf{n}\rangle\\
        &+\sideset{}{'}\sum_{qm}\sqrt{n_{qm}c_{qm}}\langle k+q|\sigma_{\mathbf{n}_{qm}^-}^{(n-1)}({\tau})|k+k_\mathbf{n}\rangle.
    \end{split}
\end{equation}
Equation~\eqref{Eq:im-time-HEOM-eq-paper} is propagated in imaginary time $\tau$ from 0 to $\beta$ with the infinite-temperature initial condition $\langle k|\sigma_\mathbf{n}^{(n)}(0)|k+k_\mathbf{n}\rangle=\delta_{n,0}$.
The initial condition for Eq.~\eqref{Eq:real-time-HEOM-before-closing-paper} is finally
\begin{equation}
\label{Eq:init_cond_Cjj}
\begin{split}
    &\langle k|{\iota}_\mathbf{n}^{(n)}(0)|k+k_\mathbf{n}\rangle=
    Z_e^{-1}\times\\&(-2J)\sin(k)\langle k|\sigma_\mathbf{n}^{(n)}(\beta)|k+k_\mathbf{n}\rangle,
\end{split}
\end{equation}
where the so-called electronic partition sum reads as
\begin{equation}
\label{Eq:el_part_sum}
    Z_\mathrm{e}=\sum_p\langle p|\sigma_\mathbf{0}^{(0)}(\beta)|p\rangle.
\end{equation}
A more detailed derivation is provided in Appendix~\ref{App:real-time-heom}.
Here, let us emphasize that the structure of the imaginary-time HEOM in Eq.~\eqref{Eq:im-time-HEOM-eq-paper} is fully compatible with the structure of the real-time HEOM in Eq.~\eqref{Eq:real-time-HEOM-before-closing-paper}.
This is different from existing approaches, in which the structures of the imaginary-time and real-time HEOM are not manifestly identical~\cite{JChemPhys.141.044114,JChemPhys.156.244102} and which may require appropriate rearrangement steps to obtain the initial condition for the real-time HEOM.~\cite{JChemPhys.141.044114}

\subsection{Closing the HEOM for the real-time current--current correlation function}
When truncated at a finite maximum depth $D$, the real-time HEOM in Eq.~\eqref{Eq:real-time-HEOM-before-closing-paper} suffers from numerical instabilities appearing at sufficiently long times, which are commonly ascribed to the discrete and undamped nature of phonons.~\cite{JChemPhys.150.184109}
Such instabilities have been observed even for not too strong couplings and at not too low temperatures.
The instabilities are particularly detrimental to evaluations of $\mu_\mathrm{dc}$, for which we need $C_{jj}(t)$ up to times so long so that it has decayed almost to zero [Eq.~\eqref{Eq:mu_from_Im_C_jj_t}].
They may be eliminated by projecting out the unstable eigenmodes of the truncated HEOM,~\cite{JChemPhys.150.184109} which requires numerically complicated filtration algorithms, or deriving a new hierarchy of equations,~\cite{JChemPhys.153.204109} whose in general nontrivial relation to the original hierarchy may complicate evaluations of physically relevant quantities.
The numerical instabilities reported in Ref.~\onlinecite{JChemPhys.150.184109} were observed under the so-called time-nonlocal (TNL) truncation scheme, which sets all ADOs at depths $n>D$ to zero.
One may thus hope that an appropriate closing of the HEOM at the maximum depth could eliminate numerical instabilities.
While a number closing schemes have been proposed recently,~\cite{JChemPhys.142.104112,MolPhys.116.780,JChemPhys.157.054108} we find that none of them stabilizes Eq.~\eqref{Eq:real-time-HEOM-before-closing-paper}.
A possible reason behind our observation is that these schemes were tried and tested for the electron coupled to a phonon bath, i.e., when the spectral density of the electron--phonon interaction is a continuous function of the energy exchanged.
Here, on the other hand, we deal with a discrete spectral density consisting of a finite number $\delta$ peaks at $\pm\omega_q$.
In Appendix~\ref{App:closing}, we build on previous density-matrix studies~\cite{RevModPhys.70.145,RevModPhys.74.895,PhysRevB.92.235208} and derive in detail a specific closing of Eq.~\eqref{Eq:real-time-HEOM-before-closing-paper} that permits us to overcome the instabilities in many (but not all!) parameter regimes.
Here, let us only mention that we eliminate the ADOs at depth $D+1$ from the equations at the maximum depth $D$ by (i) setting the ADOs with $n\geq D+2$ to zero, (ii) solving the resulting equations at depth $D+1$, which then contain only the ADOs at depth $D$, in the Markov and adiabatic approximations.
The structure of the resulting equations at depth $D$ is, however, more involved than the structure of Eq.~\eqref{Eq:real-time-HEOM-before-closing-paper} because the ADOs at depth $D$ become mutually coupled.
We eliminate these equal-depth couplings by resorting to the random phase approximation, which neglects momentum-averaged matrix elements of the ADOs at depth $D$ due to random phases at different momenta.
The above-described procedure of closing the HEOM results in equations for maximum-depth ADOs ($n=D$) that feature exponential damping terms
\begin{equation}
\label{Eq:closing_strategy}
\begin{split}
    &\left[\partial_{{t}}\langle k|{\iota}_\mathbf{n}^{(n)}({t})|k+k_\mathbf{n}\rangle\right]_\mathrm{close}=\\&-\delta_{n,D}\frac{1}{2}\left({\tau}_k^{-1}+{\tau}_{k+k_\mathbf{n}}^{-1}\right)\langle k|{\iota}_\mathbf{n}^{(n)}({t})|k+k_\mathbf{n}\rangle,
\end{split}
\end{equation}
where $\tau_k$ is the carrier scattering time in the second-order perturbation theory and the long-chain limit [see Eq.~\eqref{Eq:def_tau_k} and Ref.~\onlinecite{PhysRevB.99.104304}].

In our numerical computations, we assume that phonons are dispersionless, $\omega_q\equiv\omega_0$. For $\omega_0/J\geq 2$, the closing scheme in Eq.~\eqref{Eq:closing_strategy} cannot fully remove the numerical instabilities inherent to the undamped-mode HEOM because the carrier scattering times then become infinite for $k$ states in the vicinity of $\pm\pi/2$.~\cite{PhysRevB.99.104304,arxiv.2212.13846}
We thus obtain HEOM results for $\omega_0/J\geq 2$ only at sufficiently high temperatures and for sufficiently (but not excessively) strong interactions.
For $\omega_0/J<2$, our results summarized in Secs.~\ref{SSec:closing_reliability} and~\ref{SSec:mu_vs_T} show that the closing scheme in Eq.~\eqref{Eq:closing_strategy} removes the instabilities of Eq.~\eqref{Eq:real-time-HEOM-before-closing-paper} for not too strong $g$ or at not too low $T$.
The instabilities remain for strong $g$ and at low $T$, while their relatively early appearance prevents us from reliably computing $\mu_\mathrm{dc}$ in such parameter regimes, see Sec.~\ref{SSec:instable_strong_g}.

\subsection{HEOM for the imaginary-time current--current correlation function}
In order for $C_{jj}(t)$ to be representative of the long-chain limit and take all relevant phonon-assisted processes into account, both $N$ and $D$ should be sufficiently large.
The first proxy to how large $N$ and $D$ should be follows from analyzing the current--current correlation function $C_{jj}(\tau)$ in imaginary time.

While $C_{jj}(\tau)$ is directly accessible in QMC simulations,~\cite{PhysRevB.107.184315,JChemPhys.142.174103,PhysRevLett.114.146401} its evaluation using the HEOM method has not been considered so far, to the best of our knowledge.
The appropriate imaginary-time HEOM is obtained from Eq.~\eqref{Eq:real-time-HEOM-before-closing-paper} by performing the Wick's rotation ${t}\to-i{\tau}$.
We employ the TNL truncation of the HEOM thus obtained, i.e., we simply set all the ADOs with $n>D$ to zero.
$C_{jj}(\tau)$ is to be determined on the interval $[0,\beta]$, on which it is symmetric with respect to $\beta/2$.
To enable as accurate as possible evaluation of $C_{jj}(\tau)$, we find it useful to consider the symmetrized correlation function
\begin{equation}
    C_{jj}^\mathrm{sym}(\tau)=\frac{1}{Z}\mathrm{Tr}\left\{e^{-\beta H/2}e^{H\tau}je^{-H\tau}e^{-\beta H/2}j\right\}
\end{equation}
on the interval $[-\beta/2,\beta/2]$, which is related to $C_{jj}(\tau)$ via $C_{jj}(\tau)=C_{jj}^\mathrm{sym}(\tau-\beta/2)$ for $0\leq\tau\leq\beta$.
We note that the use of the symmetrized correlation function instead of the standard one is advantageous in real-time QMC simulations,~\cite{CompPhysCommun.63.415} and is also reported to be useful in real-time HEOM computations.~\cite{JChemPhys.143.194106,JChemPhys.156.244102}
Nevertheless, we find that the numerical instabilities of the real-time HEOM [Eqs.~\eqref{Eq:real-time-HEOM-before-closing-paper} and~\eqref{Eq:closing_strategy}] are reflected on both $C_{jj}(t)$ and $C_{jj}^\mathrm{sym}(t)$ in the same manner, which is the reason why we consider the non-symmetrized correlation function [Eq.~\eqref{Eq:def_C_jj_t}] in all our real-time computations.

We determine $C_{jj}^\mathrm{sym}(\tau)$ by two independent imaginary-time propagations: one forward from 0 to $\beta/2$, and the other backward from 0 to $-\beta/2$. The initial condition (at $\tau=0$) for both propagations is obtained from the imaginary-time HEOM for the equilibrium state of the coupled electron--phonon system [Eq.~\eqref{Eq:im-time-HEOM-eq-paper}], which we (i) propagate from 0 to $\beta/2$, (ii) multiply by $j$ from the left, and (iii) propagate once again from 0 to $\beta/2$.
Since $C_{jj}^\mathrm{sym}(\tau)$ is symmetric around 0, the results of the forward and backward propagation in imaginary time should coincide.
We use this fact to gain insight into the maximum hierarchy depth $D$ that is needed to obtain converged results.
In more detail, we find that the relative deviation ($-\beta/2\leq\tau\leq\beta/2$)
\begin{equation}
\label{Eq:delta_jj_sym}
 \delta_{jj}^\mathrm{sym}(\tau)=2\frac{|C_{jj}^\mathrm{sym}(\tau)-C_{jj}^\mathrm{sym}(-\tau)|}{C_{jj}^\mathrm{sym}(\tau)+C_{jj}^\mathrm{sym}(-\tau)}   
\end{equation}
decreases with increasing $D$.
Furthermore, following how $C_{jj}(\tau)=[C_{jj}^\mathrm{sym}(\tau-\beta/2)+C_{jj}^\mathrm{sym}(\beta/2-\tau)]/2$ ($0\leq\tau\leq\beta$) changes with the chain length $N$ provides information about the minimum chain length $N$ needed to obtain results representative of the thermodynamic limit.
We, however, emphasize that the estimates for $D$ and $N$ that stem from imaginary-axis data may differ from the corresponding estimates originating from real-time data.
Whenever possible, we also perform real-time simulations for multiple $N$ and $D$ to ensure that our real-time results are also representative of the thermodynamic limit.

\subsection{Sum rules}
\label{SSec:theory_sum_rules}
To gain additional confidence in our HEOM results, we check that certain sum rules for frequency-resolved quantities are satisfied with sufficient accuracy.~\cite{PhysRevB.106.155129,PhysRevB.72.104304}
We compare the moments of frequency-dependent quantities evaluated by (i) numerical integration over frequency, and (ii) averaging an appropriate operator in the equilibrium state of the coupled electron and phonons.
We consider the OSR
\begin{equation}
\label{Eq:OSR}
    \int_{0}^{+\infty}d\omega\:\mathrm{Re}\:\mu_\mathrm{ac}(\omega)=-\frac{\pi}{2}\langle H_\mathrm{e}\rangle
\end{equation}
and the sum rules
\begin{equation}
\label{Eq:freq_sum_rules}
    \int_{-\infty}^{+\infty}\frac{d\omega}{2\pi}\:\omega^n\:C_{jj}(\omega)=M_n
\end{equation}
for the first three moments ($n=0,1,2$) of the real-frequency current--current correlation function.
In Eq.~\eqref{Eq:OSR}, we use the dynamical mobility computed using the first equalities in Eqs.~\eqref{Eq:def_Re_mu_ac} (for $\omega\neq 0$) and~\eqref{Eq:mu_from_Im_C_jj_t} (for $\omega=0$).
The electron's kinetic energy entering Eq.~\eqref{Eq:OSR} is evaluated using the quantities defined in Eqs.~\eqref{Eq:im-time-HEOM-eq-paper} and~\eqref{Eq:el_part_sum}:~\cite{PhysRevB.105.054311}
\begin{equation}
\label{Eq:avg_H_e}
    \langle H_\mathrm{e}\rangle=\frac{1}{Z_\mathrm{e}}\sum_k\varepsilon_k \langle k|\sigma_\mathbf{0}^{(0)}(\beta)|k\rangle.
\end{equation}
We note that the initial conditions $\iota_\mathbf{n}^{(n)}(t=0)$ [Eq.~\eqref{Eq:init_cond_Cjj}] for the real-time HEOM [Eq.~\eqref{Eq:real-time-HEOM-before-closing-paper}] do not enter Eq.~\eqref{Eq:avg_H_e}.
On the other hand, the quantities $M_n$ entering Eq.~\eqref{Eq:freq_sum_rules} are expressed in terms of the initial conditions $\iota_\mathbf{n}^{(n)}(t=0)$ for the real-time HEOM, and we derive the corresponding relations in Appendix~\ref{App:sum_rules}.
Strictly speaking, Eq.~\eqref{Eq:OSR} holds only in the long-chain limit, and in Appendix~\ref{App:OSR_finite_N} we demonstrate that the corresponding finite-size corrections decrease with increasing $N$.
On the contrary, for any given $N$ and $D$, Eq.~\eqref{Eq:freq_sum_rules} (with the expressions obtained in Appendix~\ref{App:sum_rules}) is exact.
Having all these things considered, we conclude that checking the sum rules in Eq.~\eqref{Eq:freq_sum_rules} provides an important self-consistency check that all the numerical procedures (e.g., the numerical Fourier transformation) are properly implemented, while it provides limited information on the adequacy of $N$ and $D$ employed.
In addition to a test for numerical implementation procedures, checking the OSR constitutes a nontrivial test for the adequacy of both $D$ and $N$ employed, which is in line with previous studies.~\cite{PhysRevB.72.104304}
It is for this reason that we focus our discussion in Sec.~\ref{Sec:Numerical_results} on the OSR.
As a general trend, we observe that the sum rules in Eq.~\eqref{Eq:freq_sum_rules} are satisfied with a better relative accuracy than the OSR.

\section{Numerical results}
\label{Sec:Numerical_results}
We limit ourselves to dispersionless optical phonons, $\omega_q\equiv\omega_0$, and perform HEOM computations for three different values of $\omega_0/J$ spanning the range from the adiabatic regime of slow phonons ($\omega_0/J=1/3$) to the extreme quantum regime ($\omega_0/J=1$) and the antiadiabatic regime of fast phonons ($\omega_0/J=3$).
Since most of our results are obtained for $\omega_0/J\leq 1$, we use the dimensionless interaction parameter
\begin{equation}
    \lambda=\frac{g^2}{2J\omega_0}
\end{equation}
that is appropriate to describe the zero-temperature transition from free electrons ($\lambda<1$) to polarons ($\lambda>1$) at such phonon frequencies.

We devote Secs.~\ref{SSec:closing_reliability}--\ref{SSec:numerical_sum_rules} to discussing the performance of the closing strategy embodied in Eq.~\eqref{Eq:closing_strategy} (Sec.~\ref{SSec:closing_reliability}) and the effects of finite $N$ and $D$ on the HEOM results in imaginary and real time (Sec.~\ref{SSec:finite_N_D_effects}), as well as on the accuracy with which the OSR is satisfied (Sec.~\ref{SSec:numerical_sum_rules}).
Section~\ref{SSec:signatures_interaction} summarizes the most representative HEOM results for the time evolution of $C_{jj}$ and the dynamical-mobility profile.
Our most significant results, which concern the temperature dependence of $\mu_\mathrm{dc}$ for different values of $\omega_0/J$, are presented in Sec.~\ref{SSec:mu_vs_T}.
To further illustrate the capabilities and limitations of our approach, we present HEOM results for $C_{jj}(t)$ for strong electron--phonon couplings and at different temperatures in Sec.~\ref{SSec:instable_strong_g}.

We provide HEOM data on $C_{jj}(t)$, its Fourier transformation $C_{jj}(\omega)$ computed using the FFTW3 software package,~\cite{FFTW05} as well as $\mathrm{Re}\:\mu_\mathrm{ac}(\omega)$, in different parameter regimes as a freely available dataset. For more details, see Ref.~\onlinecite{veljko.jankovic.2023}, which contains all our numerical data, and the supplementary material of this manuscript, which contains their detailed description.
Here, let us only mention that the HEOM in Eq.~\eqref{Eq:real-time-HEOM-before-closing-paper} supplemented with the closing in Eq.~\eqref{Eq:closing_strategy} is propagated using the algorithm proposed in Ref.~\onlinecite{JChemTheoryComput.11.3411} with the time step $\omega_0\Delta t=(1-2)\times 10^{-2}$.
While the maximum propagation time $t_\mathrm{max}$ generally shortens with increasing $g$ and/or $T$, it is a highly nontrivial task to give its \emph{a priori} estimate based on the values of model parameters only.
Fortunately, in contrast to some other numerically "exact" methods, such as the DMRG~\cite{PhysRevB.106.155129} or real-time QMC,~\cite{PhysRevB.107.184315} the computational demands of the HEOM method do not increase with time $t$, and are completely determined by $N,D,\Delta t$, and the propagation algorithm.
We thus propagate the HEOM up to real times that are sufficiently long so that the integral $\int_0^t ds\:\mathrm{Re}\:C_{jj}(s)/T$ [see Eq.~\eqref{Eq:mu_from_Im_C_jj_t}] as a function of $t$ enters into saturation, see, e.g., Figs.~\ref{Fig:finite_N_effect_123_160223}(b) and~\ref{Fig:finite_N_effect_123_160223}(c).
The frequency resolution in the optical response is increased by continuing $C_{jj}(t)$ symmetrically for negative times $-t_\mathrm{max}\leq t\leq 0$ using $C_{jj}(-t)=C_{jj}(t)^*$, which results in the frequency step $\Delta\omega/\omega_0=\pi/(\omega_0t_\mathrm{max})$.
Therefore, if one wants both a reliable result for $\mu_\mathrm{dc}$ and a finely resolved optical response, the maximum propagation time should be sufficiently long.
In practice, we used $\omega_0t_\mathrm{max}\gtrsim 500$ for small $g$ and at relatively low $T$, $\omega_0t_\mathrm{max}\simeq 300$ at intermediate values of $g$ and $T$, and $\omega_0t_\mathrm{max}\lesssim 100$ for large $g$ or at high $T$.

\subsection{Effectiveness and reliability of our closing strategy}
\label{SSec:closing_reliability}
\begin{figure}[htbp!]
    \centering
    \includegraphics[width=0.95\columnwidth]{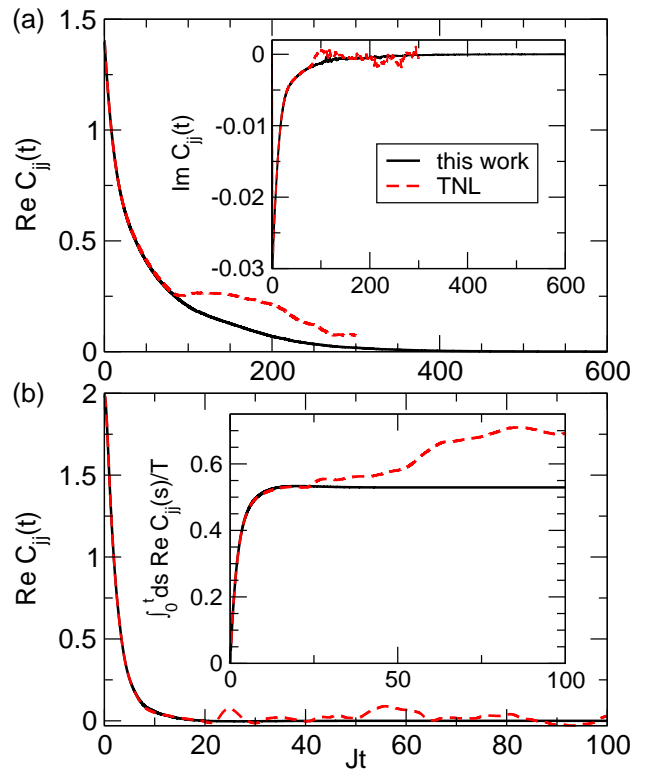}
    \caption{Time dependence of $\mathrm{Re}\:C_{jj}$ (in units of $J^2$) employing the closing in Eq.~\eqref{Eq:closing_strategy} (full black line, label "this work") or the time-nonlocal truncation (dashed red line, label "TNL"). The values of model parameters are $\omega_0/J=1$, $\lambda=0.01$, and (a) $T/J=1$, $N=160$, $D=2$, (b) $T/J=10$, $N=40$, $D=3$. The insets show the time dependence of (a) $\mathrm{Im}\:C_{jj}$ (in units of $J^2$), (b) the quantity $\int_0^t ds\:\mathrm{Re}\:C_{jj}(s)/T$, which tends to $\mu_\mathrm{dc}$ as $t\to+\infty$.}
    \label{Fig:stability}
\end{figure}

We offer numerical examples demonstrating that the closing in Eq.~\eqref{Eq:closing_strategy} actually stabilizes the real-time HEOM [Eq.~\eqref{Eq:real-time-HEOM-before-closing-paper}] without compromising the results for $\mu_\mathrm{dc}$.

Figure~\ref{Fig:stability}(a) (its inset) shows the evolution of $\mathrm{Re}\:C_{jj}$ ($\mathrm{Im}\:C_{jj}$) with the closing in Eq.~\eqref{Eq:closing_strategy} and the TNL truncation in the weak-coupling regime and at a relatively low temperature.
The beneficial effects of our closing strategy on the HEOM stability are apparent.
Moreover, the HEOM [with closing in Eq.~\eqref{Eq:closing_strategy}] estimate for $\mu_\mathrm{dc}$ using $Jt_\mathrm{max}=800$, $\mu_\mathrm{dc}^\mathrm{HEOM}=68.0$, agrees well with the estimate $\mu_\mathrm{dc}^\mathrm{wcl}=72.5$ emerging from the weak-coupling limit [see Eq.~(49) in Ref.~\onlinecite{PhysRevB.99.104304}].
The relative difference between the two results is under 10\%, which eventually emerges as the relative error that is to be associated with $\mu_\mathrm{dc}^\mathrm{HEOM}$, see Sec.~\ref{SSec:finite_N_D_effects}.
With TNL truncation, it is much more difficult to obtain a reliable estimate of $\mu_\mathrm{dc}$.
This is evident when $\mu_\mathrm{dc}$ is computed using only $\mathrm{Re}\:C_{jj}(t)$ [the first equality in Eq.~\eqref{Eq:mu_from_Im_C_jj_t}], which under TNL truncation develops a pronounced hump for $Jt\gtrsim 100$.
The same applies to the computation using only $\mathrm{Im}\:C_{jj}(t)$ [the second equality in Eq.~\eqref{Eq:mu_from_Im_C_jj_t}], when the small-amplitude long-time oscillations of $\mathrm{Im}\:C_{jj}$ around zero are amplified by multiplication with time.

One may still argue that at higher temperatures, when carrier scattering rates entering Eq.~\eqref{Eq:closing_strategy} become large, our hierarchy closing may underestimate $\mu_\mathrm{dc}$.
Such an effect may be particularly pronounced for not too strong coupling, when the maximum depth $D$ is not very large, so that the exponentially damping terms in Eq.~\eqref{Eq:closing_strategy} may appreciably affect the quantity at the hierarchy root, i.e., $C_{jj}(t)$.
The inset of Fig.~\ref{Fig:stability}(b) shows that the estimate for $\mu_\mathrm{dc}$ using the closing in Eq.~\eqref{Eq:closing_strategy} and propagating the HEOM to sufficiently long times (we took $Jt_\mathrm{max}=100$) is approximately the same as the one using the TNL truncation and propagating the HEOM to times before the instabilities arise (up to $Jt\approx 20$).
Both of these estimates ($\mu_\mathrm{dc}^\mathrm{HEOM}=0.53$) agree reasonably well with the estimate $\mu_\mathrm{dc}^\mathrm{wcl}=0.61$ emerging from the weak-coupling limit.

\subsection{Effects of finite $N$ and $D$ on the current--current correlation function and dc mobility}
\label{SSec:finite_N_D_effects}
\begin{figure}[htbp!]
    \centering
    \includegraphics{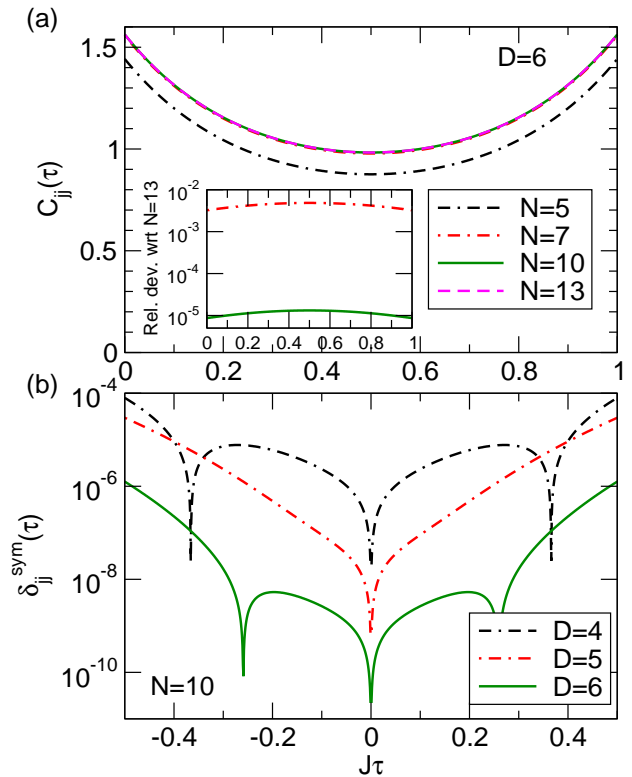}
    \caption{(a) Current--current correlation function (in units of $J^2$) in imaginary time $J\tau\in[0,\beta J]$ for $D=6$ and different chain lengths $N$. The inset shows the quantity $\left|C_{jj}^N(\tau)-C_{jj}^{N=13}(\tau)\right|/C_{jj}^{N=13}(\tau)$ for $N=7$ and 10. (b) The quantity $\delta_{jj}^\mathrm{sym}$ [Eq.~\eqref{Eq:delta_jj_sym}] as a function of imaginary time $J\tau\in[-\beta J/2,\beta J/2]$ for a 10-site chain and different maximum hierarchy depths $D$. Note the logarithmic scale on the vertical axis in the inset of (a) and in (b). The model parameters are $g/J=\omega_0/J=T/J=1$.}
    \label{Fig:im_time_data_160223}
\end{figure}

We first analyze $C_{jj}(\tau)$.
We fix $g/J=\omega_0/J=T/J=1$ and discuss the importance of finite-size effects for maximum hierarchy depth $D=6$.
Figure~\ref{Fig:im_time_data_160223}(a) shows that $C_{jj}(\tau)$ steadily approaches its long-chain limit with increasing $N$.
The approach to that limit is quite fast because the relative deviation with respect to the results for the longest chain studied ($N=13$) decreases by almost three orders of magnitude upon increasing $N$ from 7 to 10, see the inset of Fig.~\ref{Fig:im_time_data_160223}(a).
Therefore, already $N=10$ should be sufficiently large to obtain results representative of the long-chain limit.
Figure~\ref{Fig:im_time_data_160223}(b) shows that the quality of our imaginary-time data, quantified by the relative difference $\delta_{jj}^\mathrm{sym}(\tau)$ [Eq.~\eqref{Eq:delta_jj_sym}], steadily increases with increasing $D$.

\begin{table}[htbp!]
    \centering
    \begin{tabular}{c|c|c|c}
        $(N,D)$ & $\mu_\mathrm{dc}$ from $\mathrm{Re}\:C_{jj}(t)$ & $\mu_\mathrm{dc}$ from $\mathrm{Im}\:C_{jj}(t)$ & $Jt_\mathrm{max}$\\\hline\hline
        $(10,5)$ & 1.327 & 1.308 & 400\\
        $(10,6)$ & 1.493 & 1.484 & 400\\
        $(10,7)$ & 1.336 & 1.325 & 70\\
        $(10,8)$ & 1.548 & 1.537 & 200\\
        \hline
        $(7,6)$ & 1.520 & 1.513 & 400\\
        $(13,6)$ & 1.472 & 1.459 & 400\\
        $(15,6)$ & 1.473 & 1.452 & 300
    \end{tabular}
    \caption{Effects of finite $N$ and $D$ on HEOM results for the dc mobility. The HEOM is propagated up to time $Jt_\mathrm{max}$. Model parameters are $g/J=\omega_0/J=T/J=1$.}
    \label{Tab:finite_N_D}
\end{table}

\begin{figure}[htbp!]
    \centering
    \includegraphics{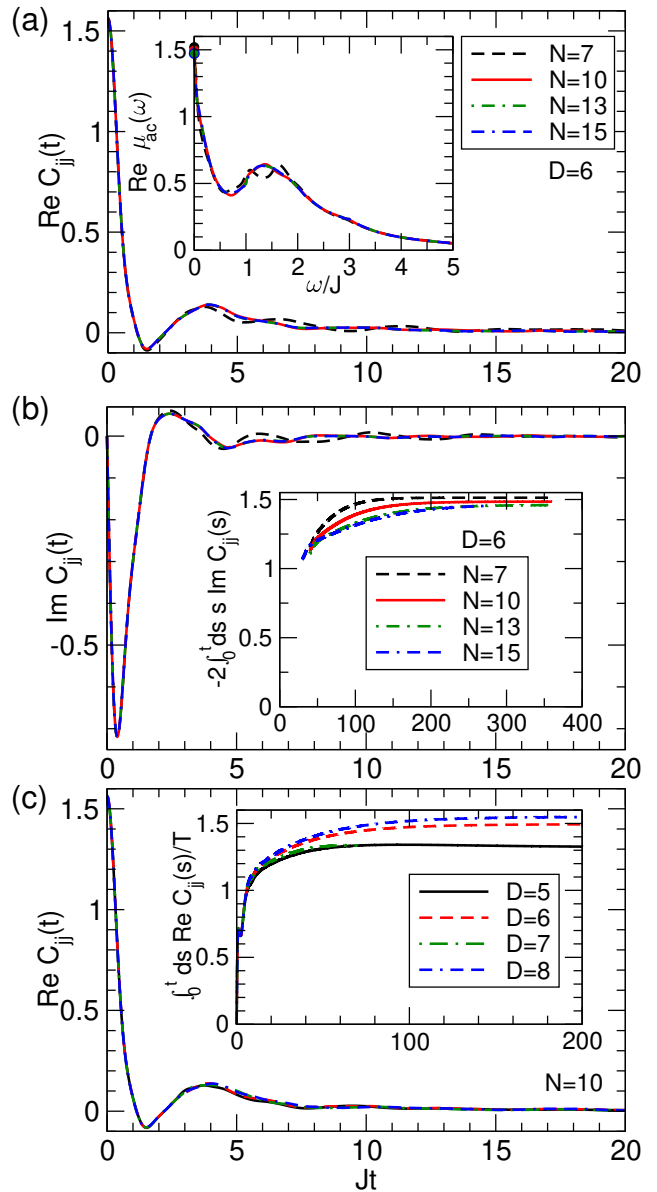}
    \caption{Time dependence of (a) $\mathrm{Re}\:C_{jj}$ and (b) $\mathrm{Im}\:C_{jj}$ (both in units of $J^2$) for $D=6$ and different chain lengths $N$. (c) Time dependence of $\mathrm{Re}\:C_{jj}$ (in units of $J^2$) for $N=10$ and different maximum hierarchy depths $D$. The inset in (a) shows the frequency profile of the dynamic mobility for $D=6$ and different $N$. Full dots at $\omega=0$ represent the results for $\mu_\mathrm{dc}$ using the first equality in Eq.~\eqref{Eq:mu_from_Im_C_jj_t}, see also Table~\ref{Tab:finite_N_D}. The inset in (b) shows how the result of the integration in Eq.~\eqref{Eq:mu_from_Im_C_jj_t} converges towards $\mu_\mathrm{dc}$ as we increase the upper integration limit. To obtain smooth curves, we perform the moving average procedure described in the text. The inset in (c) shows how the result of the integration in Eq.~\eqref{Eq:mu_from_Im_C_jj_t} converges towards $\mu_\mathrm{dc}$ as we increase the upper integration limit. The model parameters are $g/J=\omega_0/J=T/J=1$.}
    \label{Fig:finite_N_effect_123_160223}
\end{figure}

\begin{figure*}[htbp!]
    \centering
    \includegraphics[width=\textwidth]{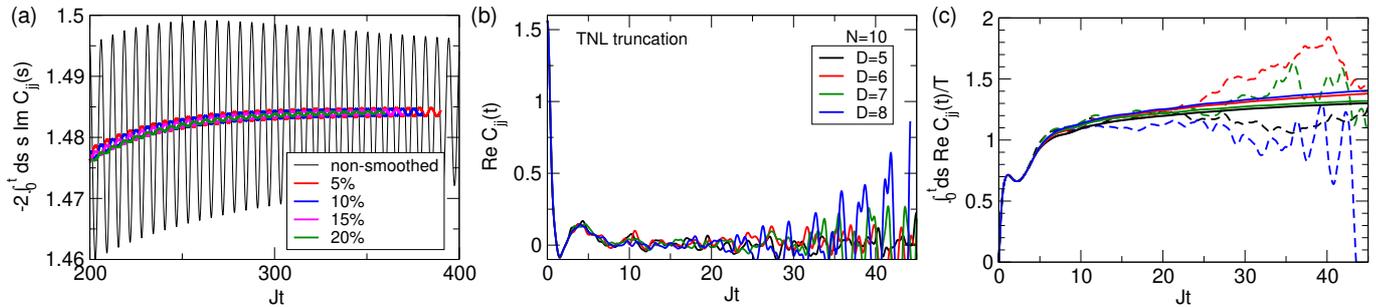}
    \caption{(a) Long-time evolution of the quantity $-2\int_0^t ds\:s\:\mathrm{Im}\:C_{jj}(s)$ [black solid line, label "non-smoothed", see Eq.~\eqref{Eq:mu_from_Im_C_jj_t}] and its moving averages computed for different sizes (determined by $N_\mathrm{move}$ defined in the text) of the averaging window. Percentages are to be taken from the total number of points in which HEOM data are available. (b) Time dependence of $\mathrm{Re}\:C_{jj}(t)$ (in units of $J^2$) under the TNL truncation for different values of $D$. (c) Time dependence of the quantity $\int_0^t ds\:\mathrm{Re}\:C_{jj}(s)/T$ [see Eq.~\eqref{Eq:mu_from_Im_C_jj_t}] under the TNL truncation (dashed lines) and our closing in Eq.~\eqref{Eq:closing_strategy} (solid lines). The values of model parameters and the color code are the same as in (b). The model parameters are $g/J=\omega_0/J=T/J=1$, $N=10$ in all three panels, while $D=6$ in panel (a).}
    \label{Fig:Fig4_revised}
\end{figure*}

While Figs.~\ref{Fig:im_time_data_160223}(a) and~\ref{Fig:im_time_data_160223}(b) demonstrate a steady convergence of the imaginary-time data towards the large-$N$ and large-$D$ limit as $N$ and $D$ are increased, the situation on the real axis is somewhat more complicated, which is summarized in Figs.~\ref{Fig:finite_N_effect_123_160223}(a)--\ref{Fig:finite_N_effect_123_160223}(c).
Fixing $D$ to 6, Fig.~\ref{Fig:finite_N_effect_123_160223}(a) [Fig.~\ref{Fig:finite_N_effect_123_160223}(b)] shows that $\mathrm{Re}\:C_{jj}(t)$ [$\mathrm{Im}\:C_{jj}(t)$] is virtually the same for $N=10,13,$ and 15.
For $N=7,10,$ and $13$, we propagated HEOM up to $Jt_\mathrm{max}=400$, while we used $Jt_\mathrm{max}=300$ for $N=15$.
At longer times (not shown here), both $\mathrm{Re}\:C_{jj}(t)$ and $\mathrm{Im}\:C_{jj}(t)$ exhibit small-amplitude oscillations around zero.  
The agreement between the results for different $N$ in the real-time domain translates to the overall profile of $\mathrm{Re}\:\mu_\mathrm{ac}(\omega)$, see the inset of Fig.~\ref{Fig:finite_N_effect_123_160223}(a).
However, the dc mobility somewhat decreases upon increasing $N$ from 7 to 10 and 13, while its values for $N=13$ and 15 are virtually the same, see the full dots in the inset of Fig.~\ref{Fig:finite_N_effect_123_160223}(a) and Table~\ref{Tab:finite_N_D}.
The relative differences between $\mu_\mathrm{dc}$ for different values of $N$ considered are of the order of percent, which is also evident from the inset of Fig.~\ref{Fig:finite_N_effect_123_160223}(b) displaying the convergence to the dc limit when only $\mathrm{Im}\:C_{jj}(t)$ is used to evaluate $\mu_\mathrm{dc}$ [the second equality in Eq.~\eqref{Eq:mu_from_Im_C_jj_t}].
The integral $-2\int_0^t ds\:s\:\mathrm{Im}\:C_{jj}(s)$ is expected to display long-time oscillations originating from the corresponding oscillations of $\mathrm{Im}\:C_{jj}(s)$ around zero.
The inset of Fig.~\ref{Fig:finite_N_effect_123_160223}(b) thus shows the smoothed data obtained using the moving-average procedure, which is employed at sufficiently long times needed to reliably compute $\mu_\mathrm{dc}$.
In more detail, the moving average of the quantity $-2\int_0^t ds\:s\:\mathrm{Im}\:C_{jj}(s)$ at instant $t$ is computed as the arithmetic average of its $N_\mathrm{move}$ values right after $t$ and its $N_\mathrm{move}$ values right before $t$.
We take $N_\mathrm{move}$ to be 10\% of the total number of points in which we have HEOM data.
Figure~\ref{Fig:Fig4_revised}(a) shows that the moving-average procedure indeed smooths out the long-time oscillations of the quantity $-2\int_0^t ds\:s\:\mathrm{Im}\:C_{jj}(s)$, while the dependence of the final result for $\mu_\mathrm{dc}$ on the averaging window is much less pronounced than, e.g., its dependence on the parity of $D$, \emph{vide infra}.
The $\mu_\mathrm{dc}$ estimates using the two equalities in Eq.~\eqref{Eq:mu_from_Im_C_jj_t} agree up to a couple of percent, see Table~\ref{Tab:finite_N_D}. 
The relative difference of the same order of magnitude is obtained when $D$ is increased from 6 to 8 when $N=10$, while decreasing $D$ from 6 to 5 leads to a decrease in $\mu_\mathrm{dc}$ by around 10\%, see Fig.~\ref{Fig:finite_N_effect_123_160223}(c) and Table~\ref{Tab:finite_N_D}.
For $N=10$ and $D=7$, $\mathrm{Re}\:C_{jj}(t)$ becomes negative at long times, which may prevent us from reliably estimating $\mu_\mathrm{dc}$.
Nevertheless, we see that following the time evolution up to $Jt_\mathrm{max}=70$ provides an estimate of $\mu_\mathrm{dc}$ that differs from the estimate for $N=10, D=5$ by a couple of percent.
We thus conclude that $\mu_\mathrm{dc}$ estimates may depend on the parity of the maximum hierarchy depth $D$.
For sufficiently large $D$, the estimates for $D$ and $D+2$ differ by a couple of percent, and those for $D$ and $D+1$ differ by around 10\%.
While we find that such a behavior of $\mu_\mathrm{dc}$ as a function of $D$ (for fixed $N$) is generic, the magnitudes of the above-mentioned relative differences generally decrease with temperature.
The preceding discussion implies that HEOM estimates for $\mu_\mathrm{dc}$ should be taken with relative uncertainties not surpassing 10\%.

Without the specific HEOM closing [Eq.~\eqref{Eq:closing_strategy}], it would be nearly impossible to obtain meaningful results for the dc or ac mobility,
which is apparent from Fig.~\ref{Fig:Fig4_revised}(b) showing HEOM data for $\mathrm{Re}\:C_{jj}(t)$ that use the TNL truncation.
Model parameters are the same as in Fig.~\ref{Fig:finite_N_effect_123_160223}(c), to which the results in Fig.~\ref{Fig:Fig4_revised}(b) are to be compared.
Our closing smooths the local maximum in $\mathrm{Re}\:C_{jj}(t)$ around $Jt\simeq 4$, and stabilizes the subsequent time evolution, whose oscillatory features under TNL truncation become more pronounced with increasing $D$.
In addition to $\mathrm{Re}\:C_{jj}(t)$, in Fig.~\ref{Fig:Fig4_revised}(c) we show the quantity $\int_0^t ds\:\mathrm{Re}\:C_{jj}(s)/T$ [see Eq.~\eqref{Eq:mu_from_Im_C_jj_t}] under TNL truncation (dashed lines) and our closing scheme (solid lines).
While the agreement between the dashed and solid lines in Fig.~\ref{Fig:Fig4_revised}(c) is good for $Jt\lesssim 10$, the results obtained under TNL truncation do not show any sign of reaching a long-time limit with increasing $t$.
One could use the HEOM data with TNL truncation up to $Jt_\mathrm{max}\simeq 20$ (before the instabilities become more pronounced) to extract the dc mobility of around 1.25, and virtually the same estimate would be obtained using the HEOM data with our closing scheme up to $Jt_\mathrm{max}\simeq 20$, see Fig.~\ref{Fig:Fig4_revised}(c).
Nevertheless, if one is to obtain the optical response with a decent frequency resolution using the HEOM data in Fig.~\ref{Fig:Fig4_revised}(b) up to $Jt_\mathrm{max}\simeq 20$, one should probably perform additional numerical procedures, such as the zero-padding (to effectively increase $t_\mathrm{max}$) possibly combined with a signal windowing.~\cite{PhysRevB.106.155129,PhysRevB.105.054311}
Such procedures, however, may affect the intensities (and to some extent the positions) of the optical-response features, mainly in its low-frequency part that corresponds to the most challenging long-time dynamics.
On the other hand, when we propagate our stabilized HEOM up to long real times, no additional numerical procedures on the HEOM data for $C_{jj}(t)$ are required to compute the optical response.
While the observed dependence of our HEOM results on the parity of $D$ could be attributed to our closing scheme, we believe that we handle this effect in an appropriate manner because we use it to estimate the relative error of our results for $\mu_\mathrm{dc}$. 
While a detailed explanation of our observations is beyond the scope of this investigation, we speculate about their possible origin as follows.
In Eq.~\eqref{Eq:real-time-HEOM-before-closing-paper}, $\mu_\mathbf{n}$ is the change in energy of the phononic subsystem due to the sequence of electron--phonon interaction events described by $\mathbf{n}$.
At depth $n$, $\mu_\mathbf{n}/\omega_0$ can assume values $\pm n,\pm (n-2),\dots$.
While at odd depths $n$ all $\mu_\mathbf{n}$s are nonzero, even depths feature some vectors $\mathbf{n}$ for which $\mu_\mathbf{n}=0$.
At even depths, there are thus ADOs having both $\mu_\mathbf{n}=0$ and $k_\mathbf{n}=0$, meaning that the resonance condition $\varepsilon_k-\varepsilon_{k+k_\mathbf{n}}+\mu_\mathbf{n}=0$ emerging from the free-rotation term in Eq.~\eqref{Eq:real-time-HEOM-before-closing-paper} is perfectly satisfied.
On the other hand, at odd depths, the resonance condition is almost never perfectly met: even if $k_\mathbf{n}=0$ for a particular $\mathbf{n}$, $\mu_\mathbf{n}$ is certainly nonzero.
When the closing scheme in Eq.~\eqref{Eq:closing_strategy} is applied, the behavior of all ADOs at an odd maximum depth can be described by damped oscillations (in the lowest approximation that neglects links to other ADOs).
In the same approximation, at an even maximum depth, there are ADOs that are exponentially suppressed with time without displaying any oscillatory behavior.
The influence of such ADOs on the overall HEOM dynamics might be less pronounced than the influence of ADOs exhibiting damped oscillations.

We finish this section by briefly discussing the decrease in the number of active variables of our momentum-space HEOM with respect to existing real-space formulations.
Namely, for an $N$-site chain, the number of ADOs upon hierarchy truncation at depth $D$ is $n_\mathrm{ADO}^\mathrm{rs}=\binom{2N+D}{D}$ when working in real space ($q=0$ phonon mode is considered) and $n_\mathrm{ADO}^\mathrm{ms}=\binom{2(N-1)+D}{D}$ when working in momentum space ($q=0$ phonon mode is not considered).
As the number of entries in each ADO is at the same time reduced by a factor of $N$, the relative decrease in the number of variables that have to be propagated upon transferring from real to momentum space is
\begin{equation}
    \frac{N^2n_\mathrm{ADO}^\mathrm{rs}-Nn_\mathrm{ADO}^\mathrm{ms}}{N^2n_\mathrm{ADO}^\mathrm{rs}}=1-\frac{2(2N-1)}{(2N+D)(2N+D-1)}.
\end{equation}
For fixed $N$, the savings in computer memory increase with $D$.
While the savings in computer memory for fixed $D$ decrease with $N$, they can be substantial on relatively short chains that are still sufficiently long so that the corresponding results are representative for the long-chain limit.
For example, for $N=10$ and $D=6$ (as in Fig.~\ref{Fig:finite_N_effect_123_160223}), the relative decrease in the number of active variables with respect to HEOM formulations in real space is around 40\%.

\subsection{Numerical examples concerning sum rules}
\label{SSec:numerical_sum_rules}
\begin{figure}[htbp!]
    \centering
    \includegraphics{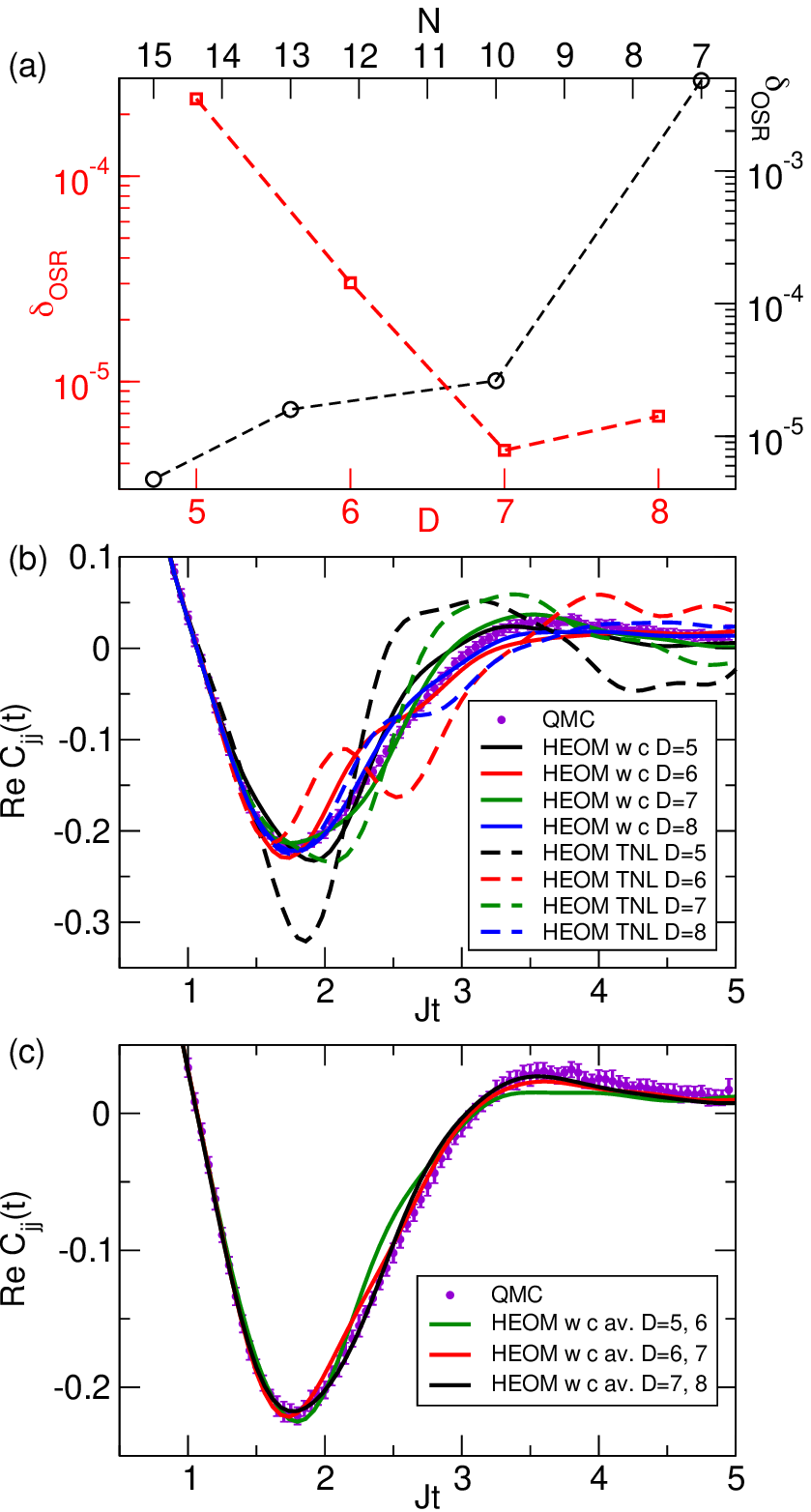}
    \caption{(a) Relative accuracy $\delta_\mathrm{OSR}$ [Eq.~\eqref{Eq:delta_OSR}] as a function of $D$ for  $N=10$ (red squares, left vertical and bottom horizontal axes) and as a function of $N$ for $D=6$ (black circles, right vertical and top horizontal axes). The dashed lines connecting the symbols serve as guides to the eye. The model parameters are $g/J=\omega_0/J=T/J=1$ ($\lambda=0.5$). For $N=10$, we use $C_{jj}(t)$ up to $Jt_\mathrm{max}=200$ for $D=5,6,$ and 8, while for $D=7$ we take $Jt_\mathrm{max}=70$, see also Table~\ref{Tab:finite_N_D}. For $D=6$, we use $C_{jj}(t)$ up to $Jt_\mathrm{max}=300$ for $N=7, 10, 13,$ and 15. (b) Comparison of HEOM (lines) and QMC (symbols with error bars) results for $\mathrm{Re}\:C_{jj}(t)$ (in units of $J^2$) for $N=10$, $g/J=1/\sqrt{3}$, $\omega_0/J=1/3$, $T/J=1$ ($\lambda=0.5$), and different values of $D$. Solid lines (label "w c") employ our closing [Eq.~\eqref{Eq:closing_strategy}], while dashed lines (label "TNL") use the TNL truncation. (c) Comparison of HEOM results averaged over depths $D-1$ and $D$ (solid lines) with QMC results (symbols with error bars) for $D=6,7,$ and 8, while other parameters assume the same values as in (b).
    QMC data are the courtesy of N. Vukmirovi\'c.}
    \label{Fig:potential_graph_osr}
\end{figure}

Figure~\ref{Fig:potential_graph_osr}(a) summarizes how the relative accuracy
\begin{equation}
\label{Eq:delta_OSR}
    \delta_\mathrm{OSR}=\frac{\left|\int_0^{+\infty}d\omega\:\mathrm{Re}\mu_\mathrm{ac}(\omega)-\frac{\pi}{2}|\langle H_\mathrm{e}\rangle|\right|}{\frac{\pi}{2}|\langle H_\mathrm{e}\rangle|}
\end{equation}
with which the OSR is satisfied depends on $N$ and $D$ for the same values of model parameters as in Figs.~\ref{Fig:im_time_data_160223},~\ref{Fig:finite_N_effect_123_160223} and~\ref{Fig:Fig4_revised}.
We observe that, in the ranges of $N$ and $D$ considered, $\delta_\mathrm{OSR}$ generally decreases with both $N$ [assuming fixed $D$, see black circles, right vertical and top horizontal axes in Fig.~\ref{Fig:potential_graph_osr}(a)] and $D$ [assuming fixed $N$, see red squares, left vertical and bottom horizontal axes in Fig.~\ref{Fig:potential_graph_osr}(a)].

Fixing $D$ to 6 and varying $N$ from 7 to 15, we observe the steepest decrease in $\delta_\mathrm{OSR}$ upon increasing $N$ from 7 to 10, while further increase in $N$ from 10 to 13 and 15 results in a much milder decrease in $\delta_\mathrm{OSR}$.
This observation is consistent with both the imaginary-axis data shown in Fig.~\ref{Fig:im_time_data_160223}(a) and the real-time data presented in Figs.~\ref{Fig:finite_N_effect_123_160223}(a) and~\ref{Fig:finite_N_effect_123_160223}(b) and Table~\ref{Tab:finite_N_D}.
Namely, the largest (the smallest) variation in $C_{jj}(\tau)$, $C_{jj}(t)$, and $\mu_\mathrm{dc}$ upon increasing $N\in\{7,10,13\}$ to the subsequent value from the sequence $[7,10,13,15]$ is observed for $N=7$ ($N=13$).
The decrease in $\delta_\mathrm{OSR}$ with increasing $N$ is, however, ultimately limited by the fact that $\delta_\mathrm{OSR}$ depends on quantities that are themselves calculated numerically and thus bring their own numerical errors into the final expression.
The integral over frequencies is computed using the trapezoidal rule, while $\langle H_\mathrm{e}\rangle$ is computed for finite values of $N$ and $D$ (we do not use its "exact" value in the limit $N,D\to\infty$ that could be obtained using our~\cite{PhysRevB.105.054311} or some other~\cite{PhysRevB.102.165155} method).
The error incurred when the integral $\int_0^{+\infty}d\omega\:\mathrm{Re}\mu_\mathrm{ac}(\omega)$ is evaluated using the trapezoidal rule is of the order of $(\Delta\omega/J)^3$,~\cite{PresTeukVettFlan92} where the frequency step $\Delta\omega$ is related to the maximum propagation time $t_\mathrm{max}$ by $\Delta\omega/J=\pi/(Jt_\mathrm{max})$.
Fixing $D$ and varying $N$ we use $Jt_\mathrm{max}=300$, meaning that the numerical error of the integral $\int_0^{+\infty}d\omega\:\mathrm{Re}\mu_\mathrm{ac}(\omega)$ is of the order of $(\pi/300)^3\sim 10^{-6}$.
In other words, $N=15$ is sufficiently large that $\delta_\mathrm{OSR}$ is most probably not dominated by finite-size effects in $N$, but rather by the error of numerical integration.
This is further corroborated by the fact that the kinetic energies for $N=13$ and $N=15$ differ on the seventh decimal place, see Table~\ref{Tab:kinetic}, meaning that the error of the kinetic energy for $N=15$ is at least an order of magnitude below the numerical integration error.
On the other hand, for $N=7$, $\delta_\mathrm{OSR}$ is most probably limited by the finite-size effects in $\langle H_\mathrm{e}\rangle$.
The data in Table~\ref{Tab:kinetic} suggest that the error of $|\langle H_\mathrm{e}\rangle|_{N=7,D=6}$, which can be inferred from its deviation from $|\langle H_\mathrm{e}\rangle|_{N=10,D=6}$, is of the order of $10^{-3}$.
This is consistent with the value of $\delta_\mathrm{OSR}$ reported in Fig.~\ref{Fig:potential_graph_osr}(a).

\begin{table}[htbp!]
    \centering
    \begin{tabular}{c|c}
        $(N,D)$ & $|\langle H_\mathrm{e}\rangle|/J$ \\
        \hline\hline
        $(7,6)$ & 1.1553082795\\
        $(10,6)$ & 1.1546239955\\
        $(13,6)$ & 1.1546229149\\
        $(15,6)$ & 1.1546230078
    \end{tabular}
    \caption{Kinetic energy of the electron as a function of $N$ and $D$ for $g/J=\omega_0/J=T/J=1$.}
    \label{Tab:kinetic}
\end{table}

A similar analysis can be repeated by fixing $N$ and varying $D$.
In practice, our criterion for choosing $N$ and $D$ is that $\delta_\mathrm{OSR}\lesssim 10^{-4}$.
As mentioned in Sec.~\ref{SSec:theory_sum_rules} and as can be seen from the supplementary material, the relative accuracy with which the sum rules in Eq.~\eqref{Eq:freq_sum_rules} are satisfied is generally better than $\delta_\mathrm{OSR}$.

At higher temperatures ($T/\omega_0\gtrsim 3$), we find that HEOM data for fixed $N$ and two sufficiently large consecutive depths $D-1$ and $D$ generally have almost the same $\delta_\mathrm{OSR}$ [as in Fig.~\ref{Fig:potential_graph_osr}(a)], while there are some differences between them already at relatively short times.
For example, for $g/J=1/\sqrt{3}$, $\omega_0/J=1/3$, $T/J=1$ ($\lambda=0.5$), we find that the differences between HEOM results for different values of $D$ (we fix $N=10$) appear already for $Jt\simeq 2$ or $\omega_0 t\simeq 2/3$, see Fig.~\ref{Fig:potential_graph_osr}(b), in contrast to the situation in Figs.~\ref{Fig:finite_N_effect_123_160223}(c) and~\ref{Fig:Fig4_revised}(c), where the differences appear at somewhat longer times.
We perform HEOM computations with TNL truncation and our specific closing and find that the short-time differences observed in Fig.~\ref{Fig:potential_graph_osr}(b) cannot be exclusively attributed to our closing scheme as they are also present under the TNL truncation.
Actually, our closing stabilizes the evolution of $\mathrm{Re}\:C_{jj}(t)$ and lowers the differences between HEOM results for different values of $D$ (with respect to the TNL truncation).
To reveal whether our HEOM results are reliable, we compare them with QMC data obtained using the methodology developed in Ref.~\onlinecite{PhysRevB.107.184315}.
QMC results are numerically "exact" as their convergence with respect to all control parameters of the simulation has been carefully checked.
In Figs.~\ref{Fig:potential_graph_osr}(b) and~\ref{Fig:potential_graph_osr}(c), we show QMC results with their statistical error bars.
Figure~\ref{Fig:potential_graph_osr}(b) suggests that the difference between HEOM (with our closing) and QMC results decreases relatively slowly with increasing $D$, while HEOM dynamics for $D$s of the same parity ($D=5,7$ and $D=6,8$) deviate from the QMC results in a similar manner (both deviations are positive/negative).
It is known that the convergence of a slowly converging sequence can be improved by performing an appropriate sequence transformation, such as the Shanks or Richardson transformation.~\cite{Bender-Orszag-book}
Here, inspired by the reasoning behind the Shanks transformation, we want to improve the convergence of the sequence $C_{jj}(D;t)$ in $D$ (for fixed $N$ and at each instant $t$) by handling the term whose decay (as a function of $D$) towards zero is the slowest.
Based on Fig.~\ref{Fig:potential_graph_osr}(b), one may imagine that the dependence of $C_{jj}(D;t)$ on $D$ can be represented as
\begin{equation}
\label{Eq:shanks-1}
    C_{jj}(D;t)=C_{jj}^{D\to\infty}(t)+\alpha(t)\frac{(-1)^D}{D^a},
\end{equation}
where $a>0$, $\alpha(t)$ is a complex number, $C_{jj}^{D\to\infty}(t)$ is the sought large-$D$ limit, while the alternating term $(-1)^D$ mimics the observed alternation of HEOM results with respect to QMC results with the parity of $D$.
We use Eq.~\eqref{Eq:shanks-1} to arrive at
\begin{equation}
\label{Eq:shanks-2}
    \frac{C_{jj}(D-1;t)-C_{jj}^{D\to\infty}(t)}{C_{jj}(D;t)-C_{jj}^{D\to\infty}(t)}=-\left(\frac{D-1}{D}\right)^a\approx -1,
\end{equation}
where the last approximate equality holds for sufficiently large $D$ (the smaller is $a$, the smaller is the minimum value of $D$ for which the approximation is good).
Equation~\eqref{Eq:shanks-2} implies that
\begin{equation}
    C_{jj}^{D\to\infty}(t)\approx\frac{C_{jj}(D;t)+C_{jj}(D-1;t)}{2},
\end{equation}
which suggests that the large-$D$ limit of $C_{jj}(D;t)$ can be approached more rapidly by considering the transformed sequence $[C_{jj}(D;t)+C_{jj}(D-1;t)]/2$ instead of the original sequence $C_{jj}(D;t)$.
While our arguments are not fully mathematically rigorous, they produce plausible results, as revealed in Fig.~\ref{Fig:potential_graph_osr}(c) showing how the HEOM data averaged over two consecutive depths (5 and 6, 6 and 7, 7 and 8) compare to the QMC data.
We observe that the agreement with the QMC data improves with increasing the depths over which the averaging is performed. 
As a result, when averaging over depths 7 and 8, HEOM results are within the QMC error bars in the largest portion of the time window displayed.
It then seems reasonable that our final HEOM result in this parameter regime be the average of the results for $D=7$ and 8.
Finally, whenever our dataset~\cite{veljko.jankovic.2023} provides raw data for $C_{jj}(t)$ for two consecutive depths, we use the average $C_{jj}(t)$ when computing physical quantities.

\subsection{Signatures of the electron--phonon interaction in time and frequency domains}
\label{SSec:signatures_interaction}

\begin{figure*}
    \centering
    \includegraphics[width=0.9\textwidth]{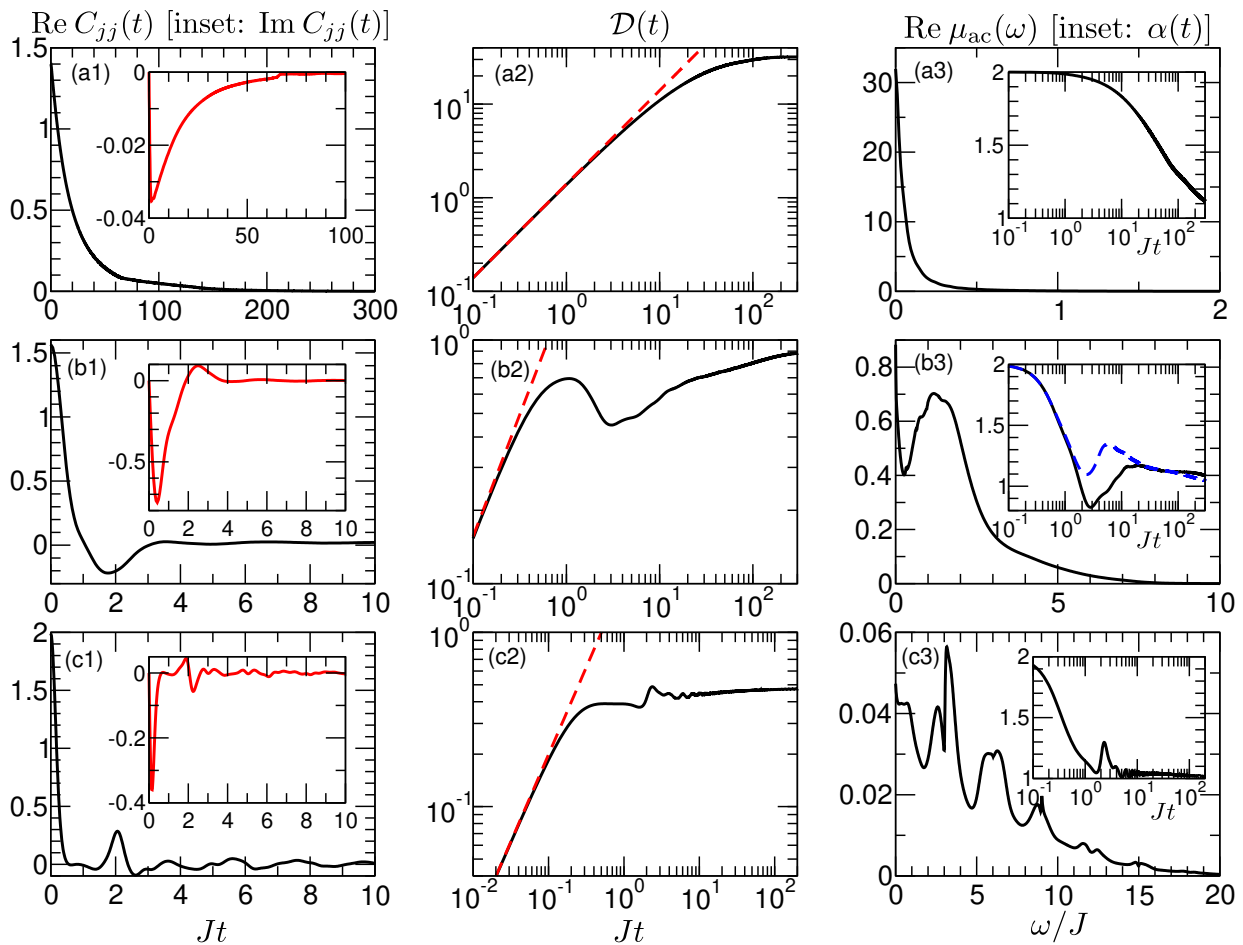}
    \caption{Time dependence of $\mathrm{Re}\:C_{jj}$ [in units of $J^2$, (a1)--(c1)], $\mathcal{D}$ [in units of $J$, (a2)--(c2)], and the frequency dependence of $\mathrm{Re}\:\mu_\mathrm{ac}$ [(a3)--(c3)] for (a) $\omega_0/J=1/3,\lambda=0.01,T/J=1$, (b) $\omega_0/J=1/3,\lambda=0.5,T/J=1$, and (c) $\omega_0/J=3,\lambda=0.5,T/J=10$. The insets of panels (a1)--(c1) [(a3)--(c3)] display the time dependence of $\mathrm{Im}\:C_{jj}$ in units of $J^2$ ($\alpha$). The dashed lines in (a2)--(c2) represent the short-time (ballistic) limit of the diffusion constant, $\mathcal{D}_\mathrm{bal}(t)=C_{jj}(0)t$. The dashed line in the inset of (c2) represents $\alpha(t)$ for the parameter regime examined in Fig.~\ref{Fig:finite_N_effect_123_160223}. The results in (a1)--(a3) emerge from HEOM computations using $N=128,D=2$, while the remaining results are the arithmetic average of HEOM computations using: (b1)--(b3) $N=10,D=7$ and $N=10,D=8$; (c1)--(c3) $N=5,D=18$ and $N=5,D=19$.}
    \label{Fig:Fig_signatures}
\end{figure*}
Figures~\ref{Fig:Fig_signatures}(a1)--\ref{Fig:Fig_signatures}(c3) present selected results concerning the dynamics of $j-j$ correlation function, diffusion constant, and diffusion exponent, together with the frequency profile of the dynamical mobility.
Our analysis of the parameter regimes in which HEOM computations are free of numerical instabilities [see Figs.~\ref{Fig:slika_mu_vs_T}(a)--\ref{Fig:slika_mu_vs_T}(c)] identifies three typical behaviors of the aforementioned quantities.

In the weak-coupling regime ($\lambda=0.01$) for $\omega_0/J=1/3$ and 1, we find a smooth crossover from the ballistic electronic motion at short times towards the diffusive motion at long times.
The representative results for $\omega_0/J=1$ are shown in Figs.~\ref{Fig:stability}(a) and~\ref{Fig:stability}(b), while Figs.~\ref{Fig:Fig_signatures}(a1)--\ref{Fig:Fig_signatures}(a3) show representative results for $\omega_0/J=1/3$.
$\mathrm{Re}\:C_{jj}(t)$, $\mathcal{D}(t)$, and $\alpha(t)$ are all monotonic functions of time, while the dynamical-mobility profile has only the Drude peak at $\omega=0$ and bears an overall resemblance to the Drude model.

Already in the intermediate-coupling regime ($\lambda=0.5$) for $\omega_0/J=1/3$ and 1, the ballistic-to-diffusive crossover is not smooth.
The representative results for $\omega_0/J=1$ are presented in Figs.~\ref{Fig:finite_N_effect_123_160223}(a)--\ref{Fig:finite_N_effect_123_160223}(c), whereas Figs.~\ref{Fig:Fig_signatures}(b1)--\ref{Fig:Fig_signatures}(b3) show representative results for $\omega_0/J=1/3$.
$\mathrm{Re}\:C_{jj}(t)$, $\mathcal{D}(t)$, and $\alpha(t)$ are non-monotonic functions of time.
We observe that $\mathrm{Re}\:C_{jj}(t)<0$ on intermediate time scales, on which the diffusion constant decreases with time.
This is clear from Eq.~\eqref{Eq:def_mathcal_D_t}, which establishes a connection between the sign of $\mathrm{Re}\:C_{jj}(t)$ and the intervals of monotoncity of $\mathcal{D}(t)$.
$\alpha(t)$ reaches a pronounced local minimum on the very same time scales.
In other words, intermediate time scales in the intermediate-interaction regime witness a temporally limited slow-down of the electronic motion.
Depending on the parameter regime (typically for strong $g$ and at high $T$), that slow-down may be so pronounced that the electronic motion changes its character from superdiffusive ($\alpha>1$) to subdiffusive ($\alpha<1$) over a limited time frame.
While such a change occurs for model parameters studied in Figs.~\ref{Fig:Fig_signatures}(b1)--\ref{Fig:Fig_signatures}(b3), it is absent for model parameters studied in Figs.~\ref{Fig:finite_N_effect_123_160223}(a)--\ref{Fig:finite_N_effect_123_160223}(c), see the dashed line in the inset of Fig.~\ref{Fig:Fig_signatures}(b3).
Apart from the Drude peak at $\omega=0$, the dynamical mobility features a prominent finite-frequency peak.

The available results for $\omega_0/J=3$, which are obtained for sufficiently strong interactions and at sufficiently high temperatures, also display a non-monotonic behavior of $\mathrm{Re}\:C_{jj}(t)$, $\mathcal{D}(t)$, and $\alpha(t)$.
In contrast to the results for $\omega_0/J=1/3$ and 1, $\mathcal{D}(t)$ increases in almost regular steps centered around integer multiples of $2\pi/\omega_0$.
The first peak of $\mathrm{Re}\:C_{jj}(t)$ is indeed centered at $2\pi/\omega_0$, while subsequent peaks are somewhat displaced towards earlier times, see Fig.~\ref{Fig:Fig_signatures}(c1).
The dynamic-mobility profile is characterized by relatively broad peaks centered around integer multiples of $\omega_0$.
These features suggest that we are close to the genuine small-polaron limit.~\cite{SovPhysJETP.16.1301,PhysRevB.74.075101}

\subsection{Temperature dependence of the dc mobility}
\label{SSec:mu_vs_T}
\begin{figure*}
    \includegraphics{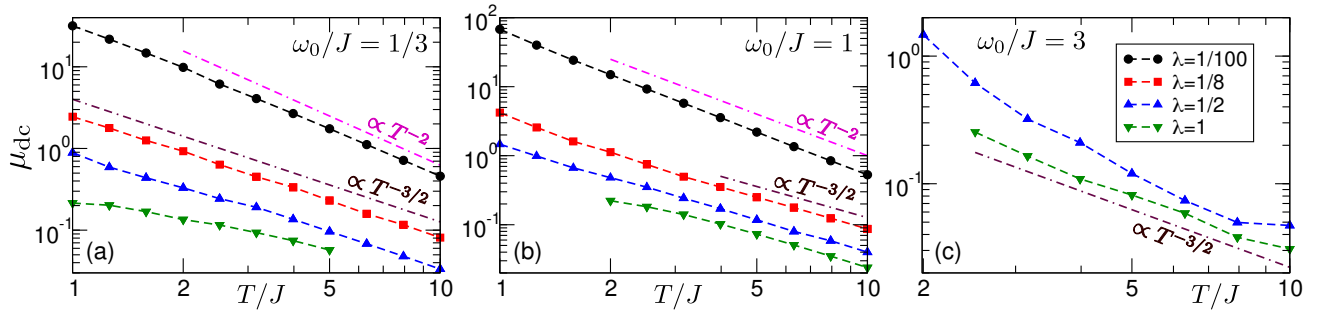}
    \caption{HEOM-method results for the temperature-dependent dc mobility for different interaction strengths (different values of parameter $\lambda$) and (a) $\omega_0/J=1/3$, (b) $\omega_0/J=1$, and (c) $\omega_0/J=3$. Full symbols are HEOM-method results, while dashed lines connecting them serve as guides to the eye. Double dash-dotted lines show the theoretically predicted~\cite{arxiv.2212.13846} power-law scaling $\mu_\mathrm{dc}\propto T^{-2}$ or $\mu_\mathrm{dc}\propto T^{-3/2}$ in appropriate limiting cases (no fitting procedures have been performed on the HEOM data).}
    \label{Fig:slika_mu_vs_T}
\end{figure*}

Figures~\ref{Fig:slika_mu_vs_T}(a)--\ref{Fig:slika_mu_vs_T}(c) present the central result of our study, the HEOM-method results for the temperature dependence of $\mu_\mathrm{dc}$ for three different values of $\omega_0/J$.
As discussed in Sec.~\ref{SSec:finite_N_D_effects}, our results for $\mu_\mathrm{dc}$ should be assigned relative errors not surpassing $10\%$.
The values reported in Figs.~\ref{Fig:slika_mu_vs_T}(a)--\ref{Fig:slika_mu_vs_T}(c) are obtained by averaging $\mu_\mathrm{dc}$ computed using only $\mathrm{Re}\:C_{jj}(t)$ or only $\mathrm{Im}\:C_{jj}(t)$ [Eq.~\eqref{Eq:mu_from_Im_C_jj_t}].
The smoothing procedure described in Sec.~\ref{SSec:finite_N_D_effects} is applied when considering only $\mathrm{Im}\:C_{jj}(t)$.
At elevated temperatures ($T/\omega_0\gtrsim 3$), we additionally average HEOM results for two consecutive depths, as discussed in Sec.~\ref{SSec:numerical_sum_rules}.

Section~\ref{SSec:instable_strong_g} gives an example of numerical instabilities appearing for sufficiently strong $g$ or at sufficiently low $T$, which prevent us from obtaining reliable HEOM results for, e.g.,  $\lambda>1$ or $T/J<2$ and $\lambda=1$ in Fig.~\ref{Fig:slika_mu_vs_T}(b).
Our data for $\lambda=1$ in Figs.~\ref{Fig:slika_mu_vs_T}(a) and~\ref{Fig:slika_mu_vs_T}(b) suggest that $\mu_\mathrm{dc}$ enters into saturation on the low-temperature side.
If we could lower the temperature further, we would enter the regime of thermally activated transport, in which $\mu_\mathrm{dc}$ grows with $T$, while we obtain HEOM results only in regimes where $\mu_\mathrm{dc}$ decreases with $T$. 
Concerning the temperature dependence of $\mu_\mathrm{dc}$, we find that the HEOM mobilities at low $g$ and for $\omega_0/J=1/3$ and 1 are consistent with the recently found scaling $\mu_\mathrm{dc}\propto T^{-2}$,~\cite{arxiv.2212.13846} which is shown as a double dash-dotted line.
For stronger coupling and at sufficiently high temperatures, the HEOM mobilities are consistent with the scaling $\mu_\mathrm{dc}\propto T^{-3/2}$, which is again shown as a double dash-dotted line.

\subsection{Numerical instabilities for strong electron--phonon interactions and at low temperatures}
\label{SSec:instable_strong_g}
\begin{figure}
    \centering
    \includegraphics{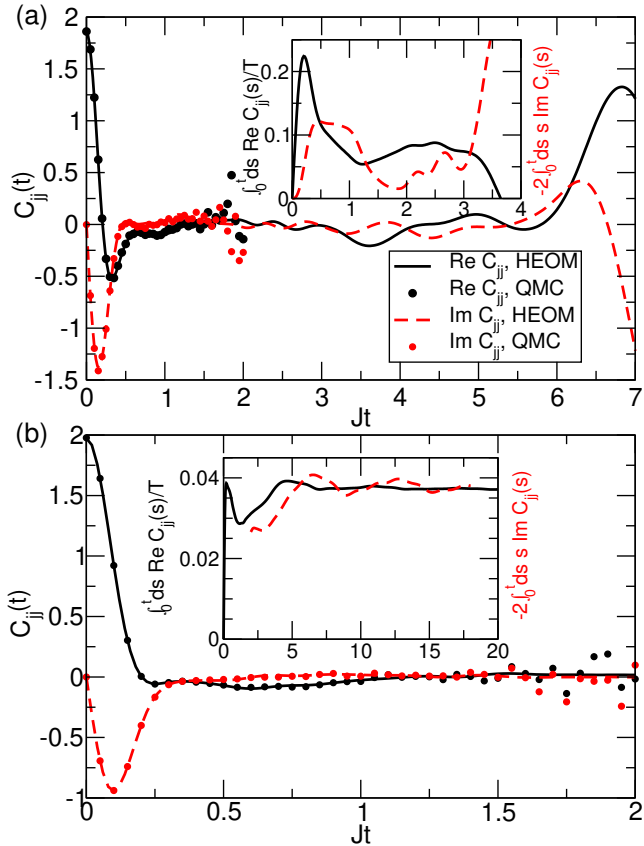}
    \caption{Time dependence of $C_{jj}$ (in units of $J^2$) computed using the HEOM method (lines) and the real-time QMC method developed in Ref.~\onlinecite{PhysRevB.107.184315} (symbols) for $\omega_0/J=1$, $\lambda=2$, and (a) $T/J=1$, (b) $T/J=5$. Solid (dashed) lines display $\mathrm{Re}\:C_{jj}(t)$ [$\mathrm{Im}\:C_{jj}(t)$]. The insets display time evolution of $\int_0^t ds\:\mathrm{Re}\:C_{jj}(s)/T$ (solid lines) and $-2\int_0^t ds\:s\mathrm{Im}\:C_{jj}(s)$ (dashed lines). The HEOM computations in (a) use $N=10$ and $D=8$, while (b) shows the arithmetic average of HEOM results obtained for $N=7,D=10$ and $N=7,D=11$. QMC simulations in both (a) and (b) use $N=10$ and the respective data are the courtesy of N. Vukmirovi\'c.}
    \label{Fig:Fig_strong_g}
\end{figure}

For stronger interactions, at lower temperatures, and for larger $\omega_0/J$, the numerical instabilities appearing already at relatively short real times (despite employing our specific closing strategy) prevent us from computing the long-time dynamics of $C_{jj}(t)$, and thus $\mu_\mathrm{dc}$. 
As an example, in Figs.~\ref{Fig:Fig_strong_g}(a) and~\ref{Fig:Fig_strong_g}(b) we present our HEOM results in the strong-coupling regime ($\lambda=2$) for $\omega_0/J=1$ at a relatively low temperature $T/J=1$ [Fig.~\ref{Fig:Fig_strong_g}(a)] and at an elevated temperature $T/J=5$ [Fig.~\ref{Fig:Fig_strong_g}(b)].

For $T/J=1$, numerical instabilities appear already for $Jt\gtrsim 3$, and become more pronounced at longer times.
The maximum time up to which reliable HEOM results for $C_{jj}(t)$ can be obtained is thus of the same order of magnitude as the maximum time that can be reached in real-time QMC simulations, whose results are shown as full symbols (QMC error bars are omitted here for visual clarity).
The inset of Fig.~\ref{Fig:Fig_strong_g}(a) shows how the integrals $\int_0^t ds\:\mathrm{Re}\:C_{jj}(s)/T$ and $-2\int_0^t ds\:s\mathrm{Im}\:C_{jj}(s)$, whose $t\to+\infty$ limit determines $\mu_\mathrm{dc}$, evolve as functions of their upper limit $t$.
While the numerical instabilities become amplified when $\mu_\mathrm{dc}$ is computed using only $\mathrm{Im}\:C_{jj}(t)$, the data using only $\mathrm{Re}\:C_{jj}(t)$ may suggest that the maximum time $Jt_\mathrm{max}=3$ is sufficiently long to capture the electron's diffusive motion [$\int_0^t ds\:\mathrm{Re}\:C_{jj}(s)/T$ reaches a plateau for $2\leq Jt\leq 3$].
Nevertheless, the corresponding mobility estimate may be unreliable.
Namely, as discussed in Ref.~\onlinecite{PhysRevB.99.104304}, for strong interactions and at not too high temperatures, $\mathrm{Re}\:C_{jj}(t)$ exhibits a series of peaks whose envelope decays over many phonon periods.
Our HEOM computations do capture the most prominent peak centered at $t=0$.
Still, the numerical instabilities arise well before the completion of the first phonon period $2\pi/\omega_0$, meaning that the HEOM results fully miss the contributions to $\mu_\mathrm{dc}$ from the peaks at later times.
Such contributions may be appreciable because of the slowly decaying envelope, and one may thus expect that the HEOM result underestimates the dc mobility.

For strong interactions and at higher temperatures, we do not observe numerical instabilities, see Fig.~\ref{Fig:Fig_strong_g}(b), and our HEOM method can again reach much longer times than real-time QMC methods.
Higher temperatures, in combination with our closing strategy, generally stabilize the HEOM.
Since the amplitude of the aforementioned later-time peaks is strongly suppressed at higher temperatures,~\cite{PhysRevB.99.104304} the HEOM method may provide reliable results for $\mu_\mathrm{dc}$.
The inset of Fig.~\ref{Fig:Fig_strong_g}(b) shows that mobility computations using either $\mathrm{Re}\:C_{jj}(t)$ or $\mathrm{Im}\:C_{jj}(t)$ [the latter in conjunction with the moving-average procedure, see the discussion of Fig.~\ref{Fig:finite_N_effect_123_160223}(b)] lead to virtually the same HEOM result, $\mu_\mathrm{dc}^\mathrm{HEOM}=0.0375$.

Finally, the very good agreement between the HEOM and QMC results in Figs.~\ref{Fig:Fig_strong_g}(a) and~\ref{Fig:Fig_strong_g}(b) strongly suggest that the values of $D$ employed in the corresponding HEOM computations are sufficiently large.
One may ask themselves whether a larger value of $D$ could mitigate the instabilities observed in Fig.~\ref{Fig:Fig_strong_g}(a).
Unfortunately, the results of Ref.~\onlinecite{JChemPhys.150.184109} show that the undamped-mode HEOM cannot be stabilized by further increasing $D$.

\section{Conclusion and outlook}
\label{Sec:Conclusion}
We develop the momentum-space HEOM method to compute real-time current--current correlation function at finite temperature for the 1d Holstein model, which provides us with a direct access to its transport properties.
By exploiting the decoupling of the $q=0$ phonon mode and transferring to the momentum representation, we greatly reduce the number of variables with respect to existing real-space HEOM implementations.
This circumstance has enabled us to obtain results that take into account all important phonon-assisted processes (i.e., $D$ is sufficiently large) on chains that are sufficiently long (i.e., $N$ is sufficiently large) so that the finite-size effects are minimized.
Another distinctive feature of our formalism is its specific hierarchy closing strategy [Eq.~\eqref{Eq:closing_strategy}], which has enabled us to overcome the numerical instabilities inherent to the undamped-mode HEOM formulated on a finite chain and truncated at a finite maximum hierarchy depth.
We are thus in position to provide reliable results for the dc mobility by computing $C_{jj}(t)$ up to quite long real times by which it has almost decayed to zero, i.e., the electronic motion has become diffusive.
A detailed analysis of how $C_{jj}(t)$, and thus $\mu_\mathrm{dc}$, depends on $N$ and $D$ suggests that our results for the temperature-dependent mobility should be considered with uncertainties typically below 10\%.
Still, the instabilities remain for strong couplings, at low temperatures, and for large phonon frequencies.
In such parameter regimes, we can obtain $C_{jj}(t)$ only up to relatively short real times.

Our momentum-space HEOM method, which has already achieved substantial memory savings with respect to existing real-space HEOM implementations, see Sec.~\ref{SSec:finite_N_D_effects}, may be combined with advanced propagation techniques based on the tensor formalism~\cite{JChemPhys.148.174102,JChemPhys.153.204109,WIREsComputMolSci.11.e1539,JChemPhys.154.194104,mangaud2023survey} or with on-the-fly filtering techniques~\cite{JChemPhys.130.084105} to treat longer chains or larger maximum depths.
This could pave the way towards HEOM computations at lower temperatures or for stronger interactions.
In such parameter regimes, accurate results for electronic dynamics can be obtained using the hierarchy of Davydov's ans\"{a}tze.~\cite{JChemPhys.158.080901,WIREsComputMolSci.12.1589}
However, its current implementations assume that the initial state of the electron--phonon system is factorized, and that phonons are in thermal equilibrium characteristic for the system with no electrons.
While following the electron's dynamics from such a state can provide information on the dc mobility,~\cite{JChemPhys.132.081101,JPhysChemB.115.5312,JChemPhys.147.214102} it provides no information on the frequency-dependent mobility, which follows from the current--current correlation function, see Sec.~\ref{SSec:model_definitions}.
Its computation, in turn, necessitates an appropriate representation of the thermal equilibrium state of the interacting electron and phonons, see Sec.~\ref{SSec:HEOM_real_time_theory}, which has still not been addressed using the hierarchy of Davydov's ans\"{a}tze, to the best of our knowledge.
Our study may motivate an extension of the Davydov's ans\"{a}tze approach to compute equilibrium correlation functions in parameter regimes that the most challenging for the HEOM method.

We also note that, at least in the limiting case of slow phonons, an interesting physical picture for the finite-frequency peak we observe in the optical response may be constructed based on the transient localization scenario.~\cite{AdvFunctMater.26.2292}
While this approach was originally devised to study transport limited by intermolecular vibrations modulating the hopping amplitude, its more recent refinements~\cite{PhysRevResearch.2.013001} and applications~\cite{SciPostPhys.11.2.039} have considered the Holstein model in the limit of vanishing carrier density.
A more detailed study in this direction might be a subject of future work.
We finally note that the method proposed in this manuscript cannot be directly applied to study carrier transport limited by the nonlocal electron--phonon interaction (e.g., the Peierls model).~\cite{JChemPhys.100.2335}
The developments reported here crucially lean on the fact that the current operator $j$ is a purely electronic operator, see Eq.~\eqref{Eq:def_j}.
In other words, the current–current correlation function $C_{jj}(t)$ can be computed using only the electronic RDM, see Sec.~\ref{SSec:HEOM_real_time_theory}.
On the other hand, when the electron interacts linearly with intermolecular phonons modulating the hopping integral, the current operator is a mixed electron–phonon operator.~\cite{PhysRevB.69.075212,JChemPhys.142.174103}
This means that $C_{jj}(t)$ cannot be computed using only the electronic RDM.
One should thus devise a procedure to retrieve correlation functions involving phonon operators from the HEOM formalism.
While this complex issue is well beyond the scope of this study, the herein developed momentum-space representation of the HEOM will remain beneficial to the overall computational performance of such a procedure.

\section*{Supplementary material}
See the supplementary material for the details of HEOM computations (values of $N$ and $D$ employed, reference HEOM results for the electron's kinetic energy and moments of $C_{jj}(\omega)$, as well as the relative accuracy with which different sum rules are satisfied). The supplementary material contains all the information needed to use our numerical data that are deposited on Zenodo platform, see Ref.~\onlinecite{veljko.jankovic.2023}.

\acknowledgments
The author thanks Nenad Vukmirovi\'c, Darko Tanaskovi\'c, and Petar Mitri\'c for numerous insightful and stimulating discussions.
The author acknowledges funding provided by the Institute of Physics Belgrade through a grant from the Ministry of Science, Technological Development, and Innovations of the Republic of Serbia.
Computational time on the ARIS supercomputing facility (GRNET, Athens, Greece) was granted by the NI4OS-Europe network under the CoNTraSt project (Open Call 2022, project No. ni4os002).
Computations were also partially performed on the PARADOX-IV supercomputing facility at the Scientific Computing Laboratory, National Center of Excellence for the Study of Complex Systems, Institute of Physics Belgrade.

\appendix
\section{Derivation of the HEOM for the real-time current--current correlation function}\label{App:real-time-heom}
The decoupling of $q=0$ phonon mode from the rest of the system permits us to consider Hamiltonians $H'_\mathrm{ph}={\sum'_q} \omega_q b_q^\dagger b_q$ and $H'_\mathrm{e-ph}=\sum'_q V_qB_q$ without the $q=0$ term, which is signalled by primes. 
The Feynman--Vernon influence functional theory~\cite{AnnPhys.24.118} implies that the only phonon quantity that determines the dynamics of $\iota(t)$ is the equilibrium free-phonon correlation function (we assume $t>0$ and $q_2,q_1\neq 0$)
\begin{equation}
\label{Eq:def_C_ph}
\begin{split}
    \mathcal{C}_{q_2q_1}(t)&=\mathrm{Tr}'_\mathrm{ph}\left\{B_{q_2}^{(I)}(t)B_{q_1}\frac{e^{-\beta H'_\mathrm{ph}}}{Z'_\mathrm{ph}}\right\}\\
    &=\delta_{q_1,-q_2}\sum_{m=0}^1c_{q_2m}e^{-\mu_{q_2m}t}.
\end{split}
\end{equation}
The time dependence in Eq.~\eqref{Eq:def_C_ph} is in the interaction picture [signalled by the superscript $(I)$], $B_q^{(I)}(t)=e^{iH'_\mathrm{ph}t}B_qe^{-iH'_\mathrm{ph}t}$, $Z'_\mathrm{ph}=\mathrm{Tr}'_\mathrm{ph}\:e^{-\beta H'_\mathrm{ph}}$ is the free-phonon partition sum, while
\begin{eqnarray}
    c_{q0}=\left(\frac{g}{\sqrt{N}}\right)^2\frac{1}{1-e^{-\beta\omega_q}},\quad\mu_{q0}=+i\omega_q,\label{Eq:def_c_q0}\\
    c_{q1}=\left(\frac{g}{\sqrt{N}}\right)^2\frac{1}{e^{\beta\omega_q}-1},\quad\mu_{q1}=-i\omega_q\label{Eq:def_c_q1}.
\end{eqnarray}
The partial trace over phonons in Eq.~\eqref{Eq:def_iota}
is performed in the same manner as in Refs.~\onlinecite{JChemPhys.141.044114,JChemPhys.143.194106,PhysRevB.105.054311,JChemPhys.156.244102}, and the final result for $\iota$ in the interaction picture reads as
\begin{equation}
\label{Eq:def_rho_j_I_general}
    \iota^{(I)}(t)=\mathcal{T}e^{-[\Phi_1(t)+\Phi_2(\beta)+\Phi_3(t,\beta)]}j\frac{e^{-\beta H_\mathrm{e}}}{Z_e}.
\end{equation}
Here, $Z_e=Z/Z'_\mathrm{ph}$ is the electronic partition sum [Eq.~\eqref{Eq:el_part_sum}], while the influence phases are given as
\begin{equation}
\label{Eq:def_Phi_1}
\begin{split}
    &\Phi_1(t)=\sideset{}{'}\sum_{qm}\int_0^t ds_2\int_0^{s_2}ds_1\:V_q^{(I)}(s_2)^\times\: e^{-\mu_{qm}(s_2-s_1)}\\&\times\left[\frac{c_{qm}+c_{q\overline{m}}}{2}\:V_{-q}^{(I)}(s_1)^\times+\frac{c_{qm}-c_{q\overline{m}}}{2}\:V_{-q}^{(I)}(s_1)^\circ\right]
\end{split}
\end{equation}
\begin{equation}
\label{Eq:def_Phi_2}
\begin{split}
    \Phi_2(\beta)=&-\sideset{}{'}\sum_{qm}\int_0^\beta d\tau_2\int_0^{\tau_2}d\tau_1\:^C\overline{V}_{-q}(\tau_1)\\&\times e^{i\mu_{qm}(\tau_2-\tau_1)}c_{qm}\:^C\overline{V}_q(\tau_2)
\end{split}
\end{equation}
\vspace{0.5mm}
\begin{equation}
\label{Eq:def_Phi_3}
\begin{split}
    \Phi_3(t,\beta)=&-i\sideset{}{'}\sum_{qm}\int_0^t ds\int_0^\beta d\tau\:V_q^{(I)}(s)^\times\\&\times e^{-\mu_{qm}s}e^{i\mu_{qm}(\beta-\tau)}c_{qm}\:^C\overline{V}_{-q}(\tau)
\end{split}
\end{equation}
The influence phase $\Phi_1(t)$ describes the pure real-time evolution, and the hyperoperators $V^\times$ and $V^\circ$ entering Eq.~\eqref{Eq:def_Phi_1} act on an arbitrary operator $O$ as $V^\times O=[V,O]$ (commutator) and $V^\circ O=\{V,O\}$ (anticommutator), respectively.
We define $\overline{m}=1$ for $m=0$ and \emph{vice versa}.
The influence phase $\Phi_2(\beta)$ represents the pure imaginary-time evolution, while $\Phi_3(t,\beta)$ takes into account the contributions mixing the real-time and imaginary-time evolutions.
The imaginary time-dependent operator in the interaction picture is defined as $\overline{V}(\tau)=e^{H_\mathrm{e}\tau}Ve^{-H_\mathrm{e}\tau}$, while the hyperoperator $^CV$ appearing in Eqs.~\eqref{Eq:def_Phi_2} and~\eqref{Eq:def_Phi_3} acts on an arbitrary operator $O$ as $^CVO=OV$. 
The time ordering symbol $\mathcal{T}$ imposes the following hyperoperator ordering: the hyperoperators depending on real time act after the imaginary time-dependent hyperoperators, the arguments of real time-dependent hyperoperators are chronologically ordered, while the arguments of imaginary-time dependent hyperoperators are anti-chronologically ordered.
This ensures that the general term in the expansion of Eq.~\eqref{Eq:def_rho_j_I_general} is of the form
\begin{equation*}
    V_{q_n}^{(I)}(s_n)^{\pi_n}\dots V_{q_1}^{(I)}(s_1)^{\pi_1}j\frac{e^{-\beta H_\mathrm{e}}}{Z_\mathrm{e}}\overline{V}_{p_m}(\tau_m)\dots\overline{V}_{p_1}(\tau_1)
\end{equation*}
where $n+m$ is even, $\pi_1,\dots,\pi_n\in\{\times,\circ\}$, $t\geq s_n\geq\dots\geq s_1\geq 0$, $\beta\geq\tau_m\geq\dots\geq\tau_1\geq 0$, and
$q_n+\dots+q_1+p_m+\dots+p_1=0$.

Starting from Eq.~\eqref{Eq:def_rho_j_I_general}, the HEOM [Eq.~\eqref{Eq:real-time-HEOM-before-closing-paper}] is formulated in the standard manner.~\cite{JChemPhys.130.234111,PhysRevB.105.054311}
The ADO $\iota_\mathbf{n}^{(n)}$ is defined (in the interaction picture) as
\begin{widetext}
\begin{equation}
\label{Eq:def_iota_adm}
    \begin{split}
        \iota_\mathbf{n}^{(I,n)}(t)=f(\mathbf{n})\mathcal{T}\sideset{}{'}\prod_{qm}\left\{i\int_0^t ds\:e^{-\mu_{qm}(t-s)}\left[\frac{c_{qm}+c_{q\overline{m}}}{2}\:V_{-q}^{(I)}(s)^\times+\frac{c_{qm}-c_{q\overline{m}}}{2}\:V_{-q}^{(I)}(s)^\circ\right]\right. \\ \left. +e^{-\mu_{qm}t}\int_0^\beta d\tau\:e^{i\mu_{qm}(\beta-\tau)}c_{qm}\:^C\overline{V}_{-q}(\tau)\right\}^{n_{qm}}e^{-[\Phi_1(t)+\Phi_2(\beta)+\Phi_3(t,\beta)]}\:j\frac{e^{-\beta H_\mathrm{e}}}{Z_\mathrm{e}}.
    \end{split}
\end{equation}
\end{widetext}
The rescaling factor $f(\mathbf{n})$ reads as~\cite{JChemPhys.130.084105}
\begin{equation}
\label{Eq:def_rescaling}
    f(\mathbf{n})=\sideset{}{'}\prod_{qm}\left[c_{qm}^{n_{qm}}n_{qm}!\right]^{-1/2}.
\end{equation}
Equation~\eqref{Eq:def_iota_adm} explicitly shows that $\iota_\mathbf{n}^{(n)}$ decreases the momentum of the electronic subsystem by $k_\mathbf{n}=\sum'_{qm}qn_{qm}$, meaning that only $N$ matrix elements $\langle k|\iota_\mathbf{n}^{(n)}|k+k_\mathbf{n}\rangle$ connecting the free-electron states whose momenta differ by $k_\mathbf{n}$ assume nonzero values.

Setting $t=0$ in Eq.~\eqref{Eq:def_iota_adm} gives the following initial condition for the real-time HEOM:
\begin{widetext}
\begin{equation}
\label{Eq:iota_adm_init_cond}
    \iota_\mathbf{n}^{(n)}(0)=jf(\mathbf{n})\mathcal{T}\sideset{}{'}\prod_{qm}\left\{\int_0^\beta d\tau\:e^{i\mu_{qm}(\beta-\tau)}c_{qm}\:^C\overline{V}_{-q}(\tau)\right\}^{n_{qm}}e^{-\Phi_2(\beta)}\:\frac{e^{-\beta H_\mathrm{e}}}{Z_\mathrm{e}}.
\end{equation}
\end{widetext}
In Eq.~\ref{Eq:iota_adm_init_cond}, we may move the current operator to the left-most position because all the hyperoperators act on the operator $je^{-\beta H_\mathrm{e}}/Z_\mathrm{e}$ from the right.
Similarly as in Ref.~\onlinecite{PhysRevB.105.054311}, one may now recognize that $\iota_\mathbf{n}^{(n)}(0)$ is the product of operator $j$ and the ADO $\sigma_\mathbf{n}^{(n)}(\beta)$, which is one of the components in the hierarchical representation of the (unnormalized) reduced density operator
\begin{equation}
\label{Eq:unnorm_rdo}
    \sigma_\mathbf{0}^{(0)}(\beta)=\mathcal{T}e^{-\Phi_2(\beta)}e^{-\beta H_\mathrm{e}}.
\end{equation}
To actually evaluate $\sigma_\mathbf{n}^{(n)}(\beta)$ and the normalization constant $Z_e=\mathrm{Tr}_\mathrm{e}\sigma_\mathbf{0}^{(0)}(\beta)$, we consider the imaginary time-dependent analog of Eq.~\eqref{Eq:unnorm_rdo}
\begin{equation}
\label{Eq:unnorm_rdo_tau_dep}
    \sigma_\mathbf{0}^{(0)}(\tau)=\mathcal{T}e^{-\Phi_2(\tau)}e^{-\tau H_\mathrm{e}},
\end{equation}
where $0\leq\tau\leq\beta$.
As discussed in detail in Ref.~\onlinecite{PhysRevB.105.054311}, Eq.~\eqref{Eq:unnorm_rdo_tau_dep} can be transformed into the imaginary-time HEOM given in Eq.~\eqref{Eq:im-time-HEOM-eq-paper}.

\section{Derivation of the closing scheme in Eq.~\eqref{Eq:closing_strategy}}\label{App:closing}
Here, we provide a detailed derivation of our strategy for hierarchy closing that is embodied in Eq.~\eqref{Eq:closing_strategy}.

Let us consider the equation for $\langle k|{\iota}_\mathbf{D}^{(D)}({t})|k+k_\mathbf{D}\rangle$ at the maximum depth $D$, which contains the ADOs ${\iota}_{\mathbf{D}_{qm}^+}^{(D+1)}$ at depth $D+1$.
The equation of motion for ${\iota}_{\mathbf{D}_{qm}^+}^{(D+1)}$ contains ADOs at depth $D+2$, which will be set to zero.
Moreover, its coupling with the ADOs at depth $D$ will be restricted by assuming it couples back only to the original ADO ${\iota}_\mathbf{D}^{(D)}$ we are considering.
With these assumptions, the equation at depth $D+1$ reads as
\begin{equation}
\label{Eq:eq-D-plus-1}
    \begin{split}
       &\partial_{{t}}\langle k|{\iota}_{\mathbf{D}_{qm}^+}^{(D+1)}({t})|k+k_\mathbf{D}+q\rangle=\\
        &-i({\varepsilon}_k-{\varepsilon}_{k+k_\mathbf{D}+q}+{\mu}_{\mathbf{D}_{qm}^+})\langle k|{\iota}_{\mathbf{D}_{qm}^+}^{(D+1)}({t})|k+k_\mathbf{D}+q\rangle\\
        &+i\sqrt{(1+D_{qm})c_{qm}}\:\langle k+q|{\iota}_{\mathbf{D}}^{(D)}({t})|k+k_\mathbf{D}+q\rangle\\
        &-i\sqrt{1+D_{qm}}\frac{c_{q\overline{m}}}{\sqrt{c_{qm}}}\:\langle k|{\iota}_{\mathbf{D}}^{(D)}({t})|k+k_\mathbf{D}\rangle.
    \end{split}
\end{equation}
Using the initial condition $\langle k|\iota_{\mathbf{D}_{qm}^+}^{(D+1)}(0)|k+k_\mathbf{D}+q\rangle=0$, which is appropriate because the imaginary-time HEOM in Eq.~\eqref{Eq:im-time-HEOM-eq-paper} is truncated at the maximum depth $D$, Eq.~\eqref{Eq:eq-D-plus-1} can be formally integrated to yield
\begin{equation}
\label{Eq:formal-solution-D-plus-1}
\begin{split}
    &\langle k|\iota_{\mathbf{D}_{qm}^+}^{(D+1)}({t})|k+k_\mathbf{D}+q\rangle=i\sqrt{1+D_{qm}}\times\\&\int_0^{{t}}dt_1\:e^{-i({\varepsilon}_k-{\varepsilon}_{k+k_\mathbf{D}+q}+{\mu}_{\mathbf{D}_{qm}^+})({t}-t_1)}\times\\
    &\biggl[\sqrt{c_{qm}}\:e^{-i({\varepsilon}_{k+q}-{\varepsilon}_{k+k_\mathbf{D}+q}+{\mu}_\mathbf{D})t_1}f_1(t_1)-\\&\frac{c_{q\overline{m}}}{\sqrt{c_{qm}}}\:e^{-i({\varepsilon}_{k}-{\varepsilon}_{k+k_\mathbf{D}}+{\mu}_\mathbf{D})t_1}f_2(t_1)\biggr].
\end{split}
\end{equation}
We have introduced the auxiliary function $f_{1}(t)$ representing the slowly changing part of the ADO at depth $D$ by factoring out the oscillating (rapidly changing) part as follows:
\begin{equation}
\begin{split}
    &\langle k+q|\iota_{\mathbf{D}}^{(D)}({t})|k+k_\mathbf{D}+q\rangle=\\&e^{-i({\varepsilon}_{k+q}-{\varepsilon}_{k+k_\mathbf{D}+q}+{\mu}_\mathbf{D}){t}}f_1({t}),
\end{split}
\end{equation}
and similarly for $f_2(t)$.
Under the integral entering Eq.~\eqref{Eq:formal-solution-D-plus-1} we introduce the variable change ${s}={t}-{t}_1$ and subsequently apply the Markovian approximation $f_{1/2}({t}-{s})\approx f_{1/2}({t})$ to the slowly changing part. As a result, we obtain
\begin{equation}
    \begin{split}
        &\langle k|\iota_{\mathbf{D}_{qm}^+}^{(D+1)}({t})|k+k_\mathbf{D}+q\rangle=i\sqrt{(1+D_{qm})}\times\\&\biggl\{\sqrt{c_{qm}}\int_0^{{t}}d{s}\:e^{-i[{\varepsilon}_k-{\varepsilon}_{k+q}+(\delta_{m0}-\delta_{m1})\omega_q]{s}}\times\\&\langle k+q|\iota_{\mathbf{D}}^{(D)}({t})|k+k_\mathbf{D}+q\rangle-\\
        &\frac{c_{q\overline{m}}}{\sqrt{c_{qm}}}\int_0^{{t}}d{s}\:e^{-i[{\varepsilon}_{k+k_\mathbf{D}}-{\varepsilon}_{k+k_\mathbf{D}+q}+(\delta_{m0}-\delta_{m1})\omega_q]{s}}\times\\&\langle k|\iota_{\mathbf{D}}^{(D)}({t})|k+k_\mathbf{D}\rangle\biggr\}.
    \end{split}
\end{equation}
Using the last result, we write one of the terms that couple the equations at depths $D$ and $D+1$ [the third term on the RHS of Eq.~\eqref{Eq:real-time-HEOM-before-closing-paper}] as follows:
\begin{equation}
\label{Eq:higher-order-2}
    \begin{split}
        &-i\sideset{}{'}\sum_{qm}\sqrt{(1+D_{qm})c_{qm}}\:\langle k|\iota_{\mathbf{D}_{qm}^+}^{(D+1)}({t})|k+k_\mathbf{D}+q\rangle=\\
        &\sideset{}{'}\sum_{qm}(1+D_{qm})c_{qm}\int_0^{{t}}d{s}\:e^{-i[{\varepsilon}_k-{\varepsilon}_{k+q}+(\delta_{m0}-\delta_{m1})\omega_q]{s}}\times\\&\langle k+q|\iota_\mathbf{D}^{(D)}({t})|k+k_\mathbf{D}+q\rangle-\\
        &\sideset{}{'}\sum_{qm}(1+D_{qm})c_{q\overline{m}}\int_0^{{t}}d{s}\:e^{-i[{\varepsilon}_{k+k_\mathbf{D}}-{\varepsilon}_{k+k_\mathbf{D}+q}+(\delta_{m0}-\delta_{m1})\omega_q]{s}}\times\\&\langle k|\iota_\mathbf{D}^{(D)}({t})|k+k_\mathbf{D}\rangle
    \end{split}
\end{equation}
The second term on the RHS of Eq.~\eqref{Eq:real-time-HEOM-before-closing-paper} can be written and analyzed in an analogous manner.

To evaluate Eq.~\eqref{Eq:higher-order-2}, we have to know the detailed structure of the vector $\mathbf{D}$ that characterizes the ADOs at the maximum depth $D$.
We also observe that the second term on the RHS of Eq.~\eqref{Eq:higher-order-2} depends only on the quantity $\langle k|\iota_\mathbf{D}^{(D)}({t})|k+k_\mathbf{D}\rangle$ whose differential equation we are considering.
On the other hand, the matrix elements of $\iota_\mathbf{D}^{(D)}$ entering the first term on the RHS of Eq.~\eqref{Eq:higher-order-2} depend on $q$.
This term may be neglected by invoking a sort of the random phase approximation: the matrix elements $\langle k+q|\iota_\mathbf{D}^{(D)}({t})|k+k_\mathbf{D}+q\rangle$ are typically oscillatory functions of $q$, and one may then argue that their average over all momenta $q$ is close to zero.
A similar approximation is also at the heart of the momentum-average approximation.~\cite{PhysRevLett.97.036402,PhysRevB.74.245104}
In the remaining terms, we: (i) exploit that for dispersionless optical phonons ($\omega_q\equiv\omega_0$) $c_{qm}$ does not depend on $q$, (ii) neglect the contributions containing $D_{qm}$, and (iii) evaluate the remaining integral in Eq.~\eqref{Eq:higher-order-2} by invoking the so-called adiabatic approximation, in which the upper limit ${t}$ of the integral is replaced by $+\infty$.
We furthermore keep only the real part of the integral evaluated, so that
    \begin{equation}
    \begin{split}
        \int_0^{{t}}d{s}\:e^{-i{\Omega}{s}}\approx
        \int_0^{+\infty}d{s}\:e^{-i{\Omega}{s}}\approx\pi\delta({\Omega}).
    \end{split}
    \end{equation}
This way, we introduce physically motivated damping of the ADOs at the terminal level of the hierarchy.
The damping effectively replaces higher-order phonon-assisted processes that are not explicitly taken into account, while phonon-assisted processes involving at most $D$ phonons are completely taken into account through the HEOM.
In the resulting expressions, we recognize the second-order approximation for carrier scattering time out of the free-electron state $|k\rangle$
\begin{equation}
\label{Eq:def_tau_k}
\begin{split}
    \frac{1}{\tau_k}=2\pi\frac{g^2}{N}\sideset{}{'}\sum_q\left[(1-e^{-\beta\omega_0})^{-1}\delta(\varepsilon_k-\varepsilon_{k-q}-\omega_0) \right. \\ \left. +(e^{\beta\omega_0}-1)^{-1}\delta(\varepsilon_k-\varepsilon_{k-q}+\omega_0)\right]
\end{split}
\end{equation}
for which an analytical expression in the limit of infinite $N$ has been derived in Refs.~\onlinecite{PhysRevB.99.104304} and~\cite{arxiv.2212.13846}.
Even though we perform HEOM computations on finite chains, in Eq.~\eqref{Eq:closing_strategy} we use the value of $\tau_k$ obtained in the infinite-chain limit.
We believe that such a procedure is appropriate because we invested great efforts to perform computations with sufficiently large $N$.

Instead of neglecting the term containing $D_{qm}$ in Eq.~\eqref{Eq:higher-order-2}, we may have replaced $D_{qm}$ by its average value $D/[2(N-1)]$ at depth $D$ [the sum $\sum'_{qm}D_{qm}$, which is equal to $D$, contains $2(N-1)$ terms].
Such a replacement would lead to larger damping factors at depth $D$, which may have already been overestimated by using the value of $\tau_k$ in the infinite-chain limit.
Since the substitution $D_{qm}\to D/[2(N-1)]$ does not improve the stability of the hierarchy at low temperatures, for strong interactions, and for fast phonons, we decided not to use it in our computations.

\section{Sum rules}\label{App:sum_rules}
The zeroth-moment sum rule states that
\begin{equation}
    M_0=\int_{-\infty}^{+\infty}\frac{d\omega}{2\pi}\:C_{jj}(\omega)=
    \langle j^2\rangle.
\end{equation}
Since $j$ is a purely electronic operator, the evaluation of $M_0$ necessitates only the operator $\iota_\mathbf{0}^{(0)}(t=0)$:
\begin{equation}
    M_0=-2J\sum_k\sin(k)\langle k|\iota_\mathbf{0}^{(0)}(t=0)|k\rangle\equiv C_{jj}(t=0).
\end{equation}

On the other hand, to evaluate $M_n$ for $n\geq 1$, we need all ADOs $\iota_\mathbf{n}^{(n)}(t=0)$ at depths starting from 0 and concluding with $n$.
The first-moment sum rule is
\begin{equation}
    M_1=\int_{-\infty}^{+\infty}\frac{d\omega}{2\pi}\:\omega\:C_{jj}(\omega)=
    \left\langle\left[j,H\right]j\right\rangle.
\end{equation}
It can be shown that
\begin{equation}
\begin{split}
\label{Eq:comm_j_H}
    &[j,H]=\left[j,H_\mathrm{e-ph}\right]=\\
    &-2J\sum_{q\neq 0,p}
    \left[\sin(p+q)-\sin(p)\right]|p+q\rangle\langle p|B_q.
\end{split}
\end{equation}
Therefore,
\begin{equation}
\begin{split}
    &M_1=
    -2J\sum_{q\neq 0,p}
    \left[\sin(p+q)-\sin(p)\right]\\
    &\times\left\langle p\left|\mathrm{Tr}_\mathrm{ph}\left\{ B_q j\frac{e^{-\beta H}}{Z}\right\}\right|p+q\right\rangle.
\end{split}
\end{equation}
The second-moment sum rule is somewhat more cumbersome to evaluate
\begin{equation}
\begin{split}
    M_2=\int_{-\infty}^{+\infty}\frac{d\omega}{2\pi}\:\omega^2\:C_{jj}(\omega)=
    \left\langle\left[\left[j,H\right],H\right]j\right\rangle.
\end{split}
\end{equation}
We separately evaluate the three contributions to the double commutator
\begin{equation}
  [[j,H],H]=\underbrace{[[j,H],H_\mathrm{e}]}_{K_1}+\underbrace{[[j,H],H_\mathrm{ph}]}_{K_2}+\underbrace{[[j,H],H_\mathrm{e-ph}]}_{K_3} 
\end{equation}
starting from the result for $[j,H]$ embodied in Eq.~\eqref{Eq:comm_j_H}. We obtain
\begin{equation}
\begin{split}
    K_1=-2
    J\sum_{q\neq 0,p}
    &\left[\sin(p+q)-\sin(p)\right]
    \left(\varepsilon_p-\varepsilon_{p+q}\right)\times\\&|p+q\rangle\langle p|B_q,
\end{split}
\end{equation}
\begin{equation}
\begin{split}
    K_2=-2
    J\sum_{q\neq 0,p}
    &\left[\sin(p+q)-\sin(p)\right]
    \omega_q
    \frac{g}{\sqrt{N}}\times\\&\left(|p+q\rangle\langle p|b_q-|p+q\rangle\langle p|b_{-q}^\dagger\right),
\end{split}
\end{equation}
\begin{equation}
\begin{split}
    &K_3=-2
    J\sum_{\substack{q_1\neq 0\\q_2\neq 0}}\sum_p|p+q_1+q_2\rangle\langle p|B_{q_2}B_{q_1}\times\\&\left[\sin(p+q_1+q_2)-\sin(p+q_1)-\sin(p+q_2)+\sin(p)\right]
\end{split}
\end{equation}
In addition to the single-phonon-assisted ADO $\mathrm{Tr}_\mathrm{ph}\left\{ B_q j\frac{e^{-\beta H}}{Z}\right\}$, the evaluation of $\langle K_1j\rangle$, $\langle K_2j\rangle$, and $\langle K_3j\rangle$ necessitates its contributions describing phonon absorption and emission, which stem from the definition $B_q=\frac{g}{\sqrt{N}}\left(b_q+b_{-q}^\dagger\right)$, as well as the two-phonon-assisted ADO $\mathrm{Tr}_\mathrm{ph}\left\{ B_{q_2}B_{q_1} j\frac{e^{-\beta H}}{Z}\right\}$.
In the following, we compute these (electronic) operators using the ideas developed in Refs.~\onlinecite{JChemPhys.137.194106,JChemPhys.145.224105}.

Denoting $\frac{e^{-\beta H}}{Z}=\rho_T^\mathrm{eq}$, we can write
\begin{equation}
\begin{split}
    j\rho_T^\mathrm{eq}=&\mathcal{T}\exp\left[-\int_0^{\beta} d\tau\:\sideset{}{'}\sum_q\: ^C\overline{V}_q(\tau)^C\overline{B}_q(\tau)\right]\times\\&j\frac{e^{-\beta H_\mathrm{e}}}{Z_\mathrm{e}}\frac{e^{-\beta H'_\mathrm{ph}}}{Z'_\mathrm{ph}}.
\end{split}
\end{equation}
To evaluate phonon-assisted contributions, we introduce auxiliary fields $f_q(\tau)$ and consider their functional
\begin{equation}
\begin{split}
    j\rho_{T,f}^\mathrm{eq}=&\mathcal{T}\exp\left[-\int_0^{\beta} d\tau\:\sideset{}{'}\sum_q\:^C [\overline{V}_q(\tau)+f_q(\tau)]\:^C\overline{B}_q(\tau)\right]\times\\&j\frac{e^{-\beta H_\mathrm{e}}}{Z_\mathrm{e}}\frac{e^{-\beta H'_\mathrm{ph}}}{Z'_\mathrm{ph}}.
\end{split}
\end{equation}
Defining
\begin{equation}
    j\rho_f^\mathrm{eq}=j\mathrm{Tr}_\mathrm{ph}\rho_{T,f}^\mathrm{eq},
\end{equation}
one can show that
\begin{equation}
    \left[j\frac{\delta\rho_{f}^\mathrm{eq}}{\delta f_q(\beta)}\right]_{f=0}=-
    \mathrm{Tr}_\mathrm{ph}\left\{B_qj\rho_T^\mathrm{eq}\right\},
\end{equation}
\begin{equation}
    \left[j\frac{\delta^2\rho_{f}^\mathrm{eq}}{\delta f_{q_2}(\beta)\delta f_{q_1}(\beta)}\right]_{f=0}=
    \mathrm{Tr}_\mathrm{ph}\left\{B_{q_2}B_{q_1}j\rho_T^\mathrm{eq}\right\}.
\end{equation}
On the other hand,
\begin{equation}
    \rho_f^\mathrm{eq}=\mathcal{T}e^{-\Phi_{2,f}(\beta)}\frac{e^{-\beta H_\mathrm{e}}}{Z_\mathrm{e}},
\end{equation}
where the influence phase $\Phi_{2,f}(\beta)$ is a functional of auxiliary fields $f_q(\tau)$ [cf. Eq.~\eqref{Eq:def_Phi_2}]
\begin{equation}
\begin{split}
    &\Phi_{2,f}(\beta)=-\int_0^{\beta}d\tau_2\int_0^{\tau_2}d\tau_1\sideset{}{'}\sum_{qm}\:^C[\overline{V}_{-q}(\tau_1)+f_{-q}(\tau_1)]\times\\&e^{i\mu_{qm}(\tau_2-\tau_1)}c_{qm}\:^C[\overline{V}_q(\tau_2)+f_q(\tau_2)].
\end{split}
\end{equation}
It then follows that
\begin{equation}
    \frac{\delta\rho_f^\mathrm{eq}}{\delta f_{q_1}(\beta)}=\mathcal{T}\left\{-\frac{\delta\Phi_{2,f}(\beta)}{\delta f_{q_1}(\beta)}e^{-\Phi_{2,f}(\beta)}\right\}\frac{e^{-\beta H_\mathrm{e}}}{Z_\mathrm{e}}
\end{equation}
\begin{equation}
\begin{split}
   &\frac{\delta^2\rho_{f}^\mathrm{eq}}{\delta f_{q_2}(\beta)\delta f_{q_1}(\beta)}=\\&\mathcal{T}\left\{\left[-\frac{\delta^2\Phi_{2,f}(\beta)}{\delta f_{q_2}(\beta)\delta f_{q_1}(\beta)}+\frac{\delta\Phi_{2,f}(\beta)}{\delta f_{q_2}(\beta)}\frac{\delta\Phi_{2,f}(\beta)}{\delta f_{q_1}(\beta)}\right]e^{-\Phi_{2,f}(\beta)}\right\}\times\\&\frac{e^{-\beta H_\mathrm{e}}}{Z_\mathrm{e}}
\end{split}
\end{equation}
The explicit evaluation of the functional derivatives produces
\begin{equation}
\begin{split}
    \frac{\delta\Phi_{2,f}(\beta)}{\delta f_{q_1}(\beta)}=&-\int_0^{\beta}d\tau\sum_m c_{q_1m}\:e^{i\mu_{q_1m}(\beta-\tau)}\times\\&^C[\overline{V}_{-q_1}(\tau)+f_{-q_1}(\tau)],
\end{split}
\end{equation}
\begin{equation}
    \frac{\delta^2\Phi_f(\beta)}{\delta f_{q_2}(\beta)\delta f_{q_1}(\beta)}=
    -\delta_{q_2,-q_1}\sum_m c_{q_1m}.
\end{equation}
Using the definition of ADOs $\iota_\mathbf{n}^{(n)}$ at $t=0$ [Eq.~\eqref{Eq:def_iota_adm}] and rescaling factors $f(\mathbf{n})$ [Eq.~\eqref{Eq:def_rescaling}], we finally obtain 
\begin{equation}
    \mathrm{Tr}_\mathrm{ph}\left\{B_qj\rho_T^\mathrm{eq}\right\}=-\sum_m \frac{1}{f(\mathbf{0}_{qm}^+)}\iota_{\mathbf{0}_{qm}^+}^{(1)}(t=0),
\end{equation}
\begin{equation}
\begin{split}
    &\mathrm{Tr}_\mathrm{ph}\left\{B_{q_2}B_{q_1}\rho_T^\mathrm{eq}\right\}=\delta_{q_2,-q_1}\left(
    \sum_m c_{q_1m}\right)\iota_\mathbf{0}^{(0)}(t=0)+\\&\sum_{m_2m_1}\frac{1}{f(\mathbf{0}_{(q_1m_1),(q_2m_2)}^{++})}\iota_{\mathbf{0}_{(q_1m_1),(q_2m_2)}^{++}}^\mathrm{(2)}(t=0).
\end{split}
\end{equation}
The contributions that differentiate between absorption and emission of a single phonon are most straightforwardly obtained by explicitly setting up an equation of motion for such processes and comparing the kinetic term in the equation thus obtained with the kinetic term in Eq.~\eqref{Eq:real-time-HEOM-before-closing-paper}. This procedure yields
\begin{equation}
    \frac{g}{\sqrt{N}}
    \mathrm{Tr}_\mathrm{ph}\left\{b_qj\rho_T^\mathrm{eq}\right\}=
    -\frac{1}{f(\mathbf{0}_{q0}^+)}
    \iota_{\mathbf{0}_{q0}^+}^{(1)}(t=0)
\end{equation}
\begin{equation}
    \frac{g}{\sqrt{N}}
    \mathrm{Tr}_\mathrm{ph}\left\{b_{-q}^\dagger j\rho_T^\mathrm{eq}\right\}=
    -\frac{1}{f(\mathbf{0}_{q1}^+)}
    \iota_{\mathbf{0}_{q1}^+}^{(1)}(t=0).
\end{equation}
The final result for the first-moment sum rule reads as
\begin{equation}
\begin{split}
    M_1=2J\sum_{q\neq 0,p}\sum_m&\frac{\sin(p+q)-\sin(p)}{f(\mathbf{0}_{qm}^+)}\times\\&\langle
    p|\iota_{\mathbf{0}_{qm}^+}^{(1)}(t=0)|p+q\rangle.
\end{split}
\end{equation}
The final result for the second-moment sum rule reads as
\begin{equation}
    M_2=\langle K_1j\rangle+\langle K_2j\rangle+\langle K_3j\rangle,
\end{equation}
where
\begin{equation}
\begin{split}
    \langle K_1j\rangle=2J\sum_{q\neq 0,p}\sum_m&\frac{\sin(p+q)-\sin(p)}{f(\mathbf{0}_{qm}^+)}(\varepsilon_p-\varepsilon_{p+q})\times\\&\langle p|\iota_{\mathbf{0}_{qm}^+}^{(1)}(t=0)|p+q\rangle,
\end{split}
\end{equation}
\begin{equation}
\begin{split}
    \langle K_2j\rangle=2J\sum_{q\neq 0,p}\sum_m&(-1)^m\frac{\sin(p+q)-\sin(p)}{f(\mathbf{0}_{qm}^+)}\omega_q\times\\&\langle p|\iota_{\mathbf{0}_{qm}^+}^{(1)}(t=0)|p+q\rangle,
\end{split}
\end{equation}
and
\begin{equation}
\begin{split}
&\langle K_3j\rangle=-2J\sum_{q\neq 0,p}\sum_m 4\sin(p)\sin^2(q/2)c_{qm}\times\\
&\langle p|\iota_\mathbf{0}^{(0)}(t=0)|p\rangle-\\
&-2J\sum_{\substack{q_2,q_1\neq 0\\p}}\sum_{m_2m_1}\frac{\langle p|\iota_{\mathbf{0}_{(q_1m_1),(q_2m_2)}^{++}}^{(2)}(t=0)|p+q_1+q_2\rangle}{f(\mathbf{0}_{(q_1m_1),(q_2m_2)}^{++})}\times\\
&\left[\sin(p+q_1+q_2)-\sin(p+q_1)-\sin(p+q_2)+\sin(p)\right]
\end{split}
\end{equation}

\section{Optical sum rule and finite-size effects}
\label{App:OSR_finite_N}
We find that the relative accuracy with which the OSR is satisfied increases with increasing $N$, see Sec.~\ref{SSec:numerical_sum_rules}.
This observation suggests that the OSR formulated in Eq.~\eqref{Eq:OSR} is strictly valid only in the long-chain limit, while on finite chains there are finite-size corrections that vanish as $N\to+\infty$.
In the following, we provide a derivation of the OSR for finite $N$, identify the finite-size corrections, and demonstrate that they vanish in the long-chain limit.

Let us start by noting that the current operator on an $N$-site chain, defined in Eq.~\eqref{Eq:def_j}, may be expressed in the site basis as
\begin{equation}
\label{Eq:def_j_site_basis}
    j=iJ\sum_{m=0}^{N-1}\left(|m\rangle\langle m\oplus 1|-|m\oplus 1\rangle\langle m|\right).
\end{equation}
The cyclic addition $\oplus$ [$a\oplus b=(a+b)\:\mathrm{mod}\:N$] takes into account that sites 0 and $N-1$ are first neighbors because of the periodic boundary conditions.
Our derivation of the OSR makes use of the continuity equation
\begin{equation}
\label{Eq:continuity}
    \widetilde{j}=\frac{dP}{dt}=-i[P,H],
\end{equation}
where the polarization operator (in the site basis) is given as
\begin{equation}
    P=-\sum_{m=0}^{N-1}m|m\rangle\langle m|.
\end{equation}
While the current operators $\widetilde{j}$ and $j$ are identical in the infinite-chain limit, their difference on a finite chain reads as
\begin{equation}
\label{Eq:finite-size-correction}
    \widetilde{j}-j=-iJ\widehat{P}_{N-1,0},
\end{equation}
where
\begin{equation}
    \widehat{P}_{N-1,0}=N\left(|N-1\rangle\langle 0|-|0\rangle\langle N-1|\right).
\end{equation}

Our derivation starts from the integral $I=\int_{-\infty}^{+\infty}d\omega\:\mathrm{Re}\:\mu_\mathrm{ac}(\omega)$, whose integrand is expressed using the first equality in Eq.~\eqref{Eq:def_Re_mu_ac}.
Substituting $C_{jj}(\omega)$ in terms of $C_{jj}(t)$, writing the resulting expression in the eigenbasis $\{|\alpha\rangle\}$ of the full Hamiltonian $H$ ($H|\alpha\rangle=E_\alpha|\alpha\rangle$), and performing the integrations over time and frequency, we obtain
\begin{equation}
\label{Eq:I_after_transform}
\begin{split}
    I=\pi\sum_{\alpha'\alpha}\frac{\langle\alpha'|j|\alpha\rangle\langle\alpha|j|\alpha'\rangle}{E_{\alpha}-E_{\alpha'}}\frac{e^{-\beta E_{\alpha'}}-e^{-\beta E_\alpha}}{Z}.
\end{split}
\end{equation}
We can eliminate the energy difference $E_{\alpha}-E_{\alpha'}$ in the denominator of each term by expressing one of the current operators using the continuity equation~\eqref{Eq:continuity}. We should, however, take into account the finite-size correction entering Eq.~\eqref{Eq:finite-size-correction}.
We then obtain
\begin{equation}
\begin{split}
    I=&-i\pi\sum_{\alpha'\alpha}\langle\alpha'|P|\alpha\rangle\langle\alpha|j|\alpha'\rangle\frac{e^{-\beta E_{\alpha'}}-e^{-\beta E_\alpha}}{Z}\\
    &+i\pi J\sum_{\alpha'\alpha}\frac{\langle\alpha'|\widehat{P}_{N-1,0}|\alpha\rangle\langle\alpha|j|\alpha'\rangle}{E_{\alpha}-E_{\alpha'}}\frac{e^{-\beta E_{\alpha'}}-e^{-\beta E_\alpha}}{Z}\\
    &=I_1+I_2.
\end{split}
\end{equation}
The term $I_1$ can then be written as
\begin{equation}
    I_1=-i\pi\mathrm{Tr}\left\{[P,j]\frac{e^{-\beta H}}{Z}\right\}.
\end{equation}
The commutator $[P,j]$, where both operators are defined on a finite chain, is equal to
\begin{equation}
    [P,j]=-iH_\mathrm{e}-iJ\widehat{P}_{N-1,0},
\end{equation}
so that
\begin{equation}
\begin{split}
    I_1&=-\pi\mathrm{Tr}_\mathrm{e}\left\{H_\mathrm{e}\frac{\sigma_\mathbf{0}^{(0)}(\beta)}{Z_\mathrm{e}}\right\}-\pi J\mathrm{Tr}_\mathrm{e}\left\{\widehat{P}_{N-1,0}\frac{\sigma_\mathbf{0}^{(0)}(\beta)}{Z_\mathrm{e}}\right\}\\
    &=I_{1,1}+I_{1,2}.
\end{split}
\end{equation}
While the term $I_{1,1}$ alone gives the desired result for the OSR ($I=-\pi\langle H_\mathrm{e}\rangle$), the terms $I_{1,2}$ and $I_2$ are finite-size corrections.
The term $I_2$ can be transformed by applying the steps that led to Eq.~\eqref{Eq:I_after_transform} in the reverse order, which gives
\begin{equation}
\begin{split}
    I_2=&\int_{-\infty}^{+\infty}d\omega\frac{1-e^{-\beta\omega}}{2\omega}\int_{-\infty}^{+\infty}dt\:e^{i\omega t}\\&\mathrm{Tr}\left\{e^{iHt}\widehat{P}_{N-1,0}e^{-iHt}j\frac{e^{-\beta H}}{Z}\right\}\\
    =&\int_{-\infty}^{+\infty}d\omega\frac{1-e^{-\beta\omega}}{2\omega}\int_{-\infty}^{+\infty}dt\:e^{i\omega t}\mathrm{Tr}_\mathrm{e}\left\{\hat{P}_{N-1,0}\iota(t)\right\}.
\end{split}
\end{equation}
Both $I_{1,2}$ and $I_2$ reduce to expectation values of the purely electronic operator $\widehat{P}_{N-1,0}$ with respect the reduced (purely electronic) density matrix $\sigma_\mathbf{0}^{(0)}(\beta)/Z_\mathrm{e}$ (for $I_{1,2}$) or $\iota(t)$ (for $I_2$) that is diagonal in the momentum representation.
For example,
\begin{equation}
\label{Eq:iota_detailed}
\begin{split}
    &\mathrm{Tr}_\mathrm{e}\left\{\hat{P}_{N-1,0}\iota(t)\right\}=\\&-2i\sum_k\langle k|\iota(t)|k\rangle\rangle\sin\left(2\pi n_k\frac{N-1}{N}\right),
\end{split}
\end{equation}
where the sum is over $N$ allowed values of the wave number $k=2\pi n_k/N$, with $n_k$ being $N$ consecutive integers.
As $N\to+\infty$, $(N-1)/N\to 1$, and each term entering Eq.~\eqref{Eq:iota_detailed} tends to zero.
In other words, the finite-size correction $I_2$ vanishes in the infinite-chain limit.
The same reasoning may be applied to show that $I_{1,2}$ vanishes in the infinite-chain limit.
\bibliography{apssamp}

\end{document}


\title{Supplementary Material for:\\Holstein polaron transport from numerically exact real-time quantum dynamics simulations}

\author{Veljko Jankovi\'c}%
 \email{veljko.jankovic@ipb.ac.rs}
\affiliation{%
 Institute of Physics Belgrade, University of Belgrade, Pregrevica 118, 11080 Belgrade, Serbia
}%

\maketitle
\section{Structure of files}
\begin{itemize}
    \item \texttt{j\_j\_real\_time.txt} contains data on the current--current correlation function $C_{jj}(t)$ in real time ($t\geq 0$); the first column is time measured in units of $\hbar/J$, where $J$ is the electronic transfer integral, while $\hbar$ is the reduced Plack constant; the second and third columns are the real and imaginary parts of $C_{jj}(t)$, respectively, measured in units $e_0^2a_l^2J^2/\hbar^2$, where $e_0$ is the elementary charge, $a_l$ is the lattice constant.
    \item \texttt{j\_j\_real\_frequency.txt} contains data on the Fourier transformation $C_{jj}(\omega)$ of $C_{jj}(t)$; the first column is frequency measured in units $J/\hbar$; the second column is $C_{jj}(\omega)$ measured in units $e_0^2a_l^2J/\hbar$; the Fourier transformation is computed applying the appropriate routine of the FFTW3 package on $C_{jj}(t)$ that is continued to negative real time using $C_{jj}(-t)=C_{jj}(t)^*$.
    \item \texttt{dynamical\_mobility.txt} contains the frequency profile of the dynamical mobility $\mathrm{Re}\:\mu_\mathrm{ac}(\omega)$ for $\omega\geq 0$; the first column is frequency measured in units $J/\hbar$; the second column is $\mathrm{Re}\:\mu_\mathrm{ac}(\omega)$ measured in units $e_0a_l^2/\hbar$; the dc mobility is computed as
    $$\mu_\mathrm{dc}=\lim_{\omega\to 0}\mathrm{Re}\:\mu_\mathrm{ac}(\omega)=\frac{1}{T}\int_0^{+\infty} dt\:\mathrm{Re}\:C_{jj}(t),$$
    while for $\omega>0$ we use
    $$\mathrm{Re}\:\mu_\mathrm{ac}(\omega)=\frac{1-e^{-\omega/T}}{2\omega}C_{jj}(\omega);$$
    $T$ is the temperature measured in units of $J/k_B$.
    \item \texttt{diffusion\_constant.txt} contains data on the diffusion constant $\mathcal{D}(t)$ in real time ($t>0$); the first column is time measured in units $\hbar/J$; the second column is the diffusion constant measured in units $a_l^2J/\hbar$.
    \item \texttt{diffusion\_exponent.txt} contains data on the diffusion exponent $\alpha(t)$ in real time ($t\geq 0$); the first column is time measured in units $\hbar/J$; the second column is the diffusion exponent (dimensionless).
    \item \texttt{delta\_x.txt} contains data on the square root of the mean-square displacement $\sqrt{\Delta x^2(t)}$ of the electron in real time ($t\geq 0$); the first column is time measured in units $\hbar/J$; the second column is $\sqrt{\Delta x^2(t)}$ measured in units of the lattice constant $a_l$.
    \item in some cases (typically for $T/\omega_0\gtrsim 3$), we provide HEOM results for two consecutive depths, in folders \texttt{Index.1} and \texttt{Index.2}; in such cases, $\mu_\mathrm{dc}$, $\mathrm{Re}\:\mu_\mathrm{ac}(\omega)$, $\mathcal{D}(t)$, $\alpha(t)$, and $\Delta x(t)$ should be computed using $C_{jj}(t)$ that is obtained as the arithmetic average of the current--current correlation functions in folders \texttt{Index.1} and \texttt{Index.2}; the corresponding data for $\mu_\mathrm{dc}$ and $\mathrm{Re}\:\mu_\mathrm{ac}(\omega)$ can be obtained by averaging the data contained in files \texttt{RegimeIndex/Index.1/dynamical\_mobility.txt} and \texttt{RegimeIndex/Index.2/dynamical\_mobility.txt}; for convenience, we provide $\mathcal{D}(t)$, $\alpha(t)$, and $\Delta x(t)$ that use the averaged HEOM data for $C_{jj}(t)$ in files\\\texttt{RegimeIndex/diffusion\_constant.txt},\\\texttt{RegimeIndex/diffusion\_exponent.txt}, and\\\texttt{RegimeIndex/delta\_x.txt}, respectively
    \item files \texttt{mu\_vs\_T\_*.txt} report temperature dependence of $\mu_\mathrm{dc}$; the first column is the temperature in units $J/k_B$; the second column is $\mu_\mathrm{dc}$ in units $e_0a_l^2/\hbar$
    \begin{itemize}
        \item when only one set of HEOM data is reported, $\mu_\mathrm{dc}$ is obtained as the arithmetic average of dc mobilities computed using only $\mathrm{Re}\:C_{jj}(t)$ or only $\mathrm{Im}\:C_{jj}(t)$; in the latter case, the moving-average procedure is obtained to smooth out the oscillatory features of the integral $-2\int_0^t ds\:s\:\mathrm{Im}\:C_{jj}(s)$
        \item when two sets of HEOM data are reported, $\mu_\mathrm{dc}$ is obtained by first averaging $C_{jj}(t)$ over two consecutive depths $D$, and then using the averaged $C_{jj}(t)$ in the same manner as described in the previous point
    \end{itemize}
\end{itemize}
\newpage
\section{Notation}
\begin{itemize}
    \item $\langle H_\mathrm{e}\rangle$: the electron's kinetic energy
    \item 1st moment: $M_1=\langle[j,H]j\rangle$
    \item 2nd moment: $M_2=\langle[[j,H],H]j\rangle$
    \item $\displaystyle{\delta_\mathrm{OSR}=\frac{\left|\int_0^{+\infty}d\omega\:\mathrm{Re}\:\mu_\mathrm{ac}(\omega)-\pi|\langle H_\mathrm{e}\rangle|/2\right|}{\pi|\langle H_\mathrm{e}\rangle|/2}}$, where $\int_0^{+\infty}d\omega\:\mathrm{Re}\:\mu_\mathrm{ac}(\omega)$ is computed numerically (frequency resolution is typically good and the trapezoid rule is sufficient)
    \item $\displaystyle{\delta_n=\frac{\left|\int_{-\infty}^{+\infty}d\omega\:\omega^nC_{jj}(\omega)-M_n\right|}{|M_n|}}$ for $n=0,1,2$; $\int_{-\infty}^{+\infty}d\omega\:\omega^nC_{jj}(\omega)$ is computed numerically
    \item in some cases (typically for $T/\omega_0\gtrsim 3$), we provide HEOM results for two consecutive depths, the indices of which are Index.1 and Index.2; in such cases, $\mu_\mathrm{dc}$, $\mathrm{Re}\:\mu_\mathrm{ac}(\omega)$, $\mathcal{D}(t)$, and $\alpha(t)$ should be computed using $C_{jj}(t)$ obtained as the arithmetic average of the correlation functions in regimes Index.1 and Index.2
\end{itemize}
\newpage
\section{Parameter regimes studied}
\begin{table}[htbp!]
    \centering
    \begin{tabular}{c|c|c|c|c|c|c|c|c}
        Index & $\omega_0/J$ & $G/J$ & $T/J$ & $N$ & $D$ & $\langle H_\mathrm{e}\rangle/J$ & 1st moment & 2nd moment\\
        \hline\hline
        0 & 1 & $\displaystyle{\frac{\sqrt{2}}{10}}$ & $1$ & 160 & 2 & -1.3905189049e+00 & 5.6113740051e-02 & 1.4912329841e-01\\
        \hline
        0a & 1 & $\displaystyle{\frac{\sqrt{2}}{10}}$ & $10^{0.1}$ & 160 & 2 & -1.2303265809e+00 & 6.2332926524e-02 & 1.9073806605e-01\\
        \hline
        0b & 1 & $\displaystyle{\frac{\sqrt{2}}{10}}$ & $10^{0.2}$ & 160 & 2 & -1.0597281605e+00 & 6.7481959494e-02 & 2.4397080716e-01\\
        \hline
        1 & 1 & $\displaystyle{\frac{\sqrt{2}}{10}}$ & $2$ & 96 & 2 & -8.9031757673e-01 & 7.1467305022e-02 & 3.1196525090e-01\\
        \hline
        1a & 1 & $\displaystyle{\frac{\sqrt{2}}{10}}$ & $10^{0.4}$ & 96 & 2 & -7.3739761712e-01 & 7.4287329163e-02 & 3.9555582822e-01\\
        \hline
        1b & 1 & $\displaystyle{\frac{\sqrt{2}}{10}}$ & $10^{0.5}$ & 96 & 2 & -6.0163228198e-01 & 7.6260836521e-02 & 5.0125463818e-01\\
        \hline
        1c & 1 & $\displaystyle{\frac{\sqrt{2}}{10}}$ & $10^{0.6}$ & 40 & 3 & -4.8639232877e-01 & 7.7583823089e-02 & 6.3372669726e-01\\
        \hline
        2 & 1 & $\displaystyle{\frac{\sqrt{2}}{10}}$ & $5$ & 40 & 3 & -3.9170164216e-01 & 7.8444459486e-02 & 7.9798405638e-01\\
        \hline
        2a & 1 & $\displaystyle{\frac{\sqrt{2}}{10}}$ & $10^{0.8}$ & 40 & 3 & -3.1273917734e-01 & 7.9013220051e-02 & 1.0085562142e+00\\
        \hline
        2b & 1 & $\displaystyle{\frac{\sqrt{2}}{10}}$ & $10^{0.9}$ & 40 & 3 & -2.4960379593e-01 & 7.9373375447e-02 & 1.2707885406e+00\\
        \hline
        3 & 1 & $\displaystyle{\frac{\sqrt{2}}{10}}$ & $10$ & 40 & 3 & -1.9887507724e-01 & 7.9602996838e-02 & 1.6005403860e+00\\
        \hline
    \end{tabular}
    \caption{$\omega_0/J=1$ and $G/J=\sqrt{2}/10$.}
    \label{Tab:omega_1_G_0.141}
\end{table}

\begin{table}[htbp!]
    \centering
    \begin{tabular}{c|c|c|c|c}
        Index & $\delta_\mathrm{OSR}$ & $\delta_\mathrm{0}$ & $\delta_1$ & $\delta_2$\\
        \hline\hline
       0 & 1.8825149510e-06 & 4.4059141372e-07 & 5.0684645705e-07 & 2.2250606505e-06\\\hline
0a & 3.2387481405e-06 & 4.8748483051e-07 & 4.2203331944e-07 & 1.5404610697e-06\\\hline
0b & 3.7996092013e-06 & 4.1155518119e-07 & 4.7754700280e-07 & 2.6331951618e-06\\\hline
1 & 3.7023497964e-06 & 5.6355939146e-08 & 1.1317477483e-08 & 2.0490631488e-06\\\hline
1a & 3.7576382002e-06 & 5.6360670234e-08 & 1.4530379430e-08 & 2.1443871875e-06\\\hline
1b & 3.6575115179e-06 & 3.3824591372e-07 & 2.6243500998e-07 & 1.9309697083e-06\\\hline
1c & 3.6374476962e-09 & 9.3784199541e-08 & 9.8567947226e-08 & 9.2195889331e-08\\\hline
2 & 1.2934181992e-08 & 1.3922583120e-07 & 1.4441164304e-07 & 1.3754650862e-07\\\hline
2a & 2.7166252253e-09 & 1.3920601627e-07 & 1.4494207420e-07 & 1.3745143748e-07\\\hline
2b & 2.1388820000e-10 & 1.3921499721e-07 & 1.4568056119e-07 & 1.3723049159e-07\\\hline
3 & 3.4289819690e-09 & 1.8519070062e-07 & 1.9268435133e-07& 1.8291465658e-07\\\hline
    \end{tabular}
    \caption{Relative accuracy with which different sum rules are satisfied. $\omega_0/J=1$ and $G/J=\sqrt{2}/10$.}
    \label{Tab:omega_1_G_0.141_deltas}
\end{table}

\begin{table}[htbp!]
    \centering
    \begin{tabular}{c|c|c|c|c|c|c|c|c}
        Index & $\omega_0/J$ & $G/J$ & $T/J$ & $N$ & $D$ & $\langle H_\mathrm{e}\rangle/J$ & 1st moment & 2nd moment\\
        \hline\hline
        4 & 1 & 0.5 & 1 & 56 & 3 & -1.3332040298e+00 & 7.1638250111e-01 & 2.1848325977e+00\\\hline
        4a & 1 & 0.5 & $10^{0.1}$ & 56 & 3 & -1.1819781290e+00 & 7.8966903391e-01 & 2.7423722383e+00\\\hline
        4b & 1 & 0.5 & $10^{0.2}$ & 56 & 6 & -1.0218262420e+00 & 8.5027002784e-01 & 3.4379926485e+00\\\hline
        5 & 1 & 0.5 & 2 & 29 & 4 & -8.6248188447e-01 & 8.9739894880e-01 & 4.3117114955e+00\\\hline
        5a & 1 & 0.5 & $10^{0.4}$ & 29 & 4 & -7.1777859457e-01 & 9.3089702932e-01 & 5.3724636709e+00\\\hline
        5b & 1 & 0.5 & $10^{0.5}$ & 29 & 4 & -5.8832374579e-01 & 9.5451328922e-01 & 6.7047497183e+00\\\hline
        5c & 1 & 0.5 & $10^{0.6}$ & 15 & 5 & -4.7758111398e-01 & 9.7045605481e-01 & 8.3679541169e+00\\\hline
        6.1 & 1 & 0.5 & 5 & 10 & 7 & -3.8593730116e-01 & 9.8089966026e-01 & 1.0426004054e+01\\\hline
        6.2 & 1 & 0.5 & 5 & 10 & 8 & -3.8593730116e-01 & 9.8089966026e-01 & 1.0426004054e+01\\\hline
        6a.1 & 1 & 0.5 & $10^{0.8}$ & 8 & 8 & -3.0904232784e-01 & 9.8784037824e-01 & 1.3061357961e+01\\\hline
        6a.2 & 1 & 0.5 & $10^{0.8}$ & 8 & 9 & -3.0904232784e-01 & 9.8784037824e-01 & 1.3061357961e+01\\\hline
        6b.1 & 1 & 0.5 & $10^{0.9}$ & 7 & 10 & -2.4723926317e-01 & 9.9225623915e-01 & 1.6341293711e+01\\\hline
        6b.2 & 1 & 0.5 & $10^{0.9}$ & 7 & 11 & -2.4723926317e-01 & 9.9225623915e-01 & 1.6341293711e+01\\\hline
        7.1 & 1 & 0.5 & 10 & 7 & 11 & -1.9736975089e-01 & 9.9508253470e-01 & 2.0464493214e+01\\\hline
        7.2 & 1 & 0.5 & 10 & 7 & 12 & -1.9736975089e-01 & 9.9508253470e-01 & 2.0464493214e+01\\\hline
    \end{tabular}
    \caption{$\omega_0/J=1$ and $G/J=0.5$.}
    \label{Tab:omega_1_G_0.5}
\end{table}

\begin{table}[htbp!]
    \centering
    \begin{tabular}{c|c|c|c|c}
        Index & $\delta_\mathrm{OSR}$ & $\delta_\mathrm{0}$ & $\delta_1$ & $\delta_2$\\
        \hline\hline
4 & 9.1836819679e-05 & 7.7612673901e-08 & 2.5876815801e-07 & 4.0023855954e-08\\\hline
4a & 7.8248286370e-05 & 1.3312316780e-07 & 7.5662472272e-08 & 1.5808749106e-07\\\hline
4b & 5.6472832492e-05 & 1.3314581466e-07 & 1.1661663476e-07 & 1.4432163412e-07\\\hline
5 & 1.9605861324e-06 & 4.3158665735e-08 & 2.5236280189e-07 & 1.5655634446e-07\\\hline
5a & 2.8560416911e-06 & 9.7478917417e-08 & 2.7861119196e-07 & 2.3453211612e-07\\\hline
5b & 4.1530903421e-06 & 2.4245217324e-08 & 5.1895339447e-07 & 1.4872890109e-07\\\hline
5c & 7.0104707056e-06 & 2.3176240410e-07 & 9.0087203551e-07 & 1.8400352327e-09\\\hline
6.1 & 1.0200766124e-05 & 1.3922833194e-07 & 1.9694369525e-07 & 1.1953276182e-07\\\hline
6.2 & 1.1572147351e-05 & 1.3922824894e-07 & 1.9694755049e-07 & 1.1952694459e-07\\\hline
6a.1 & 5.6044346634e-06 & 1.3919825304e-07 & 2.2124191362e-07 & 1.1195577915e-07\\\hline
6a.2 & 7.4129892012e-06 & 1.3919821641e-07 & 2.2124913034e-07 & 1.1194865320e-07\\\hline
6b.1 & 3.0064241233e-06 & 1.3919773588e-07 & 2.5760345918e-07 & 1.0047009175e-07\\\hline
6b.2 & 5.9321807646e-06 & 1.3919849262e-07 & 2.5760987791e-07 & 1.0045953306e-07\\\hline
7.1 & 1.9313067836e-06 & 1.5099860961e-07 & 2.6325092118e-06 & 1.0482535512e-06\\\hline
7.2 & 6.8360067388e-07 & 1.5099847324e-07 & 2.6324914650e-06 & 1.0482083956e-06\\\hline
\end{tabular}
\caption{Relative accuracy with which different sum rules are satisfied. $\omega_0/J=1$ and $G/J=0.5$.}
\label{Tab:omega_1_G_0.5_deltas}
\end{table}

\begin{table}[htbp!]
    \centering
    \begin{tabular}{c|c|c|c|c|c|c|c|c}
        Index & $\omega_0/J$ & $G/J$ & $T/J$ & $N$ & $D$ & $\langle H_\mathrm{e}\rangle/J$ & 1st moment & 2nd moment\\
        \hline\hline
        10 & 1 & 1 & 1 & 13 & 6 & -1.1546229149e+00 & 3.0539769058e+00 & 1.3368705634e+01\\\hline
        10a & 1 & 1 & $10^{0.1}$ & 10 & 6 & -1.0340833387e+00 & 3.2883089000e+00 & 1.5947349411e+01\\\hline
        10b & 1 & 1 & $10^{0.2}$ & 10 & 6 & -9.0662044619e-01 & 3.4837021570e+00 & 1.9013152433e+01\\\hline
        11 & 1 & 1 & 2 & 10 & 8 & -7.7770607902e-01 & 3.6385183730e+00 & 2.2726226279e+01\\\hline
        11a & 1 & 1 & $10^{0.4}$ & 7 & 10 & -6.5764074405e-01 & 3.7515390863e+00 & 2.7130404436e+01\\\hline
        11b.1 & 1 & 1 & $10^{0.5}$ & 7 & 10 & -5.4717954967e-01 & 3.8333794413e+00 & 3.2577024060e+01\\\hline
        11b.2 & 1 & 1 & $10^{0.5}$ & 7 & 11 & -5.4717954967e-01 & 3.8333794413e+00 & 3.2577024060e+01\\\hline
        11c.1 & 1 & 1 & $10^{0.6}$ & 7 & 11 & -4.5010364249e-01 & 3.8900023468e+00 & 3.9311359011e+01\\\hline
        11c.2 & 1 & 1 & $10^{0.6}$ & 7 & 12 & -4.5010364249e-01 & 3.8900023468e+00 & 3.9311359011e+01\\\hline
        12.1 & 1 & 1 & 5 & 7 & 11 & -3.6781934855e-01 & 3.9279089576e+00 & 4.7598400840e+01\\\hline
        12.1 & 1 & 1 & 5 & 7 & 12 & -3.6781934855e-01 & 3.9279089576e+00 & 4.7598400840e+01\\\hline
        12a.1 & 1 & 1 & $10^{0.8}$ & 7 & 11 & -2.9734141354e-01 & 3.9535733441e+00 & 5.8177140816e+01\\\hline
        12a.2 & 1 & 1 & $10^{0.8}$ & 7 & 12 & -2.9734141354e-01 & 3.9535733441e+00 & 5.8177140816e+01\\\hline
        12b.1 & 1 & 1 & $10^{0.9}$ & 5 & 19 & -2.3973068017e-01 & 3.9698513618e+00 & 7.1311312424e+01\\\hline
        12b.2 & 1 & 1 & $10^{0.9}$ & 5 & 20 & -2.3973068017e-01 & 3.9698513618e+00 & 7.1311312424e+01\\\hline
        13.1 & 1 & 1 & 10 & 5 & 20 & -1.9256239280e-01 & 3.9807500071e+00 & 8.7823407685e+01\\\hline
        13.2 & 1 & 1 & 10 & 5 & 21 & -1.9256239280e-01 & 3.9807500071e+00 & 8.7823407685e+01\\\hline
    \end{tabular}
    \caption{$\omega_0/J=1$ and $G/J=1$.}
    \label{Tab:omega_1_G_1}
\end{table}

\begin{table}[htbp!]
    \centering
    \begin{tabular}{c|c|c|c|c}
        Index & $\delta_\mathrm{OSR}$ & $\delta_\mathrm{0}$ & $\delta_1$ & $\delta_2$\\
        \hline\hline
10 & 1.6036380825e-05 & 5.9979976426e-07 & 4.8560432118e-07 & 1.1567164309e-06\\\hline
10a & 4.5996653245e-05 & 8.3264512636e-07 & 7.4574557915e-07 & 8.7550565173e-07\\\hline
10b & 4.7684441034e-05 & 8.3264101069e-07 & 7.1762494436e-07 & 8.8693233835e-07\\\hline
11 & 3.2082964632e-07 & 1.3311664260e-07 & 2.3849669782e-06 & 1.2661667450e-06\\\hline
11a & 7.6778148020e-05 & 3.3829399181e-07 & 3.1665155940e-06 & 1.8456957748e-06\\\hline
11b.1 & 2.2976859693e-05 & 1.6077349009e-07 & 4.8433644391e-06 & 2.2194607919e-06\\\hline
11b.2 & 2.4929681453e-05 & 1.6077309411e-07 & 4.8433683662e-06 & 2.2194674746e-06\\\hline
11c.1 & 7.1599468839e-06 & 1.3311844913e-07 & 7.1479694272e-06 & 3.0058552571e-06\\\hline
11c.2 & 5.3026978378e-06 & 1.3311806797e-07 & 7.1479652595e-06 & 3.0058451238e-06\\\hline
12.1 & 4.6512888516e-07 & 1.5098845322e-07 & 1.0569646777e-05 & 4.2212905274e-06\\\hline
12.2 & 6.7854629430e-07 & 1.5098825061e-07 & 1.0569645597e-05 & 4.2212878026e-06\\\hline
12a.1 & 1.3864898871e-06 & 1.5099063354e-07 & 1.5978623124e-05 & 6.0672717077e-06\\\hline
12a.2 & 9.4325310761e-07 & 1.5099207033e-07 & 1.5978616095e-05 & 6.0672621534e-06\\\hline
12b.1 & 3.9430111124e-04 & 6.9692799582e-09 & 1.5127509267e-06 & 5.4957585754e-07\\\hline
12b.2 & 3.9527948784e-04 & 6.9689141438e-09 & 1.5127477180e-06 & 5.4956812791e-07\\\hline
13.1 & 1.9663862414e-04 & 5.3948690718e-08 & 3.7305557083e-05 & 1.3006389709e-05\\\hline
13.2 & 1.9521492638e-04 & 5.3947931594e-08 & 3.7305566392e-05 & 1.3006411650e-05\\\hline
    \end{tabular}
    \caption{Relative accuracy with which different sum rules are satisfied. $\omega_0/J=1$ and $G/J=1$.}
    \label{Tab:omega_1_G_1_deltas}
\end{table}

\begin{table}[htbp!]
    \centering
    \begin{tabular}{c|c|c|c|c|c|c|c|c}
        Index & $\omega_0/J$ & $G/J$ & $T/J$ & $N$ & $D$ & $\langle H_\mathrm{e}\rangle/J$ & 1st moment & 2nd moment\\
        \hline\hline
        15.1 & 1 & $\sqrt{2}$ & 2 & 7 & 10 & -6.7810727121e-01 & 7.3904137766e+00 & 6.0488230716e+01\\\hline
        15.2 & 1 & $\sqrt{2}$ & 2 & 7 & 11 &-6.7810727121e-01 & 7.3904137769e+00 & 6.0488230749e+01\\\hline
        15a.1 & 1 & $\sqrt{2}$ & $10^{0.4}$ & 7 & 10 & -5.8595153335e-01 & 7.5691433688e+00 & 6.9551423817e+01\\\hline
        15a.2 & 1 & $\sqrt{2}$ & $10^{0.4}$ & 7 & 11 & -5.8595153335e-01 & 7.5691433689e+00 & 6.9551423822e+01\\\hline
        15b.1 & 1 & $\sqrt{2}$ & $10^{0.5}$ & 7 & 10 & -4.9733154367e-01 & 7.7036303753e+00 & 8.0647568207e+01\\\hline
        15b.2 & 1 & $\sqrt{2}$ & $10^{0.5}$ & 7 & 11 & -4.9733154367e-01 & 7.7036303753e+00 & 8.0647568208e+01\\\hline
        15c.1 & 1 & $\sqrt{2}$ & $10^{0.6}$ & 7 & 11 & -4.1628095148e-01 & 7.8000210080e+00 & 9.4270165317e+01\\\hline
        15c.2 & 1 & $\sqrt{2}$ & $10^{0.6}$ & 7 & 12 & -4.1628095148e-01 & 7.8000210080e+00 & 9.4270165318e+01\\\hline
        16.1 & 1 & $\sqrt{2}$ & 5 & 7 & 11 & -3.4519845218e-01 & 7.8665302812e+00 & 1.1095549112e+02\\\hline
        16.2 & 1 & $\sqrt{2}$ & 5 & 7 & 12 & -3.4519845218e-01 & 7.8665302812e+00 & 1.1095549112e+02\\\hline
        16a.1 & 1 & $\sqrt{2}$ & $10^{0.8}$ & 7 & 11 & -2.8254865169e-01 & 7.9127180099e+00 & 1.3219316115e+02\\\hline
        16a.2 & 1 & $\sqrt{2}$ & $10^{0.8}$ & 7 & 12 & -2.8254865169e-01 & 7.9127180099e+00 & 1.3219316115e+02\\\hline
        16b.1 & 1 & $\sqrt{2}$ & $10^{0.9}$ & 7 & 11 &-2.3009625952e-01 & 7.9431987469e+00 & 1.5853566803e+02\\\hline
        16b.2 & 1 & $\sqrt{2}$ & $10^{0.9}$ & 7 & 12 &-2.3009625952e-01 & 7.9431987469e+00 & 1.5853566803e+02\\\hline
        17.1 & 1 & $\sqrt{2}$ & 10 & 7 & 11 & -1.8635146935e-01 & 7.9632929879e+00 & 1.9158961786e+02\\\hline
        17.2 & 1 & $\sqrt{2}$ & 10 & 7 & 12 & -1.8635146935e-01 & 7.9632929879e+00 & 1.9158961786e+02\\\hline
    \end{tabular}
    \caption{$\omega_0/J=1$ and $G/J=\sqrt{2}$.}
    \label{Tab:omega_1_G_1.41}
\end{table}

\begin{table}[htbp!]
    \centering
    \begin{tabular}{c|c|c|c|c}
        Index & $\delta_\mathrm{OSR}$ & $\delta_\mathrm{0}$ & $\delta_1$ & $\delta_2$\\
        \hline\hline
15.1 & 9.3122191191e-05 & 1.2499279304e-08 & 6.5144470140e-07 & 3.2140807939e-07\\\hline
15.2 & 1.2967356871e-04 & 6.1043898433e-09 & 1.0711762190e-05 & 4.9964218350e-06\\\hline
15a.1 & 5.5818261598e-05 & 1.5098296439e-07 & 1.4696140173e-05 & 6.9414856374e-06\\\hline
15a.2 & 1.6000585928e-05 & 1.5098461188e-07 & 1.4695839089e-05 & 6.9407594472e-06\\\hline
15b.1 & 1.4158181384e-05 & 1.5093434305e-07 & 2.0934310426e-05 & 9.4989207246e-06\\\hline
15b.2 & 1.2854647294e-05 & 1.5093443679e-07 & 2.0934303211e-05 & 9.4988949934e-06\\\hline
15c.1 & 2.9447484867e-06 & 1.5092931252e-07 & 3.0352646953e-05 & 1.3220860358e-05\\\hline
15c.2 & 1.6458676440e-06 & 1.5092844137e-07 & 3.0352642202e-05 & 1.3220836773e-05\\\hline
16.1 & 1.9621584365e-06 & 1.1244510393e-09 & 2.7761894275e-06 & 1.1496825626e-06\\\hline
16.2 & 2.0251426220e-06 & 1.1250029937e-09 & 2.7761894275e-06 & 1.1496855684e-06\\\hline
16a.1 & 8.4443130598e-06 & 1.5089148526e-07 & 6.6721988303e-05 & 2.6808535215e-05\\\hline
16a.2 & 8.4910059101e-06 & 1.5089143040e-07 & 6.6721988303e-05 & 2.6808535215e-05\\\hline
16b.1 & 1.2166796472e-05 & 1.5081192239e-07 & 1.0076458856e-04 & 3.8971413534e-05\\\hline
16b.2 & 1.2156411503e-05 & 1.5081198616e-07 & 1.0076458856e-04 & 3.8971413534e-05\\\hline
17.1 & 1.6818573698e-06 & 1.1546238409e-09 & 9.5092509542e-06 & 3.5563795428e-06\\\hline
17.2 & 1.1318765080e-06 & 1.1547980709e-09 & 9.5092550259e-06 & 3.5563907608e-06\\\hline
    \end{tabular}
    \caption{Relative accuracy with which different sum rules are satisfied. $\omega_0/J=1$ and $G/J=\sqrt{2}$.}
    \label{Tab:omega_1_G_1.41_deltas}
\end{table}

\begin{table}[htbp!]
    \centering
    \begin{tabular}{c|c|c|c|c|c|c|c|c}
        Index & $\omega_0/J$ & $G/J$ & $T/J$ & $N$ & $D$ & $\langle H_\mathrm{e}\rangle/J$ & 1st moment & 2nd moment\\
        \hline\hline
        22 & $\displaystyle{\frac{1}{3}}$ & $\displaystyle{\frac{1}{\sqrt{150}}}$ & 1 & 128 & 2 & -1.3904552941e+00 & 5.5948648407e-02 & 1.3922850834e-01\\\hline
        22a & $\displaystyle{\frac{1}{3}}$ & $\displaystyle{\frac{1}{\sqrt{150}}}$ & $10^{0.1}$ & 80 & 2 & -1.2302903746e+00 & 6.2254190822e-02 & 1.8246099806e-01\\\hline
        22b & $\displaystyle{\frac{1}{3}}$ & $\displaystyle{\frac{1}{\sqrt{150}}}$ & $10^{0.2}$ & 80 & 2 & -1.0597095374e+00 & 6.7446238925e-02 & 2.3712830051e-01\\\hline
        23 & $\displaystyle{\frac{1}{3}}$ & $\displaystyle{\frac{1}{\sqrt{150}}}$ & 2 & 40 & 3 & -8.9030879367e-01 & 7.1451853466e-02 & 3.0638100348e-01\\\hline
        23a & $\displaystyle{\frac{1}{3}}$ & $\displaystyle{\frac{1}{\sqrt{150}}}$ & $10^{0.4}$ & 40 & 3 & -7.3739362962e-01 & 7.4281605326e-02 & 3.9103039138e-01\\\hline
        23b & $\displaystyle{\frac{1}{3}}$ & $\displaystyle{\frac{1}{\sqrt{150}}}$ & $10^{0.5}$ & 40 & 3 & -6.0163056408e-01 & 7.6258675649e-02 & 4.9761031639e-01\\\hline
        23c & $\displaystyle{\frac{1}{3}}$ & $\displaystyle{\frac{1}{\sqrt{150}}}$ & $10^{0.6}$ & 40 & 3 & -4.8639161148e-01 & 7.7582702465e-02 & 6.3079800233e-01\\\hline
        24 & $\displaystyle{\frac{1}{3}}$ & $\displaystyle{\frac{1}{\sqrt{150}}}$ & 5 & 40 & 3 & -3.9170134424e-01 & 7.8444001113e-02 & 7.9563845731e-01\\\hline
        24a & $\displaystyle{\frac{1}{3}}$ & $\displaystyle{\frac{1}{\sqrt{150}}}$ & $10^{0.8}$ & 40 & 3 & -3.1273905731e-01 & 7.9013037171e-02 & 1.0066902699e+00\\\hline
        24b & $\displaystyle{\frac{1}{3}}$ & $\displaystyle{\frac{1}{\sqrt{150}}}$ & $10^{0.9}$ & 40 & 3 & -2.4960374749e-01 & 7.9373302093e-02 & 1.2693027582e+00\\\hline
        25 & $\displaystyle{\frac{1}{3}}$ & $\displaystyle{\frac{1}{\sqrt{150}}}$ & $10$ & 40 & 3 & -1.9887505779e-01 & 7.9602967493e-02 & 1.5993583595e+00\\\hline
    \end{tabular}
    \caption{$\omega_0/J=1/3$ and $G/J=1/\sqrt{150}$.}
    \label{Tab:omega_0.33_G_0.082}
\end{table}

\begin{table}[htbp!]
    \centering
    \begin{tabular}{c|c|c|c|c}
        Index & $\delta_\mathrm{OSR}$ & $\delta_\mathrm{0}$ & $\delta_1$ & $\delta_2$\\
        \hline\hline
22 & 1.2388889971e-05 & 5.5508654841e-08 & 2.6820244136e-06 & 6.6684123114e-06\\\hline
22a & 1.0303602494e-05 & 1.8518928722e-07 & 3.6363158870e-07 & 3.8733046657e-06\\\hline
22b & 1.0083847788e-05 & 1.8518179194e-07 & 3.7265852094e-07 & 4.1924268672e-06\\\hline
23 & 9.6120457789e-06 & 4.8448648808e-07 & 2.8479529286e-07 & 3.7641218380e-06\\\hline
23a & 2.7485514621e-08 & 4.8451335538e-07 & 2.8018690846e-07 & 5.5313003846e-07\\\hline
23b & 1.3022323503e-08 & 4.8449317009e-07 & 2.6455953862e-07 & 5.5623772380e-07\\\hline
23c & 1.5525164291e-08 & 4.8447229548e-07 & 2.4405953341e-07 & 5.6130122735e-07\\\hline
24 & 9.7607884751e-09 & 4.8450861400e-07 & 2.1736276448e-07 & 5.6860494537e-07\\\hline
24a & 6.1793107208e-09 & 4.8448435494e-07 & 1.8105541232e-07 & 5.7904996449e-07\\\hline
24b & 7.9183080745e-09 & 4.8446891534e-07 & 1.3264468716e-07 & 5.9337292946e-07\\\hline
25 & 6.7227379205e-09 & 4.8448079136e-07 & 6.6684536087e-08 & 6.1310366650e-07\\\hline
    \end{tabular}
    \caption{Relative accuracy with which different sum rules are satisfied. $\omega_0/J=1/3$ and $G/J=1/\sqrt{150}$.}
    \label{Tab:omega_0.33_G_0.082_deltas}
\end{table}

\begin{table}[htbp!]
    \centering
    \begin{tabular}{c|c|c|c|c|c|c|c|c}
        Index & $\omega_0/J$ & $G/J$ & $T/J$ & $N$ & $D$ & $\langle H_\mathrm{e}\rangle/J$ & 1st moment & 2nd moment\\
        \hline\hline
        26 & $\displaystyle{\frac{1}{3}}$ & $\displaystyle{\frac{1}{\sqrt{12}}}$ & 1 & 30 & 4 & -1.3324071754e+00 & 7.1486081461e-01 & 2.0611429416e+00\\\hline
        26a & $\displaystyle{\frac{1}{3}}$ & $\displaystyle{\frac{1}{\sqrt{12}}}$ & $10^{0.1}$ & 30 & 4 & -1.1815321195e+00 & 7.8897587148e-01 & 2.6394120830e+00\\\hline
        26b & $\displaystyle{\frac{1}{3}}$ & $\displaystyle{\frac{1}{\sqrt{12}}}$ & $10^{0.2}$ & 30 & 4 & -1.0215986025e+00 & 8.4997233464e-01 & 3.3530866497e+00\\\hline
        27 & $\displaystyle{\frac{1}{3}}$ & $\displaystyle{\frac{1}{\sqrt{12}}}$ & 2 & 20 & 5 & -8.6237579726e-01 & 8.9722719990e-01 & 4.2416976626e+00\\\hline
        27a & $\displaystyle{\frac{1}{3}}$ & $\displaystyle{\frac{1}{\sqrt{12}}}$ & $10^{0.4}$ & 20 & 5 & -7.1773046300e-01 & 9.3082234493e-01 & 5.3156630565e+00\\\hline
        27b & $\displaystyle{\frac{1}{3}}$ & $\displaystyle{\frac{1}{\sqrt{12}}}$ & $10^{0.5}$ & 13 & 6 & -5.8830292138e-01 & 9.5448178505e-01 & 6.6590410786e+00\\\hline
        27c.1 & $\displaystyle{\frac{1}{3}}$ & $\displaystyle{\frac{1}{\sqrt{12}}}$ & $10^{0.6}$ & 7 & 9 & -4.7757297020e-01 & 9.7044163391e-01 & 8.3312969652e+00\\\hline
        27c.2 & $\displaystyle{\frac{1}{3}}$ & $\displaystyle{\frac{1}{\sqrt{12}}}$ & $10^{0.6}$ & 7 & 10 & -4.7757297020e-01 & 9.7044163391e-01 & 8.3312969652e+00\\\hline
        28.1 & $\displaystyle{\frac{1}{3}}$ & $\displaystyle{\frac{1}{\sqrt{12}}}$ & 5 & 7 & 10 & -3.8593380979e-01 & 9.8089379977e-01 & 1.0396664010e+01\\\hline
        28.2 & $\displaystyle{\frac{1}{3}}$ & $\displaystyle{\frac{1}{\sqrt{12}}}$ & 5 & 7 & 11 & -3.8593380979e-01 & 9.8089379977e-01 & 1.0396664010e+01\\\hline
        28a.1 & $\displaystyle{\frac{1}{3}}$ & $\displaystyle{\frac{1}{\sqrt{12}}}$ & $10^{0.8}$ & 7 & 11 & -3.0904089043e-01 & 9.8783805457e-01 & 1.3038025718e+01\\\hline
        28a.2 & $\displaystyle{\frac{1}{3}}$ & $\displaystyle{\frac{1}{\sqrt{12}}}$ & $10^{0.8}$ & 7 & 12 & -3.0904089043e-01 & 9.8783805457e-01 & 1.3038025718e+01\\\hline
        28b.1 & $\displaystyle{\frac{1}{3}}$ & $\displaystyle{\frac{1}{\sqrt{12}}}$ & $10^{0.9}$ & 5 & 13 & -2.4725878319e-01 & 9.9217444285e-01 & 1.6320151578e+01\\\hline
        28b.2 & $\displaystyle{\frac{1}{3}}$ & $\displaystyle{\frac{1}{\sqrt{12}}}$ & $10^{0.9}$ & 5 & 14 & -2.4725878319e-01 & 9.9217444285e-01 & 1.6320151578e+01\\\hline
        29.1 & $\displaystyle{\frac{1}{3}}$ & $\displaystyle{\frac{1}{\sqrt{12}}}$ & 10 & 5 & 14 & -1.9737760919e-01 & 9.9504134539e-01 & 2.0448084153e+01\\\hline
        29.2 & $\displaystyle{\frac{1}{3}}$ & $\displaystyle{\frac{1}{\sqrt{12}}}$ & 10 & 5 & 15 & -1.9737760919e-01 & 9.9504134539e-01 & 2.0448084153e+01\\\hline
    \end{tabular}
    \caption{$\omega_0/J=1/3$ and $G/J=1/\sqrt{12}$.}
    \label{Tab:omega_0.33_G_0.287}
\end{table}

\begin{table}[htbp!]
    \centering
    \begin{tabular}{c|c|c|c|c}
        Index & $\delta_\mathrm{OSR}$ & $\delta_\mathrm{0}$ & $\delta_1$ & $\delta_2$\\
        \hline\hline
26 & 5.5478471914e-05 & 5.0259961848e-08 & 4.9887652799e-07 & 2.8665200083e-07\\\hline
26a & 4.9729100147e-05 & 1.2619865992e-07 & 5.5237895019e-07 & 3.9805294269e-07\\\hline
26b & 4.0880969697e-05 & 1.2618890094e-07 & 7.3468009418e-07 & 4.5087206469e-07\\\hline
27 & 2.6179653786e-06 & 5.0289040288e-08 & 1.0741565180e-06 & 4.7799436666e-07\\\hline
27a & 2.8261347670e-06 & 5.0266393159e-08 & 1.4472057297e-06 & 5.9513446578e-07\\\hline
27b & 5.5329621003e-06 & 5.0276561618e-08 & 1.9981041296e-06 & 7.6914361070e-07\\\hline
27c.1 & 2.2086001048e-06 & 5.0274556755e-08 & 2.8131613730e-06 & 1.0268467552e-06\\\hline
27c.2 & 3.1790608084e-06 & 5.0274426264e-08 & 2.8131491419e-06 & 1.0268148559e-06\\\hline
28.1 & 6.9213249585e-06 & 5.0265797814e-08 & 4.0199128963e-06 & 1.4085870068e-06\\\hline
28.2 & 8.3308185270e-06 & 5.0264645031e-08 & 4.0198892851e-06 & 1.4085456840e-06\\\hline
28a.1 & 6.6011379978e-06 & 5.0259471937e-08 & 5.8818688906e-06 & 1.9962850051e-06\\\hline
28a.2 & 7.3376167345e-06 & 5.0258670625e-08 & 5.8818565819e-06 & 1.9962601362e-06\\\hline
28b.1 & 4.0616055608e-04 & 5.0288946488e-08 & 8.6745382019e-06 & 2.8751415287e-06\\\hline
28b.2 & 4.0509076173e-04 & 5.0289388821e-08 & 8.6745125249e-06 & 2.8751060510e-06\\\hline
29.1 & 2.0254984059e-04 & 5.0286179303e-08 & 1.2993971079e-05 & 4.2294494914e-06\\\hline
29.2 & 2.0443564268e-04 & 5.0286352070e-08 & 1.2993988535e-05 & 4.2295052167e-06\\\hline
    \end{tabular}
    \caption{Relative accuracy with which different sum rules are satisfied. $\omega_0/J=1/3$ and $G/J=1/\sqrt{12}$.}
    \label{Tab:omega_0.33_G_0.287_deltas}
\end{table}

\begin{table}[htbp!]
    \centering
    \begin{tabular}{c|c|c|c|c|c|c|c|c}
        Index & $\omega_0/J$ & $G/J$ & $T/J$ & $N$ & $D$ & $\langle H_\mathrm{e}\rangle/J$ & 1st moment & 2nd moment\\
        \hline\hline
        33.1 & $\frac{1}{3}$ & $\frac{1}{\sqrt{3}}$ & 1 & 10 & 7 & -1.1517445577e+00 & 3.0503755140e+00 & 1.2851205763e+01\\\hline
        33.2 & $\frac{1}{3}$ & $\frac{1}{\sqrt{3}}$ & 1 & 10 & 8 & -1.1517444891e+00 & 3.0503779741e+00 & 1.2851252563e+01\\\hline
        33a.1 & $\frac{1}{3}$ & $\frac{1}{\sqrt{3}}$ & $10^{0.1}$ & 10 & 7 & -1.0325142700e+00 & 3.2864345657e+00 & 1.5520883612e+01\\\hline
        33a.2 & $\frac{1}{3}$ & $\frac{1}{\sqrt{3}}$ & $10^{0.1}$ & 10 & 8 & -1.0325142570e+00 & 3.2864352128e+00 & 1.5520898775e+01\\\hline
        33b.1 & $\frac{1}{3}$ & $\frac{1}{\sqrt{3}}$ & $10^{0.2}$ & 10 & 7 & -9.0582381534e-01 & 3.4827600805e+00 & 1.8664537507e+01\\\hline
        33b.2 & $\frac{1}{3}$ & $\frac{1}{\sqrt{3}}$ & $10^{0.2}$ & 10 & 8 & -9.0582381307e-01 & 3.4827602401e+00 & 1.8664542126e+01\\\hline
        34.1 & $\frac{1}{3}$ & $\frac{1}{\sqrt{3}}$ & 2 & 7 & 11 & -7.7735209565e-01 & 3.6379568093e+00 & 2.2442585771e+01\\\hline
        34.2 & $\frac{1}{3}$ & $\frac{1}{\sqrt{3}}$ & 2 & 7 & 12 & -7.7735209565e-01 & 3.6379568093e+00 & 2.2442585771e+01\\\hline
        34a.1 & $\frac{1}{3}$ & $\frac{1}{\sqrt{3}}$ & $10^{0.4}$ & 7 & 11 & -6.5746714814e-01 & 3.7513262954e+00 & 2.6902123325e+01\\\hline
        34a.2 & $\frac{1}{3}$ & $\frac{1}{\sqrt{3}}$ & $10^{0.4}$ & 7 & 12 & -6.5746714814e-01 & 3.7513262954e+00 & 2.6902123325e+01\\\hline
        34b.1 & $\frac{1}{3}$ & $\frac{1}{\sqrt{3}}$ & $10^{0.5}$ & 6 & 13 & -5.4713945099e-01 & 3.8330391599e+00 & 3.2389866321e+01\\\hline
        34b.2 & $\frac{1}{3}$ & $\frac{1}{\sqrt{3}}$ & $10^{0.5}$ & 6 & 14 & -5.4713945099e-01 & 3.8330391599e+00 & 3.2389866321e+01\\\hline
        34c.1 & $\frac{1}{3}$ & $\frac{1}{\sqrt{3}}$ & $10^{0.6}$ & 6 & 13 & -4.5008360034e-01 & 3.8898558587e+00 & 3.9162577771e+01\\\hline
        34c.2 & $\frac{1}{3}$ & $\frac{1}{\sqrt{3}}$ & $10^{0.6}$ & 6 & 14 & -4.5008360034e-01 & 3.8898558587e+00 & 3.9162577771e+01\\\hline
        35.1 & $\frac{1}{3}$ & $\frac{1}{\sqrt{3}}$ & 5 & 6 & 14 & -3.6780987651e-01 & 3.9278464690e+00 & 4.7479898693e+01\\\hline
        35.2 & $\frac{1}{3}$ & $\frac{1}{\sqrt{3}}$ & 5 & 6 & 15 & -3.6780987651e-01 & 3.9278464690e+00 & 4.7479898693e+01\\\hline
        35a.1 & $\frac{1}{3}$ & $\frac{1}{\sqrt{3}}$ & $10^{0.8}$ & 6 & 14 & -2.9733719572e-01 & 3.9535475209e+00 & 5.8083221292e+01\\\hline
        35a.2 & $\frac{1}{3}$ & $\frac{1}{\sqrt{3}}$ & $10^{0.8}$ & 6 & 15 & -2.9733719572e-01 & 3.9535475209e+00 & 5.8083221292e+01\\\hline
        35b.1 & $\frac{1}{3}$ & $\frac{1}{\sqrt{3}}$ & $10^{0.9}$ & 6 & 14 & -2.3970993868e-01 & 3.9701467574e+00 & 7.1246525555e+01\\\hline
        35b.2 & $\frac{1}{3}$ & $\frac{1}{\sqrt{3}}$ & $10^{0.9}$ & 6 & 15 & -2.3970993868e-01 & 3.9701467574e+00 & 7.1246525555e+01\\\hline
        36.1 & $\frac{1}{3}$ & $\frac{1}{\sqrt{3}}$ & 10 & 6 & 14 & -1.9255390347e-01 & 3.9809018558e+00 & 8.7770445004e+01\\\hline
        36.2 & $\frac{1}{3}$ & $\frac{1}{\sqrt{3}}$ & 10 & 6 & 15 & -1.9255390348e-01 & 3.9809018560e+00 & 8.7770445006e+01\\\hline
    \end{tabular}
    \caption{$\omega_0/J=1/3$ and $G/J=1/\sqrt{3}$.}
    \label{Tab:omega_0.33_G_0.577}
\end{table}

\begin{table}[htbp!]
    \centering
    \begin{tabular}{c|c|c|c|c}
        Index & $\delta_\mathrm{OSR}$ & $\delta_\mathrm{0}$ & $\delta_1$ & $\delta_2$\\
        \hline\hline
33.1 & 4.4972581478e-05 & 5.5360028766e-08 & 6.7737618554e-05 & 3.2596817755e-05\\\hline
33.2 & 3.8660971810e-05 & 5.5358536073e-08 & 6.7745077492e-05 & 3.2604980362e-05\\\hline
33a.1 & 1.4032040606e-05 & 1.2617190480e-07 & 5.6031661544e-06 & 2.7976583744e-06\\\hline
33a.2 & 2.2209158710e-05 & 1.2620277559e-07 & 5.6034060646e-06 & 2.7979624350e-06\\\hline
33b.1 & 6.0776884346e-06 & 1.2619803966e-07 & 7.8228870026e-06 & 3.7048645583e-06\\\hline
33b.2 & 8.9832863835e-06 & 1.2619728751e-07 & 7.8230169826e-06 & 3.7050646608e-06\\\hline
34.1 & 2.2412921689e-04 & 5.0253750110e-08 & 1.1216922423e-05 & 4.9337403605e-06\\\hline
34.2 & 2.2427668314e-04 & 5.0254948337e-08 & 1.1216922423e-05 & 4.9337449012e-06\\\hline
34a.1 & 7.5430677197e-05 & 5.0286015417e-08 & 1.6081495308e-05 & 6.7838679974e-06\\\hline
34a.2 & 7.5346239911e-05 & 5.0286515813e-08 & 1.6081492839e-05 & 6.7838659313e-06\\\hline
34b.1 & 3.8347188132e-04 & 5.4186540040e-08 & 3.8334567434e-04 & 1.5249943404e-04\\\hline
34b.2 & 3.8347404858e-04 & 5.4186983295e-08 & 3.8334567434e-04 & 1.5249943404e-04\\\hline
34c.1 & 1.1797632408e-04 & 5.3251400390e-08 & 5.7004763355e-04 & 2.1781182812e-04\\\hline
34c.2 & 1.1797811479e-04 & 5.3250851826e-08 & 5.7004763355e-04 & 2.1781182812e-04\\\hline
35.1 & 6.6869054488e-05 & 5.0170247764e-08 & 5.1970778723e-05 & 1.9527214265e-05\\\hline
35.2 & 6.7002671662e-05 & 5.0170694460e-08 & 5.1970778723e-05 & 1.9527218948e-05\\\hline
35a.1 & 2.0845029455e-05 & 5.0124507803e-08 & 7.9109449141e-05 & 2.8753692677e-05\\\hline
35a.2 & 2.1098899846e-05 & 5.0126548633e-08 & 7.9109449141e-05 & 2.8753700333e-05\\\hline
35b.1 & 7.1121360005e-07 & 4.9969900266e-08 & 1.2101154342e-04 & 4.2706292544e-05\\\hline
35b.2 & 8.9527169654e-07 & 4.9967899709e-08 & 1.2101154342e-04 & 4.2706296705e-05\\\hline
36.1 & 9.5976873649e-06 & 4.9706234402e-08 & 1.8657127367e-04 & 6.4136274805e-05\\\hline
36.2 & 9.5517018914e-06 & 4.9705357226e-08 & 1.8657129228e-04 & 6.4136274803e-05\\\hline
    \end{tabular}
    \caption{Relative accuracy with which different sum rules are satisfied. $\omega_0/J=1/3$ and $G/J=1/\sqrt{3}$.}
    \label{Tab:omega_0.33_G_0.577_deltas}
\end{table}

\begin{table}[htbp!]
    \centering
    \begin{tabular}{c|c|c|c|c|c|c|c|c}
        Index & $\omega_0/J$ & $G/J$ & $T/J$ & $N$ & $D$ & $\langle H_\mathrm{e}\rangle/J$ & 1st moment & 2nd moment\\
        \hline\hline
        37.1 & $\displaystyle{\frac{1}{3}}$ & $\displaystyle{\sqrt{\frac{2}{3}}}$ & 1 & 7 & 11 & -9.3815140221e-01 & 6.5422427647e+00 & 3.9687564660e+01\\\hline
        37.2 & $\displaystyle{\frac{1}{3}}$ & $\displaystyle{\sqrt{\frac{2}{3}}}$ & 1 & 7 & 12 & -9.3815140181e-01 & 6.5422427909e+00 & 3.9687565492e+01\\\hline
        37a.1 & $\displaystyle{\frac{1}{3}}$ & $\displaystyle{\sqrt{\frac{2}{3}}}$ & $10^{0.1}$ & 7 & 11 & -8.5908367901e-01 & 6.8706465963e+00 & 4.5409558993e+01\\\hline
        37a.2 & $\displaystyle{\frac{1}{3}}$ & $\displaystyle{\sqrt{\frac{2}{3}}}$ & $10^{0.1}$ & 7 & 12 & -8.5908367898e-01 & 6.8706465992e+00 & 4.5409559105e+01\\\hline
        37b.1 & $\displaystyle{\frac{1}{3}}$ & $\displaystyle{\sqrt{\frac{2}{3}}}$ & $10^{0.2}$ & 7 & 11 & -7.7117043040e-01 & 7.1550506370e+00 & 5.2049213099e+01\\\hline
        37b.2 & $\displaystyle{\frac{1}{3}}$ & $\displaystyle{\sqrt{\frac{2}{3}}}$ & $10^{0.2}$ & 7 & 12 & -7.7117043040e-01 & 7.1550506373e+00 & 5.2049213113e+01\\\hline
        38.1 & $\displaystyle{\frac{1}{3}}$ & $\displaystyle{\sqrt{\frac{2}{3}}}$ & 2 & 7 & 11 &  -6.7746394970e-01 & 7.3900039310e+00 & 5.9920304436e+01\\\hline
        38.2 & $\displaystyle{\frac{1}{3}}$ & $\displaystyle{\sqrt{\frac{2}{3}}}$ & 2 & 7 & 12 & -6.7746394970e-01 & 7.3900039311e+00 & 5.9920304438e+01\\\hline
        38a.1 & $\displaystyle{\frac{1}{3}}$ & $\displaystyle{\sqrt{\frac{2}{3}}}$ & $10^{0.4}$ & 7 & 11 &  -5.8564996859e-01 & 7.5689062267e+00 & 6.9093110557e+01\\\hline
        38a.2 & $\displaystyle{\frac{1}{3}}$ & $\displaystyle{\sqrt{\frac{2}{3}}}$ & $10^{0.4}$ & 7 & 12 &  -5.8564996859e-01 & 7.5689062267e+00 & 6.9093110557e+01\\\hline
        38b.1 & $\displaystyle{\frac{1}{3}}$ & $\displaystyle{\sqrt{\frac{2}{3}}}$ & $10^{0.5}$ & 6 & 14 &  -4.9722541725e-01 & 7.7031070036e+00 & 8.0273626875e+01\\\hline
        38b.2 & $\displaystyle{\frac{1}{3}}$ & $\displaystyle{\sqrt{\frac{2}{3}}}$ & $10^{0.5}$ & 6 & 15 &  -4.9722541725e-01 & 7.7031070036e+00 & 8.0273626875e+01\\\hline
        38c.1 & $\displaystyle{\frac{1}{3}}$ & $\displaystyle{\sqrt{\frac{2}{3}}}$ & $10^{0.6}$ & 6 & 14 &  -4.1623249363e-01 & 7.7997832417e+00 & 9.3972749342e+01\\\hline
        38c.2 & $\displaystyle{\frac{1}{3}}$ & $\displaystyle{\sqrt{\frac{2}{3}}}$ & $10^{0.6}$ & 6 & 15 &  -4.1623249363e-01 & 7.7997832417e+00 & 9.3972749342e+01\\\hline
        39.1 & $\displaystyle{\frac{1}{3}}$ & $\displaystyle{\sqrt{\frac{2}{3}}}$ & 5 & 6 & 14 &  -3.4517680544e-01 & 7.8664244314e+00 & 1.1071854377e+02\\\hline
        39.2 & $\displaystyle{\frac{1}{3}}$ & $\displaystyle{\sqrt{\frac{2}{3}}}$ & 5 & 6 & 15 &  -3.4517680544e-01 & 7.8664244314e+00 & 1.1071854377e+02\\\hline
    \end{tabular}
    \caption{$\omega_0/J=1/3$ and $G/J=\sqrt{2/3}$.}
    \label{Tab:omega_0.33_G_0.816}
\end{table}

\begin{table}[htbp!]
    \centering
    \begin{tabular}{c|c|c|c|c}
        Index & $\delta_\mathrm{OSR}$ & $\delta_\mathrm{0}$ & $\delta_1$ & $\delta_2$\\
        \hline\hline
37.1 & 3.0932965304e-03 & 1.3348611246e-08 & 1.9578481388e-05 & 9.2149467689e-06\\\hline
37.2 & 3.0234339773e-03 & 1.3459400941e-08 & 1.9508581633e-05 & 9.1497568077e-06\\\hline
37a.1 & 1.1449952983e-03 & 2.6181889334e-08 & 4.2842493036e-04 & 1.9739735470e-04\\\hline
37a.2 & 1.1506759399e-03 & 2.6177275669e-08 & 4.2842281596e-04 & 1.9739807233e-04\\\hline
37b.1 & 3.9306627253e-04 & 3.9537752631e-08 & 5.8890794018e-04 & 2.6730570236e-04\\\hline
37b.2 & 3.8765570892e-04 & 3.9481587202e-08 & 5.8891179334e-04 & 2.6731094220e-04\\\hline
38.1 & 1.7411756189e-04 & 5.0138749380e-08 & 5.0333053019e-05 & 2.3119411072e-05\\\hline
38.2 & 1.7447787347e-04 & 5.0138725074e-08 & 5.0333053019e-05 & 2.3119393755e-05\\\hline
38a.1 & 5.3345072958e-05 & 5.0059109895e-08 & 7.1166897846e-05 & 3.1930865675e-05\\\hline
38a.2 & 5.3419628964e-05 & 5.0060057398e-08 & 7.1166902742e-05 & 3.1930882838e-05\\\hline
38b.1 & 3.6250995335e-04 & 4.9875007120e-08 & 1.0263406888e-04 & 4.4758218212e-05\\\hline
38b.2 & 3.6251528194e-04 & 4.9875147761e-08 & 1.0263405925e-04 & 4.4758214519e-05\\\hline
38c.1 & 1.3772610090e-04 & 4.9638388170e-08 & 1.5031915347e-04 & 6.3470818513e-05\\\hline
38c.2 & 1.3772465420e-04 & 4.9637238863e-08 & 1.5031915347e-04 & 6.3470818513e-05\\\hline
39.1 & 3.6720700561e-05 & 4.9215806717e-08 & 2.2232899861e-04 & 9.0684076657e-05\\\hline
39.2 & 3.6722693322e-05 & 4.9213829817e-08 & 2.2232901745e-04 & 9.0684076657e-05\\\hline
    \end{tabular}
    \caption{Relative accuracy with which different sum rules are satisfied. $\omega_0/J=1/3$ and $G/J=\sqrt{2/3}$.}
    \label{Tab:omega_0.33_G_0.816_deltas}
\end{table}

\begin{table}[htbp!]
    \centering
    \begin{tabular}{c|c|c|c|c|c|c|c|c}
        Index & $\omega_0/J$ & $G/J$ & $T/J$ & $N$ & $D$ & $\langle H_\mathrm{e}\rangle/J$ & 1st moment & 2nd moment\\
        \hline\hline
        49.1 & 3 & $\sqrt{6}$ & 2 & 10 & 6 & -7.8094846825e-01 & 3.6424211544e+00 & 2.5178008432e+01\\\hline
        49.2 & 3 & $\sqrt{6}$ & 2 & 10 & 7 & -7.8094845896e-01 & 3.6424220256e+00 & 2.5178035758e+01\\\hline
        49a.1 & 3 & $\sqrt{6}$ & $10^{0.4}$ & 10 & 6 & -6.5914832023e-01 & 3.7534387477e+00 & 2.9138888190e+01\\\hline
        49a.2 & 3 & $\sqrt{6}$ & $10^{0.4}$ & 10 & 7 & -6.5914831827e-01 & 3.7534390162e+00 & 2.9138898605e+01\\\hline
        49b.1 & 3 & $\sqrt{6}$ & $10^{0.5}$ & 8 & 8 & -5.4785020392e-01 & 3.8342402373e+00 & 3.4203236375e+01\\\hline
        49b.2 & 3 & $\sqrt{6}$ & $10^{0.5}$ & 8 & 9 & -5.4785020392e-01 & 3.8342402373e+00 & 3.4203236377e+01\\\hline
        49c.1 & 3 & $\sqrt{6}$ & $10^{0.6}$ & 7 & 11 & -4.5039301904e-01 & 3.8903769505e+00 & 4.0619578334e+01\\\hline
        49c.2 & 3 & $\sqrt{6}$ & $10^{0.6}$ & 7 & 12 & -4.5039301904e-01 & 3.8903769505e+00 & 4.0619578334e+01\\\hline
        50.1 & 3 & $\sqrt{6}$ & 5 & 7 & 11 & -3.6794218160e-01 & 3.9280719065e+00 & 4.8648769203e+01\\\hline
        50.2 & 3 & $\sqrt{6}$ & 5 & 7 & 12 & -3.6794218160e-01 & 3.9280719065e+00 & 4.8648769203e+01\\\hline
        50a.1 & 3 & $\sqrt{6}$ & $10^{0.8}$ & 7 & 11 & -2.9739181195e-01 & 3.9536416909e+00 & 5.9014140987e+01\\\hline
        50a.2 & 3 & $\sqrt{6}$ & $10^{0.8}$ & 7 & 12 & -2.9739181195e-01 & 3.9536416909e+00 & 5.9014140987e+01\\\hline
        50b.1 & 3 & $\sqrt{6}$ & $10^{0.9}$ & 7 & 11 & -2.3973240843e-01 & 3.9701859184e+00 & 7.1988347347e+01\\\hline
        50b.2 & 3 & $\sqrt{6}$ & $10^{0.9}$ & 7 & 12 & -2.3973240843e-01 & 3.9701859184e+00 & 7.1988347347e+01\\\hline
        51.1 & 3 & $\sqrt{6}$ & 10 & 5 & 18 & -1.9257077468e-01 & 3.9807615513e+00 & 8.8354563713e+01\\\hline
        51.2 & 3 & $\sqrt{6}$ & 10 & 5 & 19 & -1.9257077468e-01 & 3.9807615513e+00 & 8.8354563713e+01\\\hline
    \end{tabular}
    \caption{$\omega_0/J=3$ and $G/J=\sqrt{3}$.}
    \label{Tab:omega_3_G_1.732}
\end{table}

\begin{table}[htbp!]
    \centering
    \begin{tabular}{c|c|c|c|c}
        Index & $\delta_\mathrm{OSR}$ & $\delta_\mathrm{0}$ & $\delta_1$ & $\delta_2$\\
        \hline\hline
49.1 & 5.1024502565e-07 & 5.9978658182e-07 & 4.9668533877e-06 & 3.5084509275e-06\\\hline
49.2 & 1.8655382625e-06 & 5.9981923800e-07 & 4.9670280103e-06 & 3.5087035748e-06\\\hline
49a.1 & 7.4936405442e-07 & 5.9982127642e-07 & 6.1605456627e-06 & 3.9229692500e-06\\\hline
49a.2 & 1.5076645051e-06 & 5.9977091879e-07 & 6.1606852893e-06 & 3.9231964372e-06\\\hline
49b.1 & 2.5706638445e-07 & 1.3313572115e-07 & 8.4477136106e-06 & 4.0837880329e-06\\\hline
49b.2 & 3.3595087862e-07 & 1.3313613601e-07 & 8.4477069662e-06 & 4.0837758443e-06\\\hline
49c.1 & 7.2951308075e-06 & 1.3312868268e-07 & 1.1176359807e-05 & 5.0175534842e-06\\\hline
49c.2 & 7.3663193707e-06 & 1.3312798543e-07 & 1.1176359807e-05 & 5.0175530281e-06\\\hline
50.1 & 8.9478755777e-07 & 1.3313448762e-07 & 1.5216704025e-05 & 6.3857093198e-06\\\hline
50.2 & 5.6265355569e-07 & 1.3313389819e-07 & 1.5216705204e-05 & 6.3857108432e-06\\\hline
50a.1 & 3.1024261291e-06 & 2.3174476710e-07 & 2.1795146392e-05 & 8.0983313176e-06\\\hline
50a.2 & 7.4611617289e-06 & 2.3174606661e-07 & 2.1795125304e-05 & 8.0982773165e-06\\\hline
50b.1 & 4.4733891460e-06 & 2.3178747837e-07 & 3.1136407522e-05 & 1.1186150460e-05\\\hline
50b.2 & 6.8414154867e-06 & 2.3178722317e-07 & 3.1136412189e-05 & 1.1186162814e-05\\\hline
51.1 & 1.9531718417e-04 & 1.3308931182e-07 & 4.5012060368e-05 & 1.6207447213e-05\\\hline
51.2 & 1.9620343870e-04 & 1.3308943001e-07 & 4.5012060368e-05 & 1.6207450569e-05\\\hline
    \end{tabular}
    \caption{Relative accuracy with which different sum rules are satisfied. $\omega_0/J=3$ and $G/J=\sqrt{3}$.}
    \label{Tab:omega_3_G_1.732_deltas}
\end{table}

\begin{table}[htbp!]
    \centering
    \begin{tabular}{c|c|c|c|c|c|c|c|c}
        Index & $\omega_0/J$ & $G/J$ & $T/J$ & $N$ & $D$ & $\langle H_\mathrm{e}\rangle/J$ & 1st moment & 2nd moment\\
        \hline\hline
        53a & 3 & $\sqrt{6}$ & $10^{0.4}$ & 10 & 8 & -5.8858232456e-01 & 7.5712894837e+00 & 7.3583530924e+01\\\hline
        53b.1 & 3 & $\sqrt{6}$ & $10^{0.5}$ & 7 & 9 & -4.9852659518e-01 & 7.7047254159e+00 & 8.3907162108e+01\\\hline
        53b.2 & 3 & $\sqrt{6}$ & $10^{0.5}$ & 7 & 10 & -4.9852659518e-01 & 7.7047254162e+00 & 8.3907162130e+01\\\hline
        53c.1 & 3 & $\sqrt{6}$ & $10^{0.6}$ & 7 & 10 & -4.1680546820e-01 & 7.8005574708e+00 & 9.6890230431e+01\\\hline
        53c.2 & 3 & $\sqrt{6}$ & $10^{0.6}$ & 7 & 11 & -4.1680546820e-01 & 7.8005574708e+00 & 9.6890230431e+01\\\hline
        54.1 & 3 & $\sqrt{6}$ & 5 & 7 & 11 & -3.4542505663e-01 & 7.8667820499e+00 & 1.1305788393e+02\\\hline
        54.2 & 3 & $\sqrt{6}$ & 5 & 7 & 12 & -3.4542505663e-01 & 7.8667820499e+00 & 1.1305788393e+02\\\hline
        54a.1 & 3 & $\sqrt{6}$ & $10^{0.8}$ & 7 & 11 & -2.8264307378e-01 & 7.9128299716e+00 & 1.3386788267e+02\\\hline
        54a.2 & 3 & $\sqrt{6}$ & $10^{0.8}$ & 7 & 12 & -2.8264307378e-01 & 7.9128299716e+00 & 1.3386788267e+02\\\hline
        54b.1 & 3 & $\sqrt{6}$ & $10^{0.9}$ & 7 & 11 & -2.3013541190e-01 & 7.9432475515e+00 & 1.5987040880e+02\\\hline
        54b.2 & 3 & $\sqrt{6}$ & $10^{0.9}$ & 7 & 12 & -2.3013541190e-01 & 7.9432475515e+00 & 1.5987040880e+02\\\hline
        55.1 & 3 & $\sqrt{6}$ & $10$ & 5 & 20 & -1.8637475540e-01 & 7.9630190253e+00 & 1.9264007414e+02\\\hline
        55.2 & 3 & $\sqrt{6}$ & $10$ & 5 & 21 & -1.8637475540e-01 & 7.9630190253e+00 & 1.9264007414e+02\\\hline
        \end{tabular}
    \caption{$\omega_0/J=3$ and $G/J=\sqrt{6}$.}
    \label{Tab:omega_3_G_1.1.449}
\end{table}

\begin{table}[htbp!]
    \centering
    \begin{tabular}{c|c|c|c|c}
        Index & $\delta_\mathrm{OSR}$ & $\delta_\mathrm{0}$ & $\delta_1$ & $\delta_2$\\
        \hline\hline
53a & 8.0452868743e-06 & 8.7853691769e-08 & 1.5250372914e-06 & 6.4345091746e-07\\\hline
53b.1 & 2.2149857760e-05 & 1.8522719156e-07 & 2.0466243450e-06 & 7.1815728622e-07\\\hline
53b.2 & 1.8887467038e-05 & 1.8522696656e-07 & 2.0466139743e-06 & 7.1812587491e-07\\\hline
53c.1 & 6.2859744317e-07 & 1.8519640415e-07 & 2.6845394070e-06 & 9.6802884524e-07\\\hline
53c.2 & 5.4585438589e-06 & 1.8519675633e-07 & 2.6845643468e-06 & 9.6807741701e-07\\\hline
54.1 & 9.6798384000e-07 & 1.8519385024e-07 & 3.6347246613e-06 & 1.3285081678e-06\\\hline
54.2 & 9.6660290065e-06 & 1.8519442030e-07 & 3.6346946322e-06 & 1.3284435985e-06\\\hline
54a.1 & 2.5409969524e-06 & 1.8518592491e-07 & 5.1061675521e-06 & 1.8701205203e-06\\\hline
54a.2 & 2.5971819356e-06 & 1.8518557745e-07 & 5.1061681375e-06 & 1.8701224580e-06\\\hline
54b.1 & 2.9012961688e-06 & 1.8520986683e-07 & 7.3332789798e-06 & 2.6672458864e-06\\\hline
54b.2 & 2.7135339947e-06 & 1.8521004903e-07 & 7.3332789798e-06 & 2.6672451910e-06\\\hline
55.1 & 1.8578575189e-04 & 1.8521146933e-07 & 1.0751310932e-05 & 3.8609713518e-06\\\hline
55.2 & 1.8746672634e-04 & 1.8521079673e-07 & 1.0751306279e-05 & 3.8609675047e-06\\\hline
    \end{tabular}
    \caption{Relative accuracy with which different sum rules are satisfied. $\omega_0/J=3$ and $G/J=\sqrt{6}$.}
    \label{Tab:omega_3_G_2.449_deltas}
\end{table}

\bibliography{apssamp}